\documentclass[amsmath,amssymb,reprint]{revtex4-1} 

\usepackage[T1]{fontenc}
\usepackage{graphicx}
\usepackage{float}
\usepackage{epsf} 
\usepackage{physics}
\usepackage{amsfonts}
\usepackage{amsthm}
\usepackage{enumitem}
\graphicspath{{figs/}}

\newcommand{\reff}[1]{(\ref{#1})}
\newcommand{\lt}{\mathopen{}\mathclose\bgroup\left}
\newcommand{\rt}{\aftergroup\egroup\right}
\newcommand{\f}[2]{\ensuremath{\frac{#1}{#2}}}
\newcommand{\e}{\ensuremath{\mathrm{e}}}
\DeclareMathOperator*{\pr}{P}

\newcommand{\deq}{\ensuremath{\overset{d}{=}}}
\newcommand{\defeq}{\mathrel{\mathop:}=}
\DeclareMathOperator{\erfc}{erfc}

\begin{document}

\title{From diffusion in compartmentalized media to non-Gaussian random walks}

\author{Jakub \'Sl\k{e}zak} \email{Corresponding author: jakub.slezak@pwr.edu.pl}
\author{Stanislav Burov} \email{stasbur@gmail.com}
\affiliation{Physics Department, Bar-Ilan University, Ramat Gan 5290002, Israel}

\date{\today}
\begin{abstract}

\textbf{ABSTRACT.} In this work we establish a link between two different phenomena that were studied in a large and growing number of biological, composite and soft media: the diffusion in compartmentalized environment and the non-Gaussian diffusion that exhibits linear or power-law growth of the mean square displacement joined by the exponential shape of the positional probability density. We explore a microscopic model that gives rise to transient confinement, similar to the one observed for \textit{hop-diffusion} on top of a cellular membrane. The compartmentalization of the media is achieved by introducing randomly placed, identical barriers. Using this model of a heterogeneous medium we derive a general class of random walks with simple jump rules that are dictated by the geometry of the compartments. Exponential decay of positional probability density is observed and we also quantify the significant decrease of the long time diffusion constant. Our results suggest that the observed exponential decay is a general feature of the transient regime in compartmentalized media.

\end{abstract}
\maketitle

\section{Introduction}
The Brownian motion is ubiquitous in applications, be it the microscopic motion of molecules \cite{feynmanBM}, search patterns of animals \cite{searchBM}, or the prices of financial options \cite{merton}. Following the classical argument of Einstein \cite{einstein}, if there exists a time scale in which the changes of the investigated variable can be treated as a result of additive, independent and homogenous fluctuations with finite second moment, then in larger time scales the observed process is the Brownian motion. This insight was later formalised as the functional central limit theorem \cite{billingsley}.

The Brownian motion became a natural start for theoretical and experimental investigations of more complicated stochastic models. For example, lowering the requirement of finite moments led to the rich theory of L{\'e}vy flights \cite{shlesinger}; lowering the assumption of independence was one of the cornerstones of the anomalous diffusion modelling \cite{guide}. The Langevin theory of diffusion investigates the time scales lower than those required by the central limit theorem \cite{coffey} which leads to the motions in which the deviations from Brownian motion appear in the memory structure; the probability density of motion is still Gaussian, only with a different scale.

In recent years the influx of experimental data proved the existence of a robust class of systems exhibiting non-Gaussian, in particular exponential, tails of the probability density \cite{wangNG,fair, bhattacharya}, together with normal or anomalous mean square displacement \cite{hapca,lampo17,wang2}. This common phenomenon is called \emph{Brownian, yet non-Gaussian diffusion} \cite{wangNG,fair}. Its raising prominence and importance in understanding the biochemical nature of the transport stimulated various attempts to provide some -- at least effective -- description. These include works using variants of the Langevin equation that use superstatististical \cite{beck2,beck3,beck4, superstatLang} or diffusing diffusivity \cite{diffDiff, diffDiffReact, diffDiffChechkin} approach. 

\begin{figure}\centering
	\includegraphics[width=0.8\columnwidth]{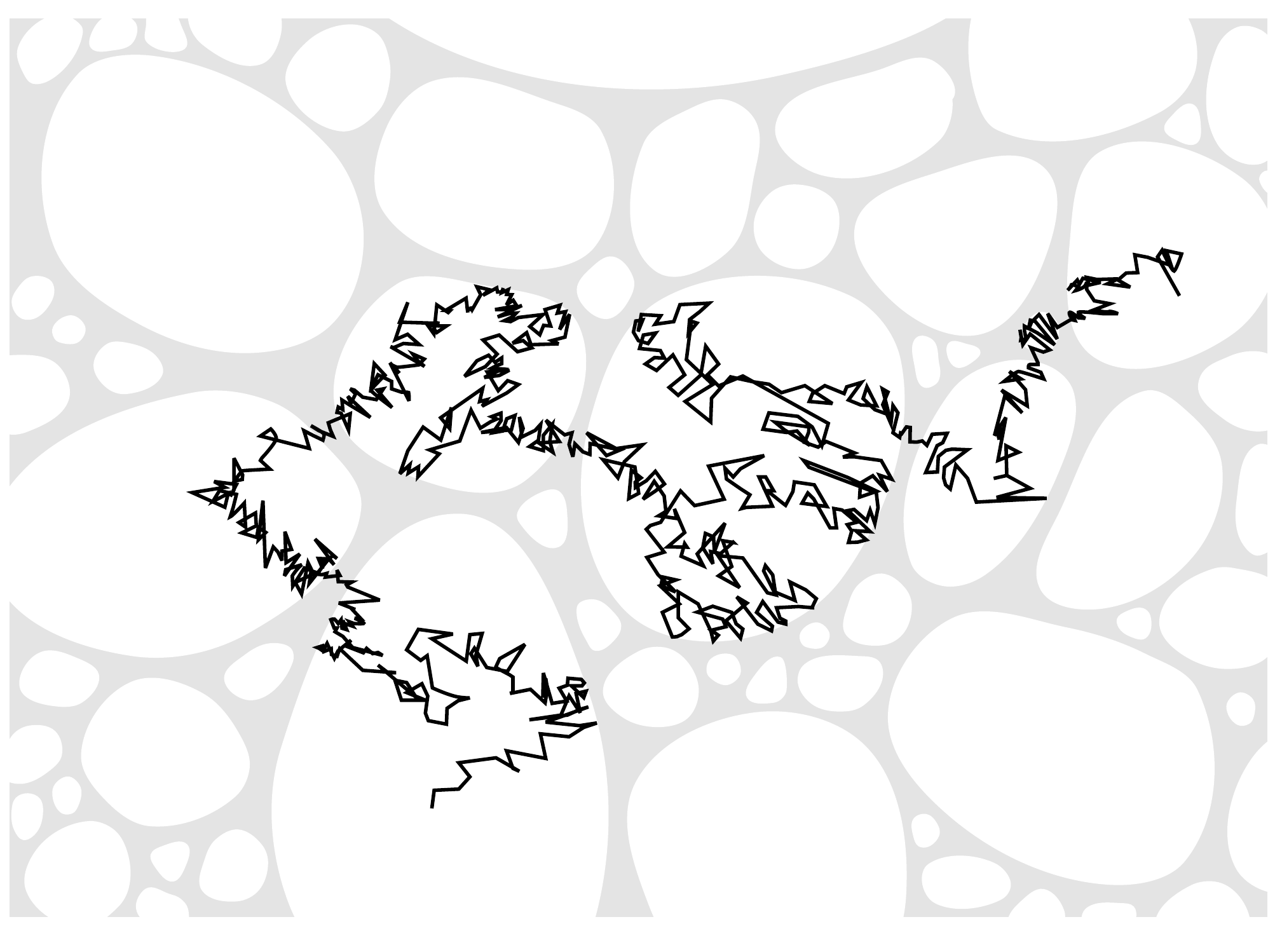}
	\caption{A schematic representation of the diffusion in a two dimensional compartmentalized medium. A particle (black trajectory) diffuses freely inside the domains (white areas) but only rarely crosses the thick divisions between them (grey area) which are thin compared to the average size of the domains and act as barriers.}\label{fig:sites}
\end{figure}

It is commonly suspected that the true source of non-Gaussianity lays in the heterogeneity of the medium \cite{diffDiffChechkin,hetero,he16, wangNG}.  It is also hypothesised that the correct description may be related to random walks with traps \cite{itoInterp,BnGPostnikov}.  Yet, formulating and solving suitable models is still a challenge far from completion. Here, we attempt to overcome it for systems in which the diffusion is locally impeded by barriers, providing a general mechanism for the appearance of the non-Gaussian diffusion; for a simplified illustration of the type of medium considered see Fig. \ref{fig:sites}.

 Such restrictions of the molecular motions are ubiquitous in nature \cite{novikov}, especially in composite or porous materials \cite{song,mair}. For example, according to the fences and pickets model the cellular membrane  is compartmentalized (likely by actin-based membrane skeleton and various transmembrane proteins \cite{confDiff,krapfComp}), a fact essential in understanding its structure and  functions \cite{flow, kasumi}. It is also widely agreed that this phenomenon is crucial in determining the material properties of the biological media, the transport of proteins, lipids and their functions \cite{sykova,cory,yablonskiy, flow}. 
The \textit{transient confinement} caused by these obstacles should result in a motion similar to a random walk, with a particle jumping between adjacent domains. This phenomenon was experimentally observed and termed \textit{hop diffusion} \cite{krapfComp2,fujiwara}. Its characteristic property is that the long-time macroscopic \textit{effective diffusion coefficient} is greatly reduced compared to the short-time in-compartment diffusion coefficient with their ratio providing important insight into the molecular properties of the system. This ratio was found to be between 10 and 100 in various experiments, see the list in \cite{murase}.  Similar results were found for porous materials \cite{latour, sen,diffEscape} and for  colloidal particles near the glass transition; in the last case what they call the \textit{cage effect} was observed together with the exponential tails of the probability density \cite{weeksCages,chaudhuri, weeks2}.

In what follows, we first make a universal observation on how the confinement results in the static form of non-Gaussianity (Section II.). The argument is quite general and shows the time scales at which the increments of the diffusing particle's position exhibit exponential tails. It does not specify the dynamics, but provides a single jump's distribution, which is a basic building block of the random walk models which we derive. We study them using one dimensional model of the system. We start from reducing the local diffusivity model to the diffusion equation with interface conditions (Section III.), which we solve domain-wise in order to derive the transition times of the jumps between domains. Consequently, we manage to describe the diffusion using environment-dependent random walk (Section IV.). It can be effectively approximated using much simpler model of continuous time random walk (CTRW), yielding the formula which links the effective diffusion coefficient to the barriers' permeability (Section V.). Next, we decouple the correlations of this process and reveal its broad-tail non-Gaussian nature using the kurtosis and the logarithm of characteristic function (Section VI.). Finally, we consider a more reductive but mathematically elegant approximating CTRW for which we calculate the exact probability density; it exhibits exponential tails characteristic for the Brownian, yet non-Gaussian diffusion. The derivations presented in Sections V. and VI. show the mathematical origin of the analytical predictions presented in Figs 8-11, but for general understanding of the results the form of the reduced model \reff{eq:redMod} and the leading exponential-tail behaviour \reff{eq:besselApprAsympt} suffice. The overview of the subject and the results is given in Section VII. We also provide a list of commonly used notation. Our simulation code is attached in the supplemental material.

\section{ Non-Gaussianity stemming from confinement}
A possible approach for understanding the prevalence of the Gaussian distribution is to look at it through the notion of entropy. For all different variables $X$ in free space with a given variance, characterised by the probability density function (PDF) $p_X$, the ones with Gaussian distributions (with different means) maximise the entropy $-\expval{\ln p_X(X)}$. Thus, if they are no other constrains (such as potential field), any coordinate coupled to a heat reservoir will converge to a Gaussian state \cite{statTher}. 

This naturally opens up the question in what conditions the Gaussianity is \emph{not} to be expected. Following the entropic argument, the simplest such case is any bounded domain in which the entropy is maximized by the uniform, not Gaussian distribution. Physically speaking, it corresponds to particles being confined in some finite area by a reflecting barrier. Such a perfect local confinement excludes the diffusion in the macroscopic scale, however in a more detailed setting one can think about a system with dynamics dominated by two time scales: the relaxation time $t_r$ of reaching the domain-wise stationary uniform state, and the average escape time $t_e\gg t_r$ of passing through the barrier. In this setting, for $t\ll t_r$ the scale of the dynamics is too small to be affected by the confinement. For $t\gg t_e$ as long as there are no ``hard'' traps it is expected that the system will be again homogenised and Brownian (even for the cases when the intra-domain dynamics was not, e.g. was subdiffusive). For the time range in-between, the motion is dominated by the confinement in one ($t\approx t_r$) or few ($t\approx t_e$) domains and always non-Gaussian. This time scale will be in the centre of our discussion.

For $t<t_e$ the situation is simple: for a given particle its distribution will be a uniform one with the shape reflecting the domain it was found in. For the whole ensemble the distribution will be a mixture of uniform distributions reflecting the domain shapes all over the medium. As we can see we are in the situation in which the observed ensemble distribution is non-Gaussian and completely determined by the random geometry of the system, to some extent independently form the details of the dynamics.

This geometry is clearly hard to determine for many real systems and may be very complex, e.g. porous media are often fractal-like. However, in many media such as cellar membranes it is reasonable to assume that the domains are mostly spherical and they vary mostly in size. Furthermore, the simplest size distribution to consider is the exponential one. The argument for this is the lack of dependence. We consider one dimensional line in the medium and ask if in any given $\dd{x}$ it crosses the barrier or not. If the crossing in $\dd{x}$ does not affect the distribution of other crossings (they do not ``see'' each other) they must be located according to the Poisson point measure \cite{kallenberg} and the distances between any two subsequent barriers on this line are exponential.

We are now ready to calculate the stationary PDF which will be observed for the whole ensemble. Let $L$ denote the diameter of a given domain. By choosing proper displacement units we can always assume that $\expval{L}=1$ and its distribution of the sizes is a simple $p_L(l)=\exp(-l)$. However, this is not the proper PDF to average over as we need to account for the average number of particles in each domain. If the domains are not correlated to any kind of a trap, this number will be proportional to the volume of the domain. This leads to the equilibrium PDF \cite{cox} $ p_L^\text{eq}(l)=\ c_d l^d\exp(-l), c_d =1,1/2,1/6$ in dimensions $d=1,2,3$ respectively (these are gamma distributions $\mathcal Gam(d+1,1)$, see the Notation section). 

\begin{figure}
	\includegraphics[width=0.7\columnwidth]{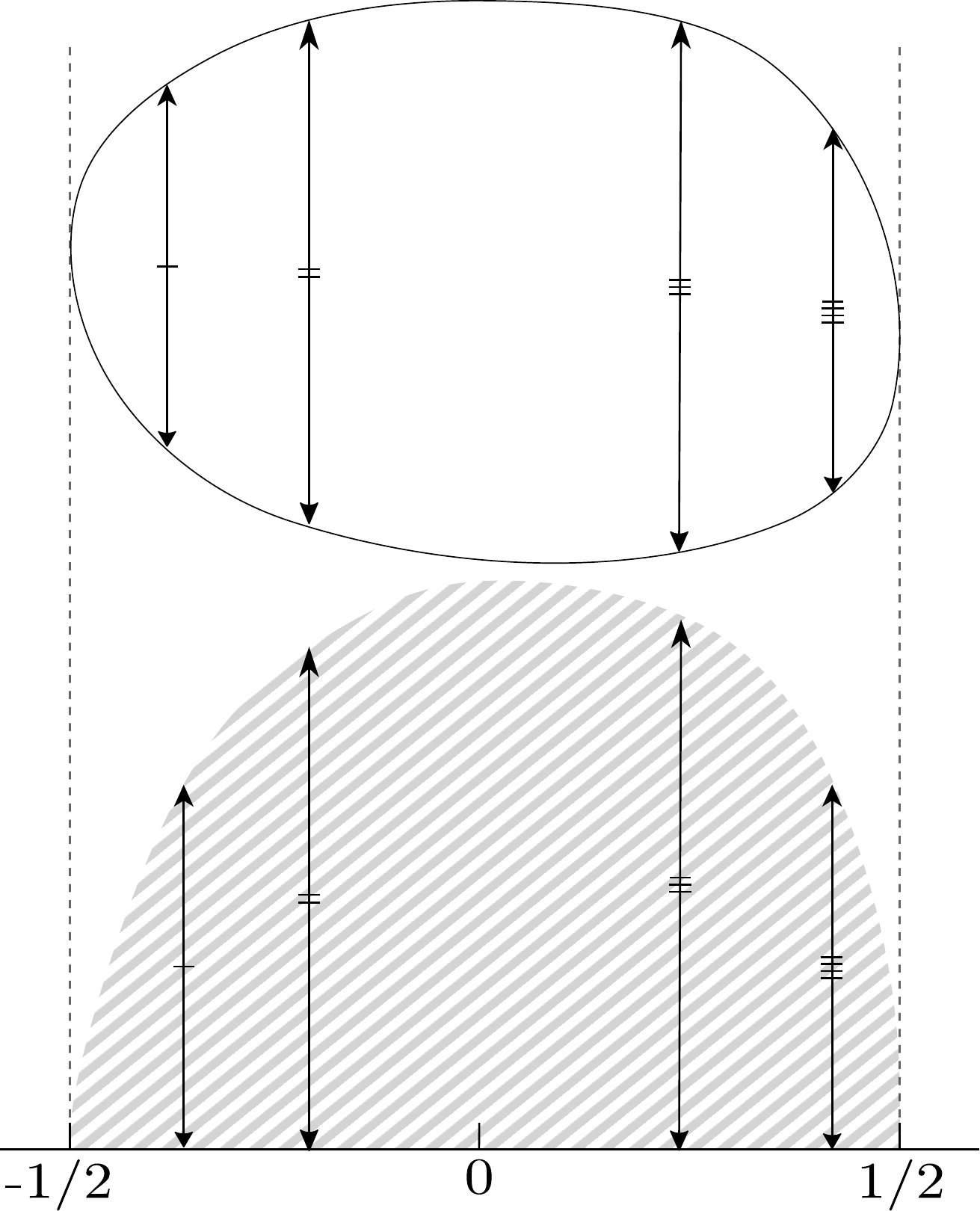}
	\caption{Projection procedure applied to an exemplary domain. All the probability mass in two dimensions (the oval shape up) is projected onto a given axis, i.e. moved down leaving lengths of the depicted arrows intact, resulting in one dimensional mass (schematically shown as grey area below); then it is renormalized.}\label{fig:proj}
\end{figure}

For a given domain with a fixed $L$ the distribution is uniform, for one dimension it is a simple $p_X(x|L)=1/L, |x|<L/2$, which yields
\begin{equation}\label{eq:pdfElementary}
	p_X(x) = \int_{|x|<l/2} \!\!\!\dd{l} \f{1}{l}p^\text{eq}_L(l) = \int_{2|x|}^\infty\!\!\dd{l}\e^{-l} = \e^{-2|x|}.
\end{equation}
This simple argument shows how the geometry of the system determines the distribution of $L$ and is reflected in the stationary PDF of particles' displacements, which is found to be the Laplace (also called ``two sided exponential'') distribution with mean 0 and variance $1/2$, symbolically $\mathcal Lap(0,1/2)$. For two and more dimensions the procedure is slightly more complicated: for each coordinate $x,y$ or $z$ we will observe only the projection of the total multidimensional probability mass, see Fig. \ref{fig:proj}. For spheres these are semicircular law for two dimensions, and parabolic law for three dimensions (as for each $\dd{x}$ we squeeze onto it a circle with surface $ \pi((1/2)^2-x^2)$). The projections and averaged PDFs are shown in the first and second row of Table \ref{tab:conf}.
\begin{table}[!h]
	\bgroup
	\def\arraystretch{2}
	\begin{tabular}{ r |c| c |c |}
		\multicolumn{1}{r}{} & \multicolumn{1}{c}{1D} & \multicolumn{1}{c}{2D} & \multicolumn{1}{c}{3D}\\
		\cline{2-4}
		$p_X(x|L)$ & $\f{1}{L}$ & $\f{4}{\pi L}\sqrt{1-\lt(\f{2x}{L}\rt)^2}$ & $\f{3}{2L}\lt(1-\lt(\f{2x}{L}\rt)^2\rt)$\\
		\cline{2-4}
		$p_X(x)$ & $\e^{-2|x|}$ & $\f{4|x|}{\pi}K_1(2|x|)$ & $\f{1}{4}(1+2|x|)\e^{-2|x|}$\\
		\cline{2-4}
		$p_{\Delta X}(x)$ & $\sim \e^{-|x|}$ & $\sim \f{8}{3\pi}\e^{-|x|}$ & $\sim\f{6}{|x|}\e^{-|x|}$\\
		\cline{2-4}
	\end{tabular}
	\egroup
	\caption{Confined one-dimensional PDFs for the particles trapped in 1, 2, 3D spheres. The conditional (i.e. with fixed trap diameter $L$) PDFs $p_X(x|L)$ in the top row are Wigner $n$-sphere distributions. Middle and bottom rows are the PDFs averaged over the trap sizes, of the displacement $X$ and the increments $\Delta X \defeq X_{t+\Delta t}-X_t, \Delta t\gg t_r$ respectively. By $K_1$ we denote the modified Bessel function of the second kind (see \cite[Eq. 10.32.8]{DLMF}) which has tails $K_1(2|x|)\sim (\pi/x)^{1/2} \exp(-2|x|)/2$. In the bottom row we show only the asymptotics, but in the case of 1D and 3D  the exact values of $p_{\Delta X}$ can be calculated by direct integration, the formulas are complicated but may be expressed using special function Ei and polynomials.}\label{tab:conf}
\end{table}

The important thing to notice there is that they all decay like $\exp(-2|x|)$ multiplied by some power-law factor. Integrating by parts shows that this a more universal behaviour, the requirements for its occurrence are only that the PDF of the sizes has a tail $\propto l^\alpha \exp(-l)$ and the stationary PDF within a fixed domain has a cut-off at $\pm L/2$. However, there is a simplifying assumption in \reff{eq:pdfElementary} which we did not immediately discuss: the centre of the domain is located at $x=0$, whereas in typical experimental data the position of the particle at the start of the measurement is taken as 0. This can be fixed with a litte effort and the corrected PDF still has exponential tails, see the footnote \footnote{ In this more realistic scenario the left end of the initial domain lays at $\Theta L$ and the right one at $(\Theta -1) L$; variable $\Theta$ has the uniform distribution $\mathcal Unif(0,1)$. Calculation analogical to \reff{eq:pdfElementary} yields incomplete gamma PDF $p_X(x) = \Gamma(0,|x|)/(2|x|)$ which has tails $\sim \exp(-|x|)/(2|x|)$, they are even thicker than $\exp(-2|x|)$. On the other hand this distribution has logarithmic singularity at $x=0$ which make the motion more constrained at short distances.}.

The limitation of the practical usefulness of the insights made above is that the stationary distributions are visible in experiments only if many data points within the range $t_r<t<t_e$ can be measured. However, we can extend our approach if the sampling rate of the measurements is sufficiently smaller than the escape time, $t_r\ll \Delta t < t_e$. Because $\Delta t<t_e$ the subsequent values $X_t$ and $X_{t+\Delta t}$ will most often  still be inside the same domain. Consequently, the series of increments $\Delta X_t\defeq X_{t+\Delta t}-X_t$ will consist of long intervals of the in-domain differences only rarely interrupted by the jumps to the adjoining domains; these rare spikes can be then neglected. Because $t_r\ll \Delta t$, the values $X_t$ and $X_{t+\Delta t}$ will be independent (as the particle will transverse the domain multiple times during $\Delta t$), so its PDF will determined by the difference of two independent variables with $p_X(x|L)$ distribution: this is triangular distribution in one dimension, complicated but elementary polynomial distribution in three dimensions and a rather involved one in two dimensions (this is because the convolution of two $\sqrt{1-x^2}$ functions is not elementary). In any case their tails can be derived using integration by parts, see the third row of Table \ref{tab:conf}. Using the ergodicity of $\Delta X_t$, this stationary distribution can be estimated using even a single long trajectory, no ensemble averaging is required. For all the cases this PDF also exhibits exponentially decaying tails.

The limit of this line of arguments is the time scale $t>t_e$ in which the dynamics becomes important. We need to know how long it takes for the particle to jump from one domain to another and for this a more detailed model of the barriers is required. This is the subject of the following sections.

\section{Transiently confined diffusion with locally Brownian dynamics}

We imagine the heterogeneous medium as the collection of domains  with regular shapes inside which the particle moves relatively freely, separated by narrow structures composed of a thick material (e.g. actin meshwork in the case of plasma membranes \cite{confDiff, krapfComp, flow, kasumi}) which impedes the diffusion within (these are grey areas in Fig. \ref{fig:sites}). We model this behaviour by making the diffusion coefficient locally small, which results in the stochastic differential equation
\begin{equation}
\dd{X}_t = \sqrt{2D(X_t)}\dd{B_t}.
\end{equation}
governed by the Brownian increments $\dd{B_t}$. The local diffusivity function $D(x)$ appearing here is a combination of two microscopic kinematic parameters: mean free path and correlation time. As such, the equation by itself is physically ambiguous and requires interpretation. It was previously established that if $D(x)$ is to describe an environment with barriers (and no traps) present, the proper choice is \textit{kinetic H{\"a}nggi–Klimontovich interpretation} in which the PDF of the diffusion solves the Fokker-Planck equation \cite{sokolovInterp}
\begin{equation}\label{eq:heat}
	\pdv{t}p_X(x;t) = \nabla \cdot \big( D(x) \nabla p_X(x;t)\big).
\end{equation}
In a bounded domain with volume $V$ the constant PDF $p_X(x)=1/V$ is the unique stationary solution ($\partial p_X/\partial t=0$) of this equation, so this, and only this, interpretation agrees with our assumption of ``no other constraints'' in the medium and consequently leads to the maximal entropy distribution being uniform.

Up to this moment our considerations were quite general, from now on we will limit ourselves to the one dimensional system, which will allow us to simplify the geometry immensely and consequently obtain quite straightforward description of the dynamics. In one dimension the barriers are thin $\Delta x$ intervals with a small local diffusivity $D_b\ll D$. Let a barrier be a layer starting at $x_k$ and ending at $x_k+\Delta x$. Preservation of the number of particles forces the flux $D(x)\partial p_X/\partial x$ to be a continuous function, in particular at the borders of the barrier
\begin{align}\label{eq:DDrel}
	 D\pdv{x}p_X(x_k^-;t) &= D_b\pdv{x}p_X(x_k^+;t),\\
	 D_b\pdv{x}p_X(x_k+\Delta x^-;t) &= D\pdv{x}p_X(x_k+\Delta x^+;t).\nonumber
\end{align}
When $D_b\ll D$ these conditions cause $p_X$ to change rapidly inside the barrier, though this function must still be continuous. For the flux to be also continuous the derivative $\partial p_X/\partial x$  must be discontinuous at $x$ and $x+\Delta x$, but with $p_X$ still being smooth inside the barrier. Now, we may express $p_X$ on the outside edges of barrier using the values inside, then make a Taylor approximation and apply \reff{eq:DDrel}
\begin{align} \label{eq:interfDer}
&p_X(x_k+\Delta x^+;t) - p_X(x_k^-;t) \nonumber\\
	 =\ &p_X(x_k+\Delta x^-;t) - p_X(x_k^+;t) \nonumber\\
	 =\ &\Delta x \pdv{x}p_X(x_k^+;t) +\order{\Delta x^2}\nonumber\\
	 =\ &D \f{\Delta x}{D_b}\pdv{x}p_X(x_k^+;t) +\order{\Delta x^2}.
\end{align}
If we go to the limit $\Delta x\to 0$ and $D_b\to 0$ in such a way that $D_b/\Delta x \to \kappa D$ we end up with a diffusion determined by the heat equation within the domains
\begin{equation}\label{eq:heat2}
	\pdv{t}p_X(x;t) = D \pdv[2]{x}p_X(x;t),\quad x\neq x_k
\end{equation}
which breaks at the locations of barriers where two \textit{interface conditions} are linking the values of the PDF on their left and right side. The first one is again the flux continuity, the second one is the reduced form of \reff{eq:interfDer}
\begin{align}\label{eq:interface}
	\pdv{x}p_X(x_k^-;t) &= \pdv{x}p_X(x_k^+;t),\\
	\pdv{x}p_X(x_k;t) &= \kappa\big(p_X(x_k^+;t)-p_X(x_k^-;t)\big).\nonumber	
\end{align}
The parameter $\kappa$ here is the barrier \textit{permeability}; mind that some authors denote $\kappa' = \kappa D$ as permeability instead. This argument can be generalized to two or three dimensional domains in a straightforward manner, see \cite{vibrTher}. It is also worth adding that the same interface conditions can be derived using the barriers modelled by the bumps of potential,  but the derivation is quite technical \cite{hardMem}.  Mathematically speaking, this diffusion has a rather peculiar PDF: by making barriers' thickness negligible we caused it to exhibit finite jump discontinuities at each $x_k$ but at the same time still has continuous directional derivatives everywhere. It is smooth only in the stationary state, which (in a finite space $V$) is uniform as required, $p_X(x;t\to\infty)=1/V$.

Conditions \reff{eq:interface} also appear in chemistry \cite{erban,andrews} and  should come as no surprise: it is nothing but a stochastic version of the \emph{Newton's law of cooling}. In the most typical form it states that the heat flux at a boundary is proportional to the temperature difference, $\partial Q/\partial t \propto -(T_\text{in}-T_\text{out})$. Combining it with the Fourier's law $\partial Q/\partial t \propto - \partial T_\text{in}/\partial x$ we end up exactly with \reff{eq:interface}. For this reason its commonly called ``Newton'' or ``convection'' boundary condition. However, our case is far less typical since we look for the solution on both sides of each barrier. This is in contrast to the thermal problems in which $T_\text{out}$ is most often taken to be a fixed state of the environment. For this reason the mathematical difficulty in solving the stochastic variant increases significantly.

\begin{figure}\centering
	\includegraphics[width=1\columnwidth]{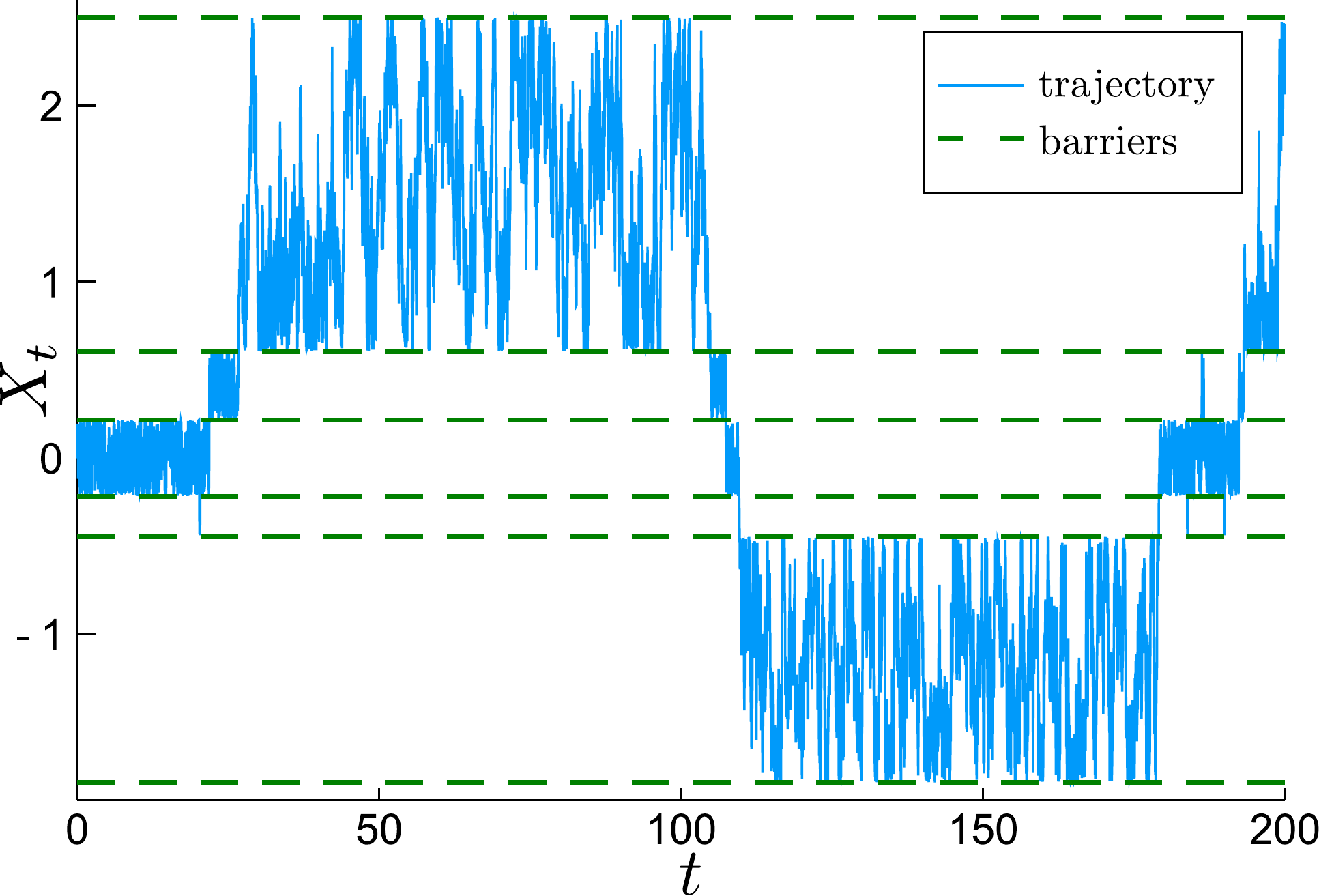}
	\caption{Exemplary simulated trajectory for $\kappa=1/50,D=1$ and discretization step $\Delta t = 0.01$. Barriers were located according to the standard Poisson random measure.}\label{fig:trajEx}
\end{figure}

If only one barrier is present, let it be at $x_0=0$, the solution is still straightforward to obtain. One only needs to split the initial condition into symmetric and antisymmetric terms, $p_X(x;0) = p_X^S(x;0)+p_X^A(x;0)= (p_X(x;0)+p_X(-x;0))/2+ (p_X(x;0)-p_X(-x;0))/2$. Then the solution itself can be split into symmetric and antisymmetric terms which evolve independently with reflective (Neumann), $\partial p_X^S(0^+,t)/\partial x = 0$, and radiation (Robin), $\partial p_X^A(0^+,t)/\partial x = 2\kappa p_X^A(0^+,t)$ boundary conditions. Both can be solved using multiple standard methods. This argument also helps to understand physical meaning of the parameter $\kappa$. For $\kappa\to 0$ the boundary conditions converge to the purely reflective one and the particle stops being able to transverse the barrier. For $\kappa \to \infty$ they reduce to the condition $p_X^A(0^+;t)=0$. But as $p_X^A$ is antisymmetric by definition, it only forces $p_X^A$ to be a continuous function; the interface conditions and the barrier disappear altogether. The in-between case of a finite $\kappa$ is a partially reflective barrier or a \textit{semi-permeable barrier}; look at Fig. \ref{fig:trajEx} for an illustration how the resulting trajectories look like.

Alas, for any system with more than one barrier we loose this convenient symmetry \footnote{In the overall valuable work of Powles \emph{et al.} \cite{powles} a system with the regularly placed barriers at $x_k \in\{ \ldots, -1,0,1,2,\ldots\}$ is considered. They claim they obtained the exact PDF for this particular case but no proof is provided and the result may be doubted.}. The heat equation and interface conditions \reff{eq:interface} are linear, but the dependence on the environment is non-linear, as expected for a heterogeneous system. For any finite number of barriers using the Fourier or Laplace transform in the position space reduces the problem to solving a system of algebraic equations, but the solutions are quite complicated and inverting the transform seems to be beyond reach for more than two barriers present.

Still, there exists a convenient stochastic representation of this diffusion which provides another layer of physical meaning and facilitates numerical simulations. In the mathematically impressive series of works Antoine Lejay derived such a representation for a system with one barrier \cite{lejaySkew,lejay,lejayMC} which extends easily to the general case. If the space would be discrete the obvious and correct representation would be a random walk for which at each visit to the barrier there is a chance of passing or being reflected, see e.g. \cite{nag,barton}. For continuous space this approach does not work directly because Brownian trajectories are so irregular they pass any threshold infinitely many times. The representation becomes valid if we restate this model as ``each time the particle spends a unit of time at a barrier there is a chance of being reflected''. Now, as we decrease the discretization mesh the occupation times converge to the \emph{local time} random field $\ell_t(x) \defeq \int_0^t\dd{\tau} \boldsymbol{1}_{X_\tau}(x)$ \cite{localTime} (this is basically a histogram calculated from the sample trajectory $\{X_t\}$). At the same time the discrete escape times which have geometric distribution converge to the smooth exponential variables.

Indeed, Lejay showed that if we take a series of independent and identically distributed (i.i.d.) random times $\tau_1,\tau_2,\ldots \deq \mathcal Exp(\kappa)$ the process which is the reflected Brownian motion on the one side of the barrier until $\ell_t(x_0)>\tau_1$, then the reflected Brownian motion on the other side until $\ell_t(x_0)>\tau_1+\tau_2$ and so on, has a PDF which solves \reff{eq:heat2} and \reff{eq:interface}. This makes sense if we come back to the derivation: for a barrier with a finite thickness $\Delta x$ the particle must explore the inside of the barrier for the time long enough it will reach the other side. For small $\Delta x$ it is no surprise the passing is a Markovian event, thus the waiting time must be exponential.

The joint distribution of the reflected Brownian motion and its local time is known, so the condition $\ell_t(x)>\tau_1$ can be directly implemented in a stochastic simulation \cite{lejay}. For multiple barriers system only the one on the left and the one on the right side of the current domain are important. Because the dynamics is Markovian and local, at each step it is only required to look for the closest barrier and check if the particle escaped through it in a given $\Delta t$ or not. This is the method used to simulate the trajectory shown in Fig. \ref{fig:trajEx}.

\section{From the transient confinement to a random walk}

It is natural to suspect that the diffusion dominated by the transient confinement is a type of random walk, which is even suggested by the term ``hop diffusion''. We will derive the random walk which corresponds to the model of the medium described in Section II.  Similar approach was used as a possible explanation for L\'evy flights \cite{levyFlight} in quenched disordered media \cite{LWquenched,superfiffQD}. However, in contrast to our system, for these phenomena the short scale motion is assumed to be ballistic and the distances between barriers have a power law distribution. 

 In our case, the process can be imagined as a particle which at each given time is localised according to a uniform distribution in the domain it is in, but after waiting a random escape time it transitions to a uniform distribution in one of the adjacent domains with the cycle starting again.

Let us consider one such domain with ends at $\pm L/2$. The escape time is determined by the solution of the diffusion equation if we remove returns to the domain, that is, after the escape event $\ell_t(\pm L/2)> \tau_1^\pm$ took place the particle becomes absorbed and cannot return. This is equivalent of putting absorbing (Dirichlet) boundary conditions just outside the domain walls. By instantly drawing out all the probability mass outside the domain this procedure reduces the interface conditions \reff{eq:interface} to the radiation boundary conditions
\begin{equation}
	\pdv{x}p_X((\pm L/2)^\mp;t) = \mp \kappa p_X((\pm L/2)^\mp;t).
\end{equation}
The resulting process is a type of \emph{partially reflected Brownian motion} \cite{grebenkovPart}. 

Using again the equivalence to thermal problems, the dynamics can be described by two dimensionless parameters: the Fourier number $Dt/L^2$ regulating the ratio of diffusive to transport motion and the Biot number $\kappa_B = \kappa L$ describing the ratio of heat resistance inside to that of the surface. Systems with small Biot number have uniform temperature, which is precisely our assumption of the dominance of local equilibria, which can now be specified to be the requirement that $\kappa_B\ll 1$. There is a nuance here as $L$ and thus $\kappa_B$ are domain-dependent, but as the distribution of $L$ has short tails, extreme values of $\kappa_B$ are improbable and we may just require that the average Biot number is small, $\expval{\kappa_B}\ll 1$.

\begin{figure}\centering
	\includegraphics[width=1\columnwidth]{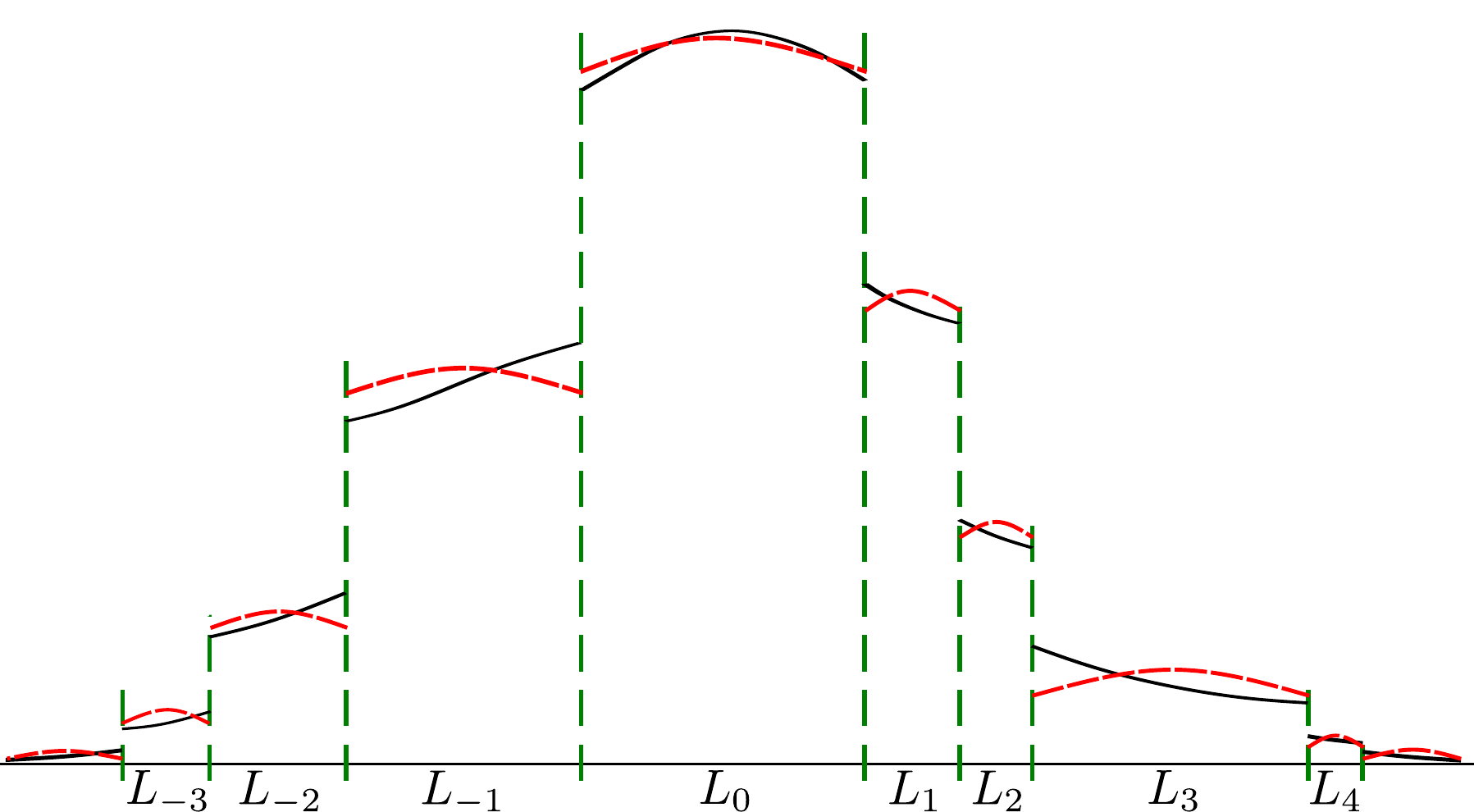}
	\caption{Schematic comparison of the original PDF (black lines) and the base mode approximation (red lines) with gradients exaggerated; for small $\kappa$ the functions are close to a step function with jumps at the barriers (green lines). The approximation has the shape $A_k(t)\cos(\sqrt{2\kappa/L_k}x)$ inside each domain, centred for the middle. The cosines are then approximated by ones, so the PDF is given by the amplitudes $A_k(t)$. The random walk approximation is equivalent to stating that $A_k$ evolve according to the master equation $\dd A_k/\dd t =2\kappa DL_k^{-1} (-A_k+A_{k-1}/2+A_{k+1}/2)$; one can also think about it as a trap model \cite{bouchaud}, though with traps having varying sizes and exponential waiting times.}\label{fig:pdfEx}
\end{figure}

By rescaling $t$ and $x$ the equation can be reduced to the dimensionless
\begin{align}
	\pdv{t}u(x;t) &= \pdv[2]{x}u(x;t),\nonumber\\
	 \pdv{x}u((\pm 1/2)^\mp;t) &= \mp \kappa_B u((\pm 1/2)^\mp;t)
\end{align}
which we expand into a series of eigenfunctions
\begin{equation}
	u = \sum_{n=0}^\infty a_n \e^{-\lambda_n^2 t}\cos(\lambda_n x)+ \sum_{n=0}^\infty b_n \e^{-\beta_n^2 t}\sin(\beta_n x).
\end{equation}
The eigenvalues solve $\tan(\lambda_n/2) = \kappa_B/\lambda_n$ and $\cot(\beta_n/2) = -\kappa_B/\beta_n$. For small $\kappa_B$ we can expand the trigonometric functions around their zeros and obtain
\begin{align}
	\lambda_n &= 2n\pi+\f{2\kappa_B}{\sqrt{n^2\pi^2+2\kappa_B}+n\pi}+\order{\kappa_B^2},\nonumber\\
	\beta_n & = \lambda_n+\pi,\qquad n\in\{0,1,2,\ldots\}.
\end{align}
Due to our assumptions we expect the distribution to be mostly uniform, and it truly is: there is a spectral gap between the smallest eigenvalue $\lambda_0 = \sqrt{2\kappa_B}+\order{\kappa_B^2}$ and all the rest, thus the corresponding eigenfunctions decay much faster. After a brief relaxation only the mode $\cos(\lambda_0 x) = 1-\order{\kappa_B}$ remains and then it also decays with the slower rate $\exp(-2\kappa_B t)$. This is the PDF of the escape times for small $\kappa_B$. After coming back to the general case with all the physical constants explicitly present this distribution is found to be $\mathcal Exp(2 \kappa D /L)$. Observe that the average time spent in the domain during a visit is proportional to its length $L$, therefore this approximation preserves the globally uniform distribution of particles. We illustrate how the resulting approximation looks like for multiple domains in Fig. \ref{fig:pdfEx}.
We are now  able specify the random walk model. First, we need to fix the locations of the domains $x_k$. In Section II. we noted that if they are independent from each other hey should be drawn from the stationary (i.e. translationally invariant) Poisson point measure. 

\begin{figure*}\centering
	\includegraphics[width=0.32\textwidth]{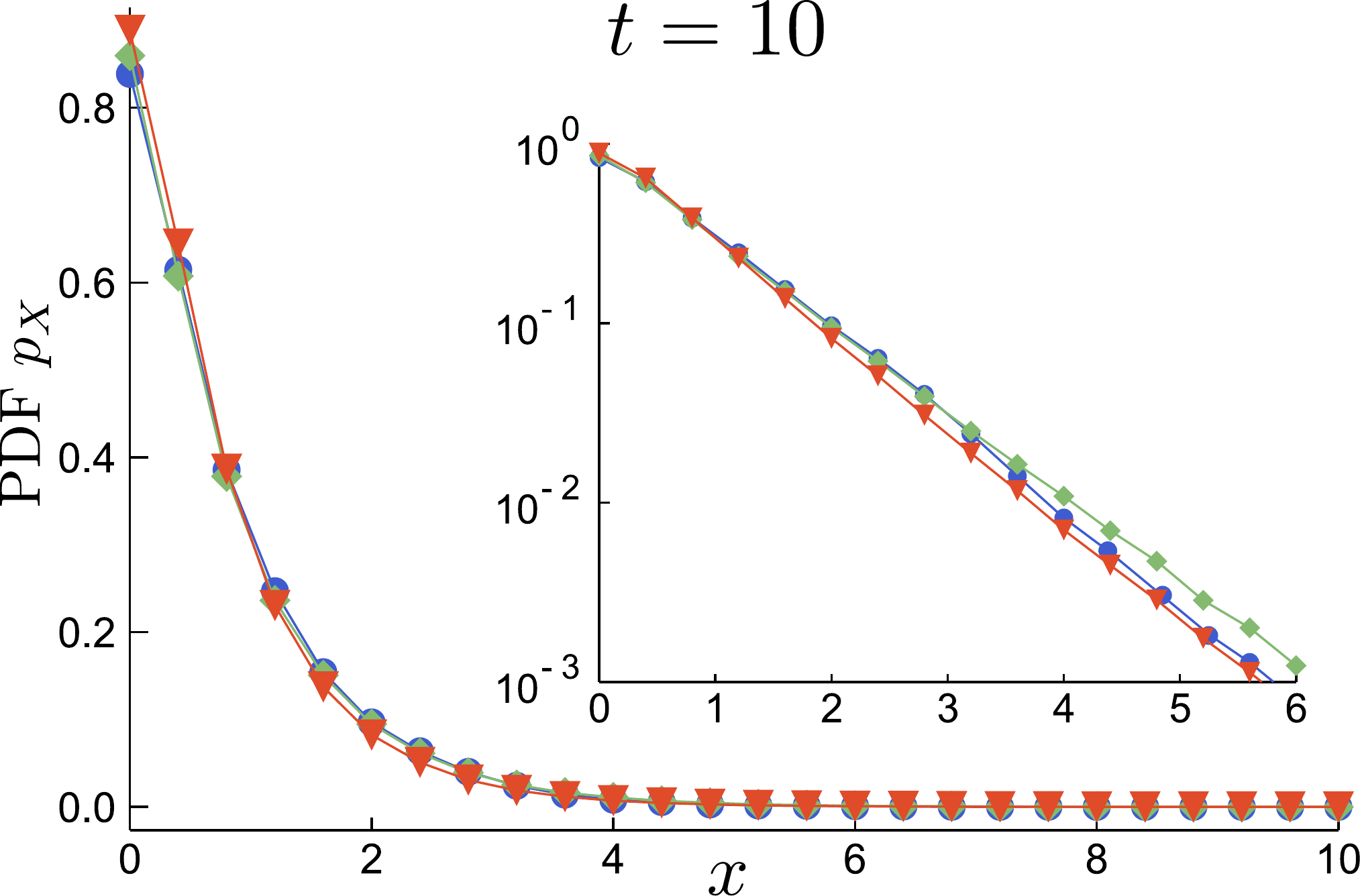}\ \ \includegraphics[width=0.32\textwidth]{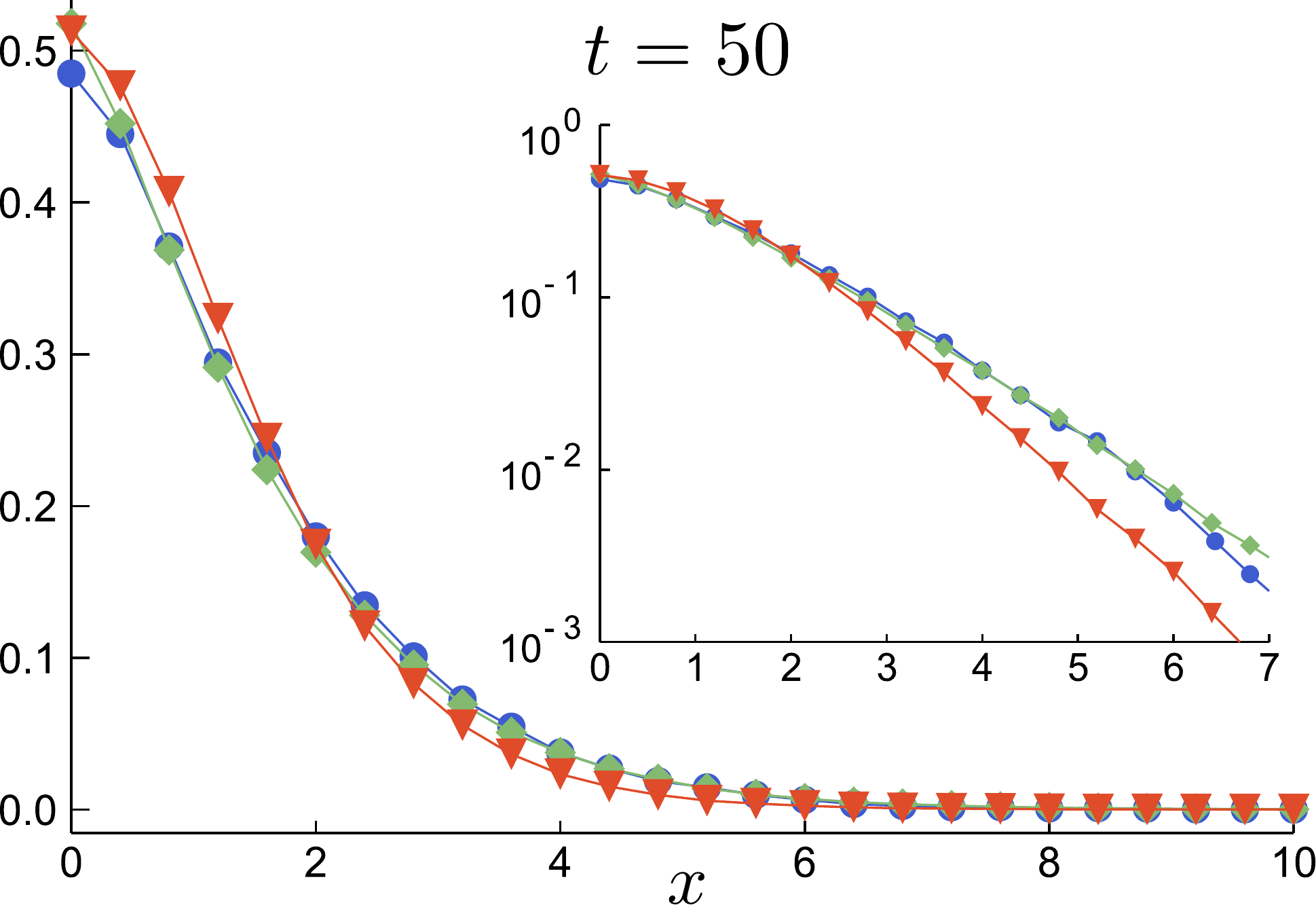}\ \ \includegraphics[width=0.32\textwidth]{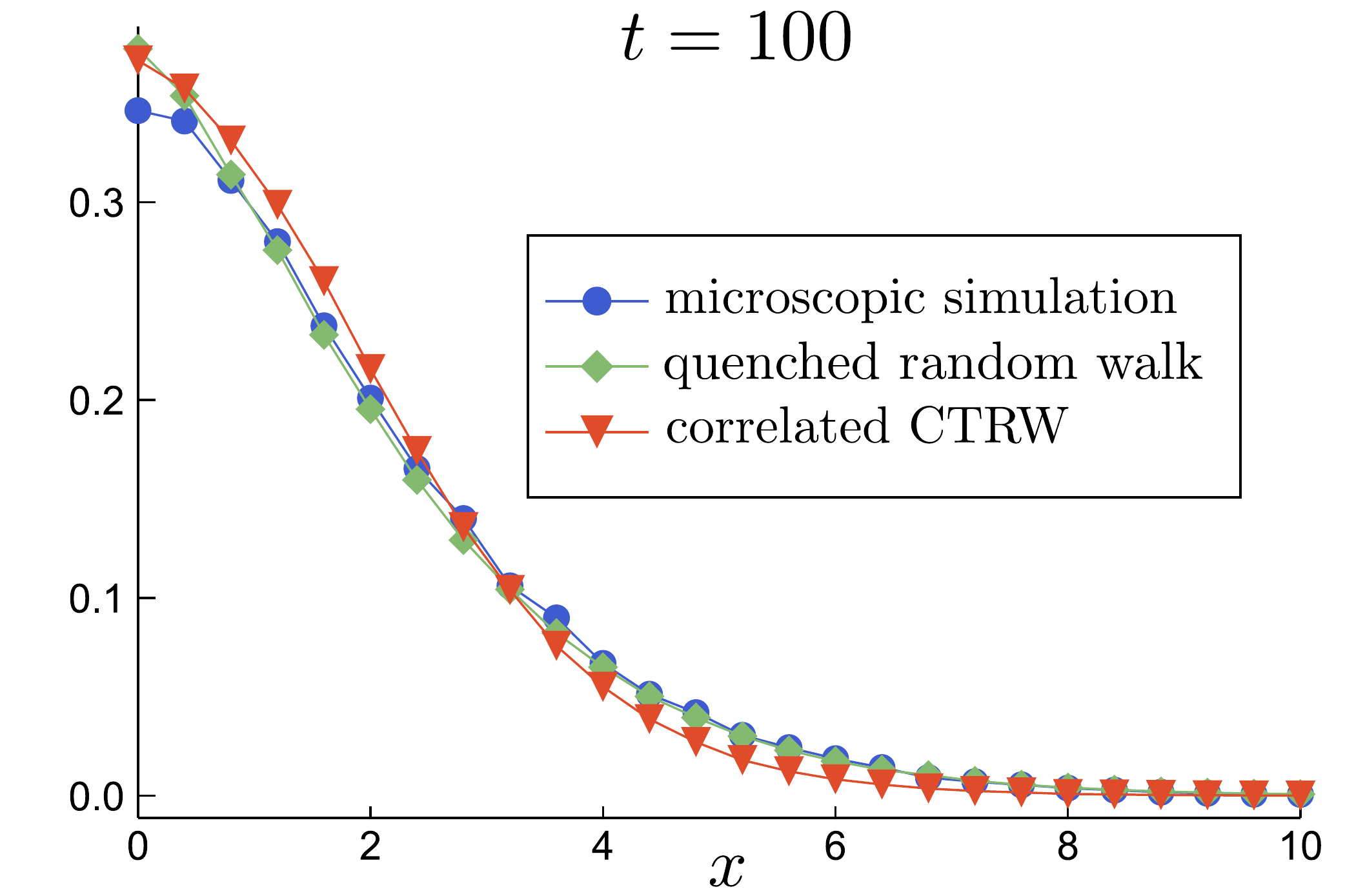}
	\caption{Comparison between the Euler scheme simulation  ($\Delta t = 0.0005$) of the diffusion process \reff{eq:heat2}, \reff{eq:interface} with $D=1,\kappa = 1/50$, the random walk approximation $X_{2\kappa D t}$ in the quenched environment, and the annealed version with the correlated jumps and waiting times. (The PDFs are symmetrical, we only show the right half.) In the regime $t=10$ the mass contained in the initial domain is dominating; for $t=50$ the majority of the PDF corresponds to few jumps, $t=100$ is the start of the Gaussian regime (in the semi-log space they are nearly parabolas).}\label{fig:simVSctrw}
\end{figure*}

 For simulations one can just use the fact that the standard Poisson measure (corresponding to the first barrier put at point $x_0=0$) converges to the stationary one rather quickly. It suffices to put barriers one after another until sufficiently large $x$ are reached (for domain size 1 few hundred is more than enough) and then move the beginning of the coordinate frame there. More elegant approach is to correct the initial domain. The proper choice was described in the footnote \cite{Note1}, we draw $L_0\deq \mathcal Gam(2,1), \Theta \deq \mathcal Unif(0,1)$ and put the left end at $(\Theta-1)L_0$, the right end at $\Theta L_0$. The rest of the domains is unaffected and they have lengths $L_k$ with i.i.d. $\mathcal Exp(1)$ distributions. 

The random walk starts at domain $k=0$, after $T_1\deq \mathcal Exp(2 \kappa D /L_0)$ it jumps with probability 1/2 to $k=1$ domain or to $k=-1$ one. The jumping with the analogical domain-dependent waiting times repeats until we reach the final domain $k_f$ for which the sum of waiting times exceeds $t$. The position of the particle is then drawn from the uniform distribution $\mathcal Unif(x_{k_f},x_{k_f+1})$ where $x_k$ denotes the left end of the $k$th domain.

Conveniently, this approximation decouples parameters $\kappa$ and $D$ from the dynamics, now they only determine the timescale. We may standardise this random walk by saying $X_t$ has $\mathcal Exp(1/L_k)$ waiting times, the physical units are can then be returned considering the process with rescaled time $X_{2\kappa D t}$.

This random walk depends on the local barriers' placement (that is \textit{quenched}) which makes the analytical analysis difficult. Both the domain sizes and waiting times are short-tailed distribution, so in this model the particle explores the space relatively freely and it may be expected that the annealing procedure should yield a good approximation. Actually, we will make another simplification at the same time by removing the correlation between the subsequent jumps and waiting times. For the quenched random walk each transition moves the probability mass into a uniform distribution inside some domain and is therefore correlated with the (domain size dependent) subsequent escape time. If we think about the quenched random walk as jumping from the middle to the middle of the domain, each transition has length $L_k/2+L_{k\pm1}/2$. One can just consider annealed random walk with this property, but it can be considerately simplified if we just divide all the transition lengths in half. 

The annealed model is thus as follows: we draw the initial domain size as  $L_0\deq \mathcal Gam(2,1)$ and the subsequent ones as i.i.d. $L_k\deq \mathcal Exp(1)$. The corresponding waiting times are drawn as $T_k\deq \mathcal Exp(2/L_k)$ or equivalently $T_k=E_k L_k/2, E_k\deq\mathcal Exp(1)$. Using them we construct a CTRW, i.e. the random sum $\sum_{k=0}^{N_t} J_k$  generated by jumps $J_k \defeq \pm L_k/2$ and the counting process $N_t\defeq\#\{k\colon T_1+\ldots + T_k\le t\}$. This process should approximate the centre of the domain the particle is in; to obtain the final position we  account for the last local equilibrium by adding $\mathcal Unif((\Theta-1)L_0,\Theta L_0)$ if $N_t=0$ or $\mathcal Unif(-L_{N_t}/2,L_{N_t}/2)$ if $N_t\ge 1$.

We show the positional PDFs of the simulated processes in Fig. \ref{fig:simVSctrw}. There we have chosen $D=1$ which determines the timescale and $\expval{L_k}=1$ which analogically fixes the position scale. The comparison proves that the random walk models are quite successful at capturing the bulk of the probability. There is a noticeable (but not huge) difference in the rate of exponential decay of the annealed process, but it is to be expected in this type of approximation.

The simulations also suggest that all the processes have similar Gaussian limit. For the annealed CTRW the Brownian limit can be derived in a formal way using the standard approach. We consider the rescaled process $X_{c t}/\sqrt{c}$ and then push the parameter $c$ to infinity. In this limit the correction at the last site gets squeezed to zero, so it may be ignored. The rescaled counting process $N_{c t}/c$ converges to $t/\expval{T_k} = 2t$ (this is the renewal theorem) and the underlying classical random walk $S_n\defeq \sum_{k=0}^{n}  J_k$ converges to the Brownian motion, $S_{cn}/\sqrt{c}\to \sqrt{\expval{J_k^2}}B_n = B_n/\sqrt{2}$. The CTRW $X_t$ is given by subordination of $S_n$ by $N_t$, so it converges to $B_{2t}/\sqrt{2}\deq B_t$ \cite{limitTher}. Important thing to notice here is because the counting process $N_t$ collapses to a deterministic function, the dependence between waiting times $T_k$ and jumps $\pm L_k/2$ becomes irrelevant in the long time limit.

\begin{figure}[h]\centering
	\includegraphics[width=1\columnwidth]{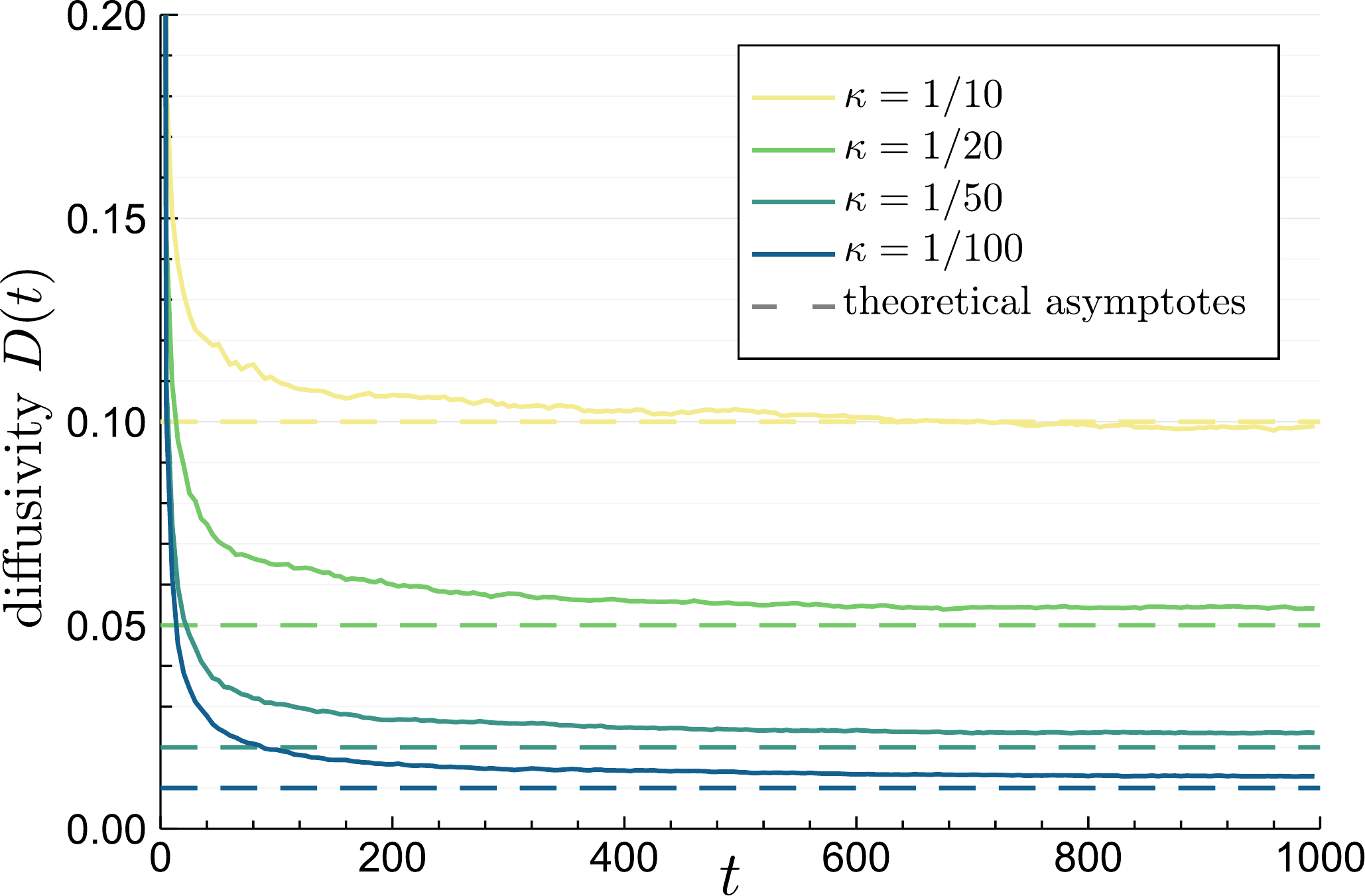}
	\caption{Time dependent diffusivity defined as $D(t)\defeq \delta_X^2(t)/(2t)$ calculated from the numerical simulation of the transiently confined process \reff{eq:heat2}, \reff{eq:interface} with short time $D=1$ and $\expval{L_k}=1$. For various permeabilities $\kappa$  the diffusivity $D(t)$ starts at $D(0)=1$ and decays, reaching the level close to the value of theoretical prediction  \reff{eq:Deff} derived for the annealed process. The range of $D_\text{eff}/D$ shown was suggested by the experiments of the hop diffusion \cite{murase}.}\label{fig:Deff}
\end{figure}

As a consequence, after accounting for all the physical constants, the long time effective diffusion coefficient $D_\text{eff}\defeq \lim_{t\to\infty} \delta_X^2(t)/(2t)$ of the annealed process is 
\begin{equation}\label{eq:Deff}
	D_\text{eff} = \expval{\kappa_B} D.
\end{equation}
This agrees with the former results which were established for the systems with periodically placed barriers \cite{tanner,powles, crick,latour, grebenkovBarrier}. A comparison of $D_\text{eff}$ observed in the simulations of the transiently confined diffusion and the theoretical value \reff{eq:Deff} is shown in Fig. \ref{fig:Deff}.

\begin{figure}[h]\centering
	\includegraphics[width=1\columnwidth]{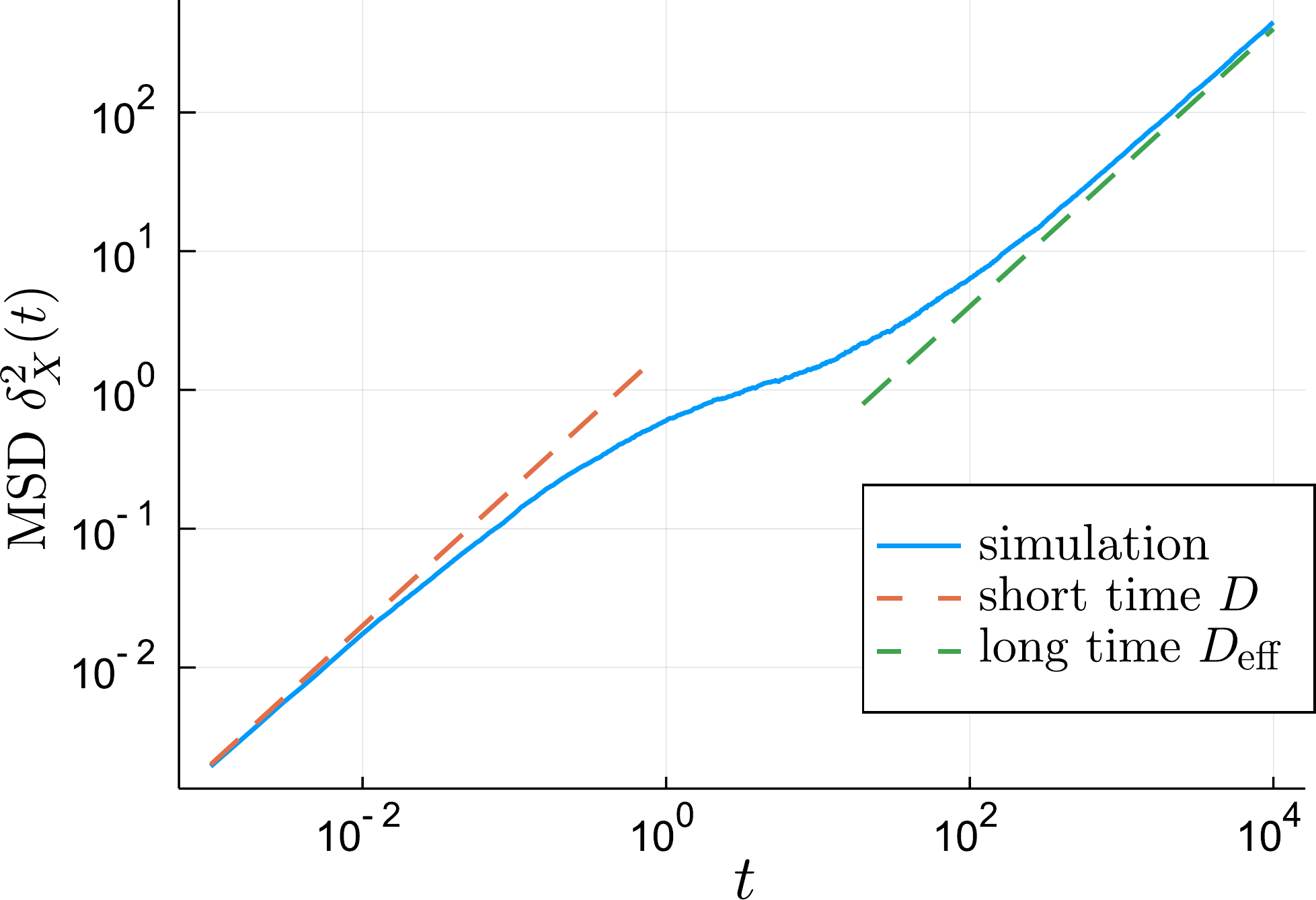}
	\caption{The MSD of the simulated transiently confined diffusion, $D=1,\expval{L_k}=1,\kappa = 1/50, \Delta t = 0.001$ (blue line) together with the short time prediction of the unconstrained Brownian motion with  $\delta_X^2(t)\sim 2Dt$ (red dashed line) and the long time prediction \reff{eq:Deff} with $\delta_X^2(t)\sim 2\kappa Dt$ (green dashed line).}\label{fig:loglogMSD}
\end{figure}

It is worth to note that this process is Fickian (has linear square displacement), but the transition from short-time diffusion constant $D$ to the long-time $D_\text{eff}= \expval{\kappa_B}D$ causes the mean square displacement to be a convex function. This behaviour can be seen in Fig. \ref{fig:loglogMSD} and bears great resemblance to contemporary experimental results for porous media \cite{murase,diffEscape}. The observed shape is easy to be mistaken for subdiffusion (MSD $\propto t^\alpha,\alpha<1$); this ``intermediate subdiffusive transport'' \cite{hoefling} suggest paying strong attention towards this possibility in the applied works as it could be misleading when observed without the entire six orders of magnitude shown on the time axis. 

Figure \ref{fig:loglogMSD} also indicates the time range in which we can expect the non-Gaussian behaviour. For short times the particle did not yet explore its initial domain and behaves like undisturbed Brownian motion with diffusivity $D$. For very long times the particle crossed a large number of domains, the environment becomes homogenised in this scale and the movements are again Brownian, but with lowered diffusivity $D_\text{eff}$. But, in the range between, where the MSD in Fig. \ref{fig:loglogMSD} bends, the particle's movements are strongly affected by the local environment, which originates the non-Gaussianity.

\section{Bessel and gamma waiting times CTRWs}
It seems reasonable that the dependence between waiting times and jumps may not be a crucial aspect of the dynamics. We will remove it and analyse the resulting CTRW revealing its non-Gaussian behaviour.

 For a particle exploring a medium with no long time correlation sources (such as strong traps) the dependence becomes irrelevant at long times and even at short times its is much weaker than in models such as L\'evy walks. For the latter class, the amplitudes of jumps and waiting times are identical and the joint distribution is degenerate \cite{LW}. In our case it has a smooth PDF $p_{J,T}(x,\tau) =\exp(-2|x|)\exp(-\tau/|x|)/|x|$ and the correlation between the jump amplitudes and waiting times has a lower value around 0.5.

Similarly, we may ignore the uniform distribution within the last domain and say that the particle just stops in the middle of the last domain. Using the same line of thought, we can say that the particle started in the middle of the first domain which therefore has form $[-L_0/2,L_0/2]$. Conveniently, it makes the distribution of the initial stationary state and of the subsequent jumps the same, that is the Laplace distribution $\mathcal Lap(0,1/2)$ derived in \reff{eq:pdfElementary}. It clearly introduces noticeable error for very short times when there is a significant probability mass in the initial domain. The uncorrelated CTRW PDF $p_X$ can be easily corrected by the formula
\begin{equation}
p_X'(x;t) = \pr(T_1<t)p_X^\text{stat}(x;t) + \pr(T_1\ge t)p_X(x;t),
\end{equation}
where $p_X^\text{stat}$ is the stationary PDF in the uncentred domain, see the footnote \cite{Note1}. For the clarity of the presentation we will ignore this correction further on, but we note it makes the agreement shown in Fig. \reff{fig:comp} better for the short times.

After all these steps, we end up with a CTRW
\begin{equation}\label{eq:redMod}
	X_t = \sum_{k=0}^{N_t}J_k,\quad N_t=\#\{k\colon T_1+\ldots+T_k\le t\}
\end{equation}
with jumps $J_k\deq \mathcal Lap(0,1/2)$ and independent waiting times $T_k= E_k L_k/2$ . The sum starting from $k=0$ accounts for the initial distribution. Conditioning and direct integration shows that $T_k$ have PDF which can be expressed by the Bessel function, $p_T(\tau) = 4 K_0(\sqrt{8\tau})$. This distribution or the distribution of $\sqrt{T_k}$ appear in various sources as the \textit{Bessel distribution} or \emph{K-distribution} \cite{Kdist,kotz}, we will use the former term. Its peculiar property is that this PDF has logarithmic singularity at $0^+$.

\begin{figure}
	\includegraphics[width=\columnwidth]{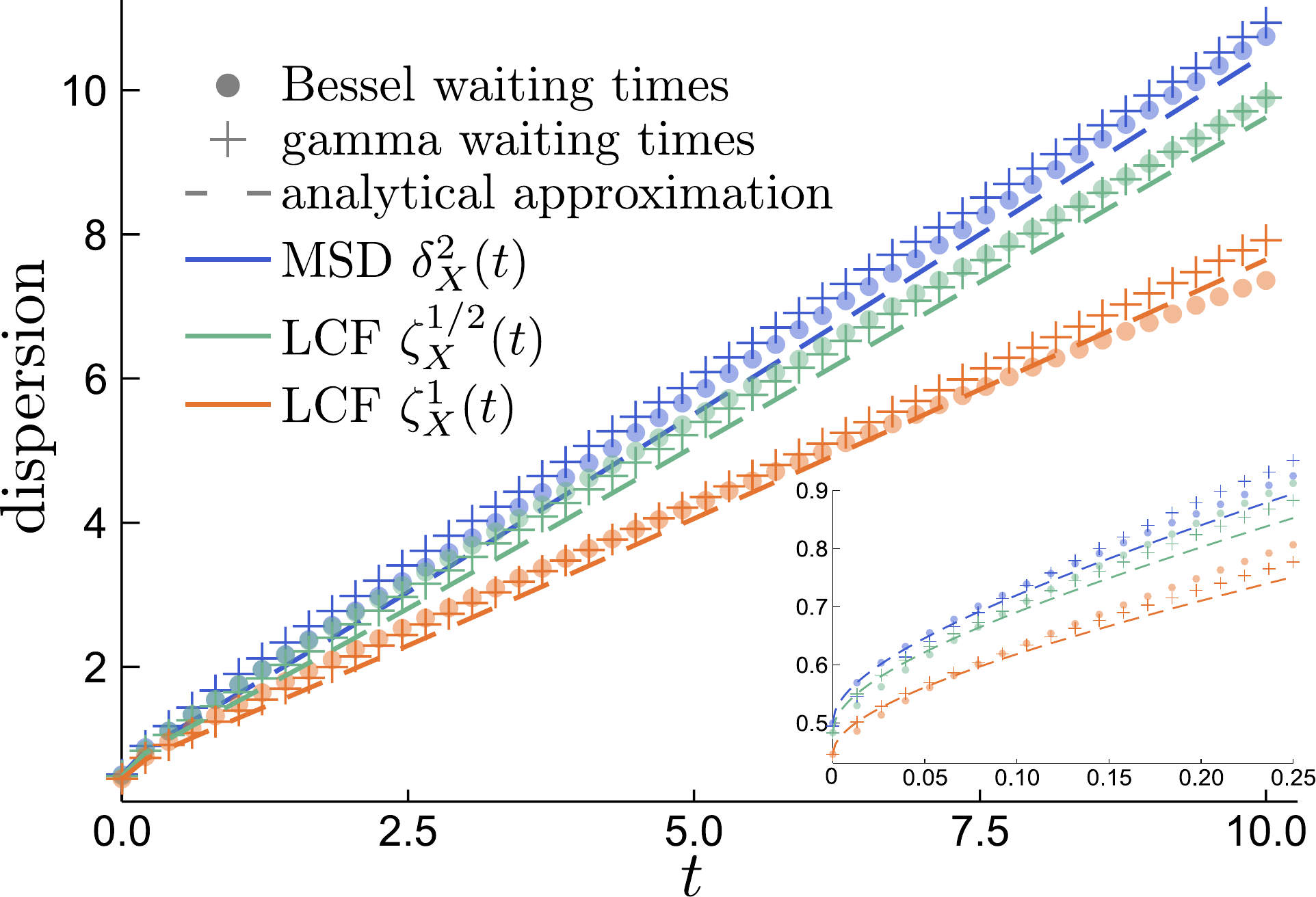}
	\caption{Comparison of the MSD and LCF with $\omega = 1/2,1$ for simulations of Bessel waiting times CTRW, gamma waiting times CTRW, and their analytical approximations \reff{eq:lcfAppr} and \reff{eq:msdAppr}. LCF being lower than MSD shows that the distribution has more spread out bulk than Gaussian.  }\label{fig:LCF}
\end{figure}

The obtained Bessel waiting times CTRW is simple enough that one can use Montroll-Weiss formula to obtain expressions for the moments and PDF in Laplace and Fourier-Laplace spaces \cite{guide}. In particular it shows that the MSD has a logarithmic cusp $\propto t\ln(1/t)$ at small times $t\to 0^+$, see blue line in Fig. \ref{fig:LCF}. However, we will not pursue this route, instead we will use a method which circumvents the use of Laplace transform and provides formulas which work globally with respect to $t$.

This is made possible if we approximate the waiting times distribution with a similar one for which the PDF of the counting process can be more easily managed. Counting processes with infinitely divisible waiting times are known to be well-studied; from this class short tailed processes are commonly   modelled using gamma distribution (especially in finance \cite{levyFinance}). From those, $\mathcal Gam(1/2,1)$ is remarkably close to the Bessel PDF which is reflexed in very small Kolmogorov distance $\approx 0.0395$. To put this number in a perspective, in standard hypothesis testing setting, we would need samples with around 2000 observations to notice the difference. Crucially, this distribution also has the same mean, which is necessary to preserve the Brownian limit of the original process.

We remark that such a significant similarity seems to stem from deeper properties of the distributions in question. Gamma distribution has wrong both $x\to0^+$ asymptotics ($1/\sqrt{x}$ instead of logarithmic) and $x\to\infty$ asymptotic (exponential instead of $\exp(-\sqrt{x})$). However, it nearly does not matter, as $\mathcal Gam(\alpha,1)$ for $\alpha = 1/2$ balances these two discrepancies so they cancel themselves out for the vast majority of the probability mass. The simple value of the parameter $\alpha = 1/2$ seems to be a coincidence, careful numerical study suggests that the value $\alpha \approx 0.494$ is marginally better. For any practical purpose the difference is insignificant but it is worth to add that the rest of the argument below does not depend on this particular value $\alpha=1/2$ but rather only on $\alpha$ being sufficiently close to 1.

We start with a counting process $N_t$ with $\mathcal G(1/2,1)$ waiting times. Knowing the exact distribution of the sums $T_1+\ldots+T_n$ we can determine the PDF of $N_t$ using the relation
\begin{align}\label{eq:survGamma}
	&\pr(N_t \ge n) = \pr(T_1+\ldots + T_n \le t)  \\
	&= \pr(\mathcal Gam(n/2,1) \le t)= \widetilde \Gamma(n/2,t);	\nonumber
\end{align}
here $\widetilde \Gamma$ is the regularised gamma function, $\widetilde \Gamma(a,t)\defeq \int_t^\infty\dd{s} s^{a-1}\exp(-s)/\Gamma(a)$. From this we get PDF expressed as a forward difference
\begin{align}\label{eq:pNgamma}
	p_N(n;t) &= \pr(N_t = n) = \pr(N_t \ge n) - \pr(N_t \ge n+1)\nonumber\\
	& = \widetilde \Gamma(n/2,t) - \widetilde \Gamma((n+1)/2,t).
\end{align}
The PDF of the whole CTRW can be linked to $p_N$ if we condition it by the number of jumps performed,
\begin{align}\label{eq:pdfGammaEx}
	p_{X}(x;t)  = \sum_{n=0}^\infty p_{S}(x;n) p_{N}(n),
\end{align}
where $S_n$ is, as before, a partial sum process, $S_n\defeq J_0+\ldots+J_n$. To obtain its PDF  we represent each jump as a difference of two independent exponential variables $J \deq \mathcal J^+-\mathcal J^-$, $\mathcal J^\pm\deq \mathcal Exp(2)$. Because of this their sum can also be split into $S_n = (\mathcal J_0^++\ldots +\mathcal J_n^+)-(\mathcal J_0^-+\ldots +\mathcal J_n^-) = \mathcal S_n^+-\mathcal S_n^-$. Now, we can use the fact that $\mathcal S_n^\pm$ have a simple distribution, i.e. $\mathcal Gam(n+1,2)$, and calculate $p_S$ as a convolution
\begin{align}\label{eq:pJconv}
	&p_{S}(x;n) = \int_{-\infty}^\infty \!\!\dd{y} p_{\mathcal S^+}(y;n)p_{\mathcal S^-}(-(|x|-y);n)\nonumber\\
	&= \int_0^\infty \!\!\dd{z} p_{\mathcal S^+}(|x|+z;n) p_{\mathcal S^-}(z; n)\\
	&= 4 \e^{-2|x|} \f{4^n}{(n!)^2}\int_0^\infty \!\!\dd{z} (z+|x|)^n z^n\e^{-4z}\nonumber\\
	&= \f{1}{n!}\sum_{k=0}^n \f{(2n-k)!}{k!(n-k)! 4^{n-k}} |x|^k \e^{-2|x|}.\nonumber
\end{align}
 Finally, \reff{eq:pNgamma} and \reff{eq:pJconv} substituted into \reff{eq:pdfGammaEx} form the exact series representation of the PDF of CTRW with gamma waiting times and Laplace jumps. It can be easy plotted and compared with the data. Alas, because of the complicated form of \reff{eq:pNgamma} and \reff{eq:pJconv} it is unwieldy for analytical investigation. To proceed, we will introduce another two approximations. The first yields the Fourier transform of the PDF and moments.

We will exploit the fact that the Poisson counting process has a simple PDF $t^n\exp(-t)/n!$ and survival function $\widetilde\Gamma(n,t)$. It has the same shape of graph as the  distribution \reff{eq:survGamma}, i.e. $\widetilde\Gamma(n/2,t)$, as we see it is only rescaled by 1/2. So, for the PDF expressed as a forward difference between this function at $n+1$ and $n$ we can replace the result by the Poisson PDF rescaled by 1/2
\begin{equation}
	p_N(n;t)\approx \f{1}{2}\f{t^{n/2} }{\Gamma(n/2+1)} \e^{-t}.
\end{equation}
As mentioned previously, the same argument works also for more general $\mathcal G(\alpha,\beta)$ waiting times for which the PDF would be rescaled by $\alpha$ instead of $1/2$. What we did here is basically a perturbation of the counting process around the Poisson one which states that as long as the power law $t^{\alpha-1}$ at $t\to 0^+$ is not to far away from $\alpha=1$ the obtained distribution is the Poisson one rescaled.

This procedure does not preserve the normalisation of the PDF. Correcting it leads to
\begin{equation}\label{eq:MLappr}
p_N(n;t)\approx  \f{t^{n/2} }{\Gamma(n/2+1)} \f{1}{E_{1/2}\big(\sqrt{t}\big)},
\end{equation}
where $E_{1/2}$ is the Mittag-Leffler function; for this particular parametrisation it can be also expressed as $E_{1/2}(x) = \exp(x^2)\erfc(-x)$. 

The error we make in this approximation is a differentiation error; it depends only on the smoothness of the underlying function. Therefore, it will become smaller when $t$ increases and the distribution spreads. So for large times this correction of normalisation is unimportant, but it helps for the small times, making the approximation globally efficient. 

To simplify the formulas, for the remainder of this section let us mark out the initial condition, i.e. decompose $X_t= \mathcal X_t + X_0$. By conditioning, we relate the distribution of $\mathcal X_t$ in the Fourier space to the transform of the jumps' PDF $\widehat p_J(\omega) = 1/(1+\omega^2/4)$,
\begin{align}\label{eq:gammaPDF}
	&\widehat p_{\mathcal X}(\omega;t) = \sum_{n=0}^\infty \big(\widehat p_J(\omega)\big)^n p_{N}(n;t) \\
	& \approx \sum_{n=0}^\infty \f{z^{n}}{\Gamma(n/2+1)} \f{1}{E_{1/2}\big(\sqrt{t}\big)}=  \f{ E_{1/2}(z)}{ E_{1/2}\big(\sqrt{t}\big)}, \nonumber
\end{align}
where for brevity we denoted $z\defeq \sqrt{t}/(1+\omega^2/4)$. This formula can be directly compared against experimental data by using the sample estimate of $\expval{\cos(\omega X_t)}$.

We want to reveal the non-Gaussian nature of the process and a convenient form to do is log-characteristic function (LCF) $\zeta^\omega_\mathcal{X}(t)\defeq -2\ln\expval{\cos(\omega \mathcal X_t)}/\omega^2$. It measures dispersion of the displacements like MSD but gives bigger emphasis of the spread of probability bulk, so it is expected to be smaller than MSD for processes with a broader tails and more peaky PDF (for any $\omega$) \cite{codiff}. Indeed, for
\begin{equation}\label{eq:lcfAppr}
\zeta^\omega_\mathcal{X}(t) \approx t\f{1+\omega^2/8}{(1+\omega^2/4)^2} +\ln\f{\erfc(-z)}{\erfc\big({-\sqrt{t}}\big)}
\end{equation}
the first linear term is dominating for the large times and the second one $\sim \sqrt{t} \pi^{-1/2}/(1+\omega^2/4)$ is dominating for the short times; both have scaling coefficients smaller than those of the MSD, which we calculate as the second derivative of \reff{eq:gammaPDF}
\begin{align}\label{eq:msdAppr}
	\delta^2_{\mathcal{X}}(t) &= \expval{\mathcal X_t^2} = -\pdv{^2}{\omega^2} \widehat p_{\mathcal X}(\omega; t)\Big|_{\omega = 0} \nonumber\\
	&\approx t +\pi^{-1/2}\f{\sqrt{t}}{\erfc\big({-\sqrt{t}}\big)}\e^{-t}.
\end{align}
Variable $\mathcal X_t$ corresponds to $X_0=0$ initial condition, but to account for the non-zero one we only need to add $\expval{X_0^2}=1/2$ to the MSD and $-2\ln\expval{\cos(\omega X_0)}/\omega^2=2\ln(1+\omega^2/4)/\omega^2$ to the LCF (it is as expected always $\le 1/2$). For an illustration of the behaviour of these dispersion measures see Fig. \ref{fig:LCF}. It is also interesting to note that the MSDs shown there prove that in practice it is very hard to see the difference between the logarithmic cusp (at $t\to 0^+$) of the Bessel waiting times CTRW and the square-root cusp of the gamma waiting times CTRW whereas the difference in asympotics might have suggested otherwise.

\begin{figure}
	\includegraphics[width=\columnwidth]{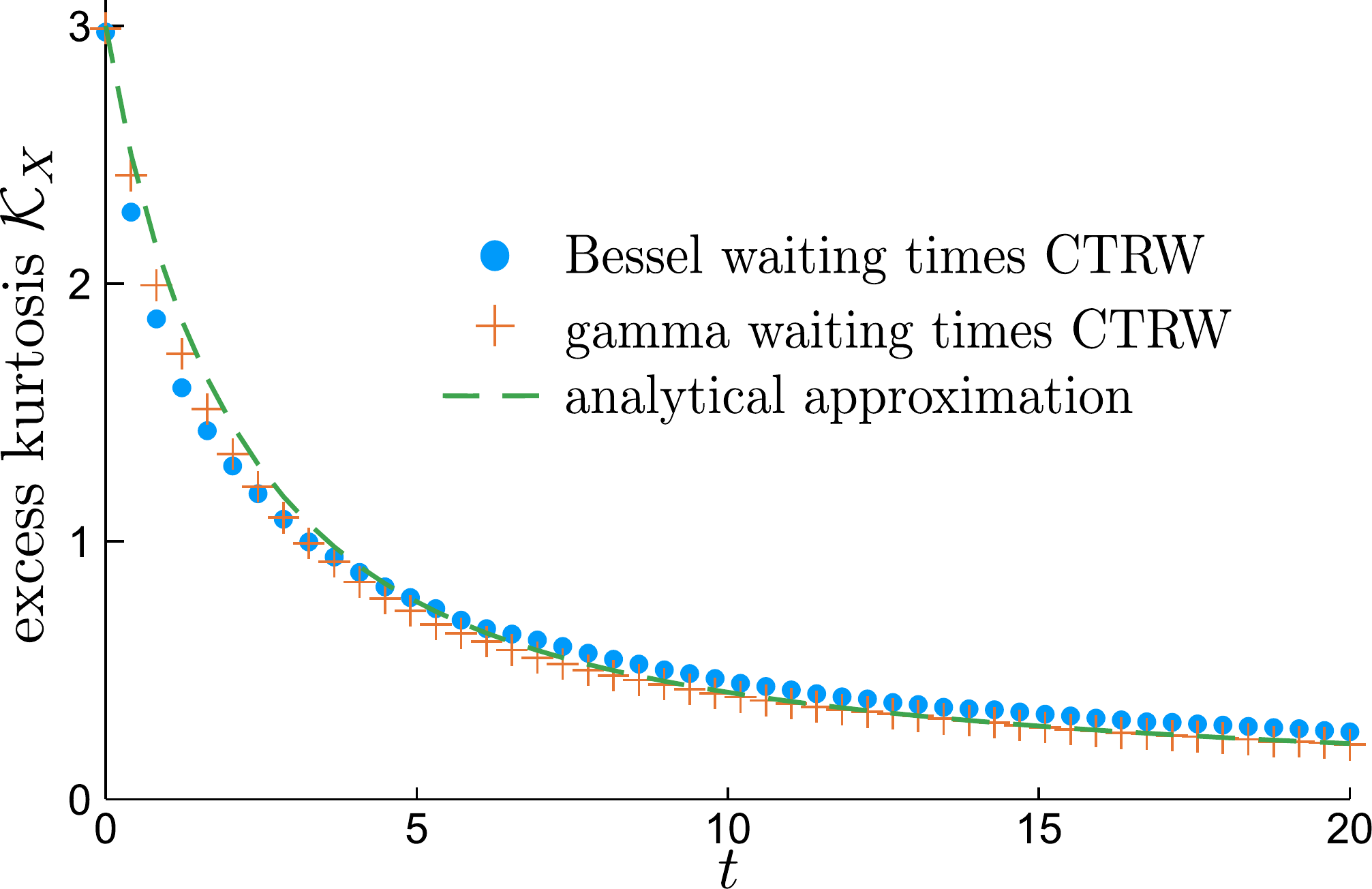}
	\caption{Excess kurtosis calculated from the simulations of CTRW with Bessel waiting times, gamma waiting times compared with the approximation \reff{eq:msdAppr},\reff{eq:4momAppr}. This type of shape is shows relaxation from Laplace to Gaussian distribution. }\label{fig:kurt}
\end{figure}

Continuing the analysis of non-Gaussianity, it can be also described using higher moments and one of particular interest is the excess kurtosis 
\begin{equation}\label{eq:kurt}
\mathcal K_X(t) \defeq \f{\expval{X_t^4}}{\expval{X_t^2}^2} - 3 = \f{\expval{\mathcal X_t^4}-3\expval{\mathcal X_t^2}^2 + \expval{X_0^4}-3\expval{ X_0^2}^2}{\big(\!\expval{\mathcal X_t^2}+\expval{X_0^2}\!\big)^2}.
\end{equation}
For any Gaussian variable it is 0; variables with tails broader than Gaussian are expected to have positive excess kurtosis (to be ``leptokurtic'') and in particular for any Laplace variable it equals 3. We get the fourth moment from the fourth derivative,
\begin{align}\label{eq:4momAppr}
	\expval{\mathcal X_t^4} &= \pdv{^4}{\omega^4} \widehat p_{\mathcal X}(\omega; t)\Big|_{\omega = 0} \nonumber\\
	&\approx  t (t+9/2) + 3 \pi^{-1/2}\f{ (t+1) \sqrt{t}}{\erfc\big({-\sqrt{t}}\big)}e^{-t}.
\end{align}
Substituting the calculated averages into \reff{eq:kurt} leads to a rather bulky, but purely elementary formula for $\mathcal K_X(t)$, which is compared against simulations in Fig. \ref{fig:kurt}. This function starts from $\mathcal K_X(0) = 3$ and then monotonically decays, at the beginning with rate $\dd \mathcal K_X/\dd t = 6(1-4/\pi)$ and as time grows the decrease becomes faster, reaching asymptotic  $\sim 9/(2t)$. This behaviour reflects the initial Laplace regime which transitions to the long-time Gaussian relaxation. 

As a side remark, we note that the kurtosis of $\mathcal X_t$ diverges at 0, $\mathcal K_{\mathcal X}(0^+) = \infty$ and then it converges to 0 in the same manner as $\mathcal K_X(t)$. This happens because kurtosis measures  the broadness of the distribution's tails and also its spikiness at $x=0$. In this case the latter is the culprit: the initial condition $\mathcal X_0 = 0$ causes the distribution of $\mathcal X_t$ to have Dirac delta at $x=0$ as $\mathcal X_t=0$ unless $T_1>t$. For a system like ours it is clearly an unphysical artefact, which shows the importance of a proper choice of the initial condition.

\section{Compound Poisson approximation}

\begin{figure*}\centering
	\includegraphics[width=0.33\textwidth]{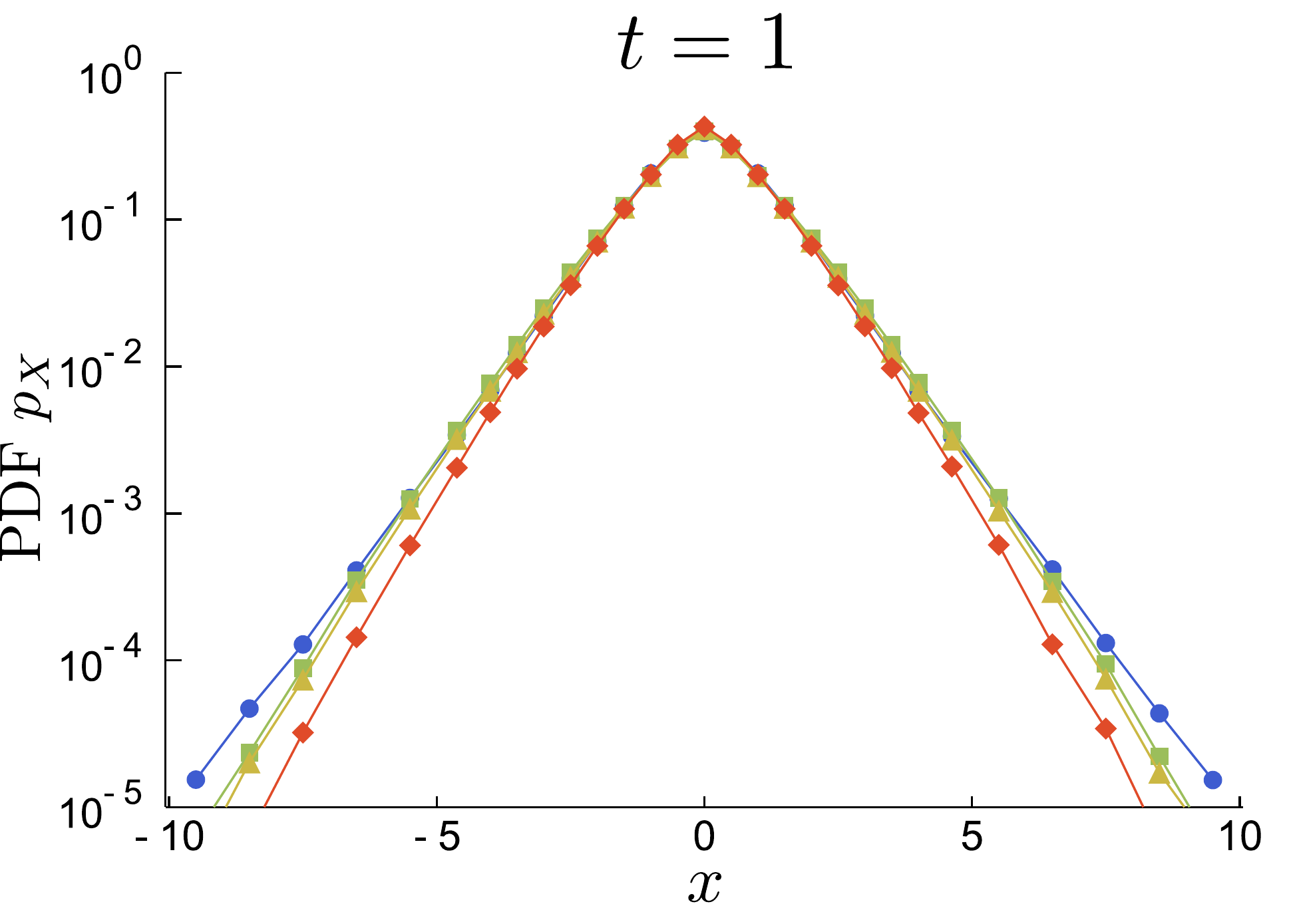}\includegraphics[width=0.33\textwidth]{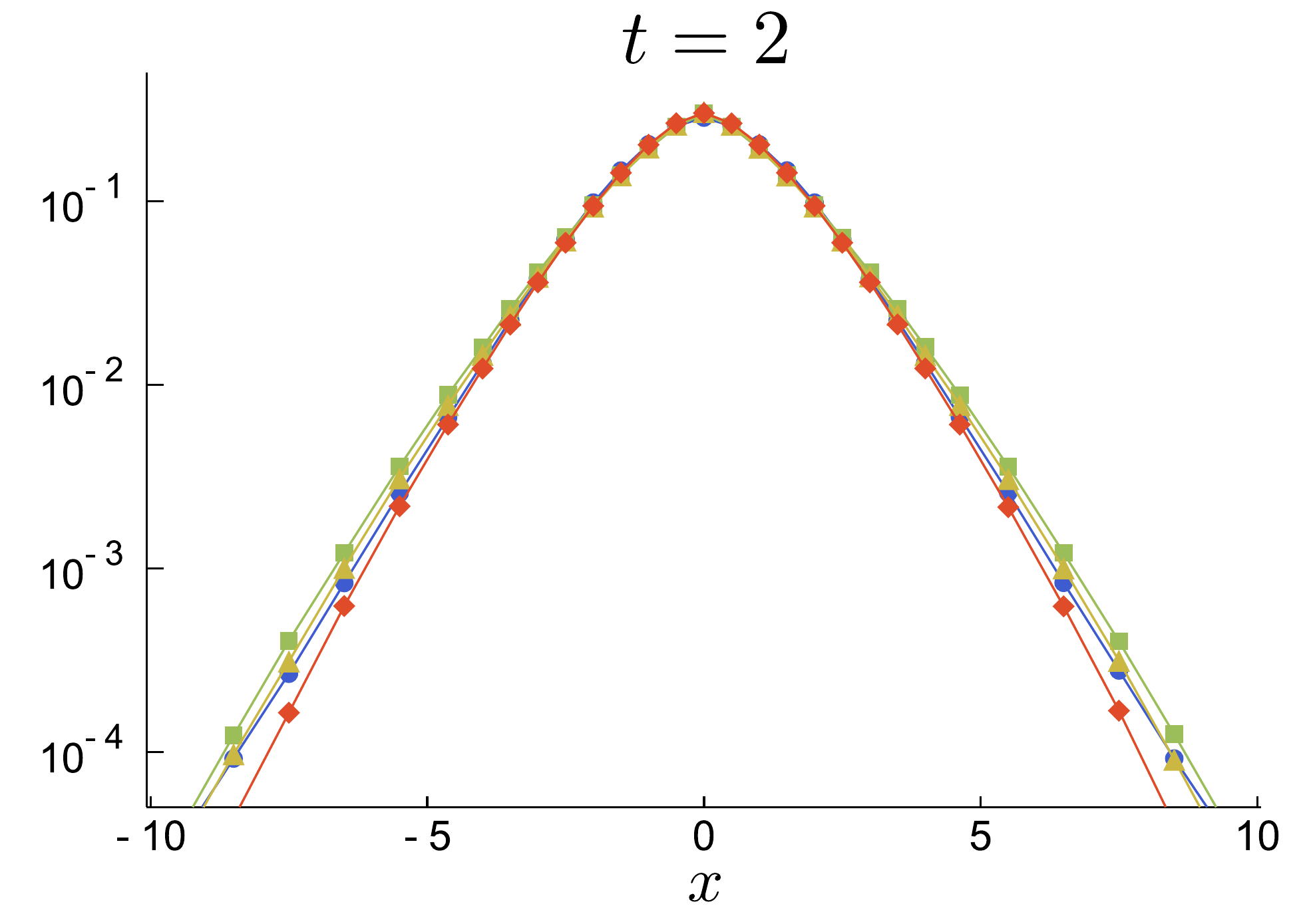}\includegraphics[width=0.33\textwidth]{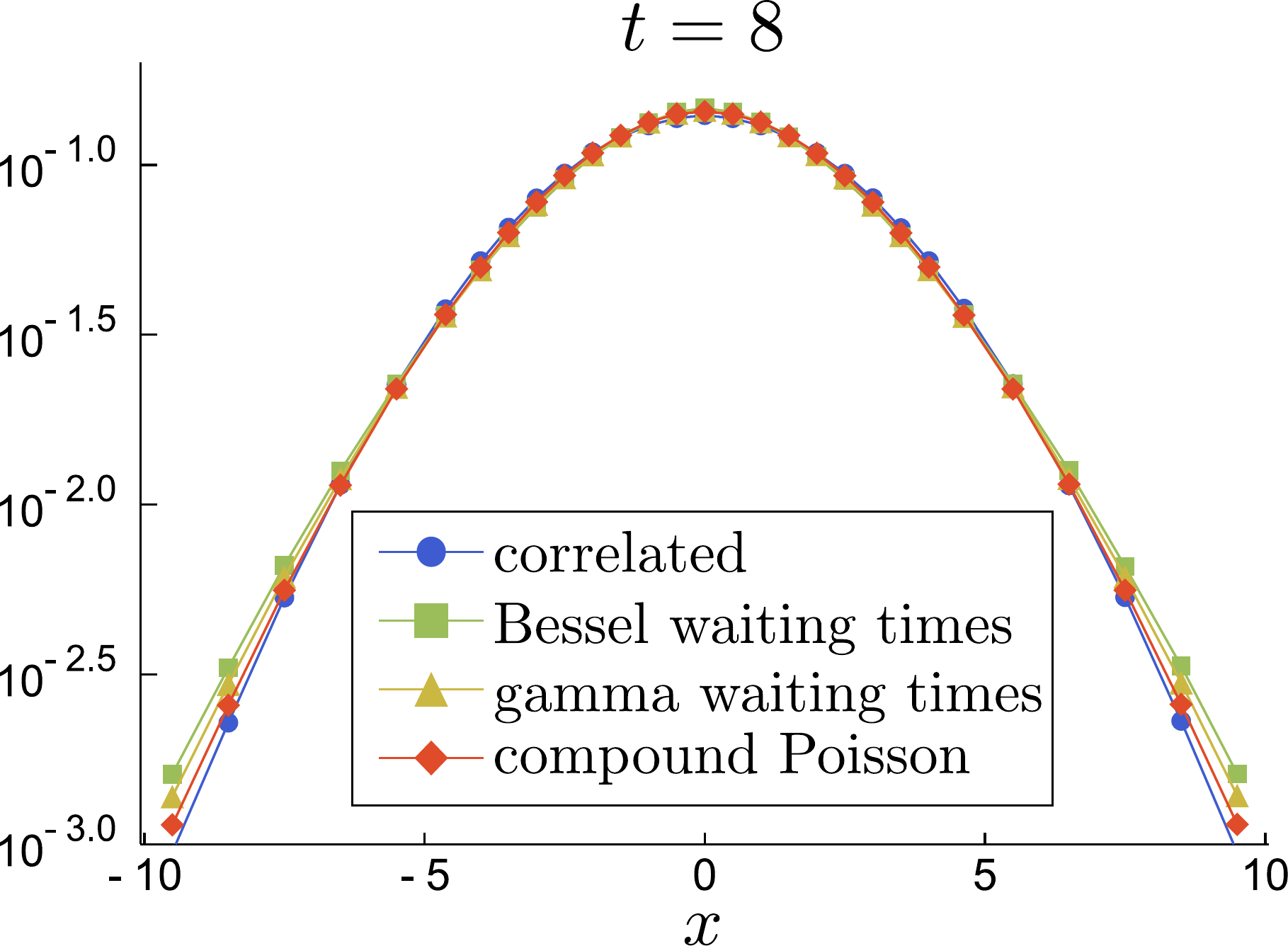}
	\caption{Comparison of the correlated CTRW and further approximations. Bessel and gamma waiting times CTRWs are nearly undistinguishable, all others are also close. These random walks have the same Gaussian limit, as shown for $t=8$.}\label{fig:comp}
\end{figure*}

The methods which we used in the former section were successful in revealing the broad tails of the studied diffusion, but it would be beneficial to show the exponential decay of the PDF directly \footnote{We note that this route will bring stronger results than using the Fourier transform \reff{eq:gammaPDF} directly, as it is unfortunately hard to inverse analytically. One can use the method of steepest descent to uncover the tail behaviour of $p_X$, but the result does not seem to be very practical, describing only a far away range not easily available in experiments.}. To achieve this we are introducing a second CTRW approximation with a very simple memory structure which will yield a full asymptotic series expansion of the PDF.

The insight that we use is a particular interpretation of the gamma waiting times CTRW. The PDF of the gamma waiting times process agrees with the Poisson counting process at even $n$. This is because a sum of any pair of $\mathcal Gam(1/2,1)$ waiting times has exponential waiting time, $T_k+T_{k+1}\deq \mathcal Gam(1/2,1)+\mathcal Gam(1/2,1)\deq \mathcal Exp(1)$. Essentially, if we ignore odd numbers of jumps, the process behaves exactly like the one with exponential waiting times which makes double jump each time. Previously, we accounted for the odd numbers of jumps by interpolating the PDF with the Poisson formula.

The second possibility is to replace the counting process with double jumps by the one with twice the intensity but only single jumps. This way the average number of jumps remains  the same (we just spread them), however we distort their variability. For this reason the second approximation is expected to be less accurate than the first one, but it makes up for it by providing useful representation of the PDF.

Thus, we have come to consider the CTRW with Laplace jumps and the Poisson counting process $\mathcal Poiss(2t)$. For a comparison of this process to the other CTRWs considered in this work see Fig. \ref{fig:comp}. In the decomposition $X_t = \mathcal X_t+X_0$ the term $\mathcal X_t$ now belongs to the class of \emph{compound Poisson processes}. These processes are very regular, being Markovian, infinitely divisible and having independent increments. As one of the consequences, $\mathcal X_t$ must have linear MSD, precisely $\delta^2_{\mathcal X}(t) = t$. It means that in this approximation we completely neglect the non-linearity of the MDS present in the more detailed models. However even in those, the linear range was appearing quickly (even when the motion was still highly non-Gaussian), see Fig. \ref{fig:LCF}, so the error caused by this is not huge.

The Fourier space representation of the compound Poisson PDF often has a sleek form; in our case
\begin{equation}
	\widehat p_{\mathcal X}(\omega; t) = \exp(2t\lt(\f{1}{1+\omega^2/4}-1 \rt)).
\end{equation}
Again, we may use it to calculate excess kurtosis, which is a simple rational function
\begin{equation}
	\mathcal K_X(t)  = \f{3 t + 3/4}{(t+1/2)^2}.
\end{equation}
This function decays to 0 even slower than in the case of the gamma waiting times, $\dd\mathcal K_X/\dd t = 0$ at $t=0^+$ and the asymptotic decay is $\sim 3/t$, but overall the shape of this function is not much different than before.

Now, to get the PDF of $X_t$ in the position space we will use a particular representation available only for the compound Poisson processes. Instead of dividing each jump into the difference of two variables like before, we separate them into two categories corresponding to which are positive and which are negative for a given trajectory; the same is made to the initial condition. This can always be done, but only for the Poisson process the two thinned counting processes obtained, $N_t^+\defeq \#\{k\colon T_1+\ldots+T_k, J_k>0\}$ and $N_t^-\defeq \#\{k\colon T_1+\ldots+T_k, J_k<0\}$ are independent and Poisson with the twice smaller intensity, $N_t^\pm\deq \mathcal Poiss(t)$ (this is a particular consequence of the process being Markovian). Thus, we may represent the total displacements as a difference of two positive, independent processes, $X_t = X_t^+-X_t^-$, where
\begin{equation}
	X_t^\pm = \sum_{k=0}^{N_t^\pm} \mathcal J_k^\pm, \quad \text{i.i.d.\ } \mathcal J_k^\pm\deq \mathcal |J_k| \deq \mathcal Exp(2).
\end{equation}
We again use the fact that $\mathcal J_0+\ldots + \mathcal J_n\deq \mathcal Gam(n+1,2)$; conditioning by the number of jumps leads to a surprisingly elegant formula for the  PDF of $X_t^\pm$ \cite{buchak},
\begin{align}
	&p_{X^\pm}(x;t)  = \sum_{n=0}^\infty p_{\mathcal Gam(n+1,2)}(x) p_{N^\pm}(n;t)\\
	&= 2\lt(\sum_{n=0}^\infty  \f{(2xt)^n}{(n!)^2}\rt)   \e^{-2x}\e^{-t} = 2 I_0\big(2\sqrt{2t x}\big)\e^{-2x}\e^{-t}.\nonumber
\end{align}
From this we immediately determine the part of $p_X$ corresponding to the series of jumps in a one direction, these far tails are given by $ \pr(X_0>0)p_{N^-}(0;t)p_{X^+}(x;t) = I_0(2\sqrt{2t x})\e^{-2x}\e^{-2t}$. Of course the full PDF is significantly broader. As in \reff{eq:pJconv} it is given by the convolution
\begin{align}
	&p_X(x;t) = \int_0^\infty \!\!\dd{z} p_{X^+}(|x|+z;t) p_{X^-}(z; t)\\
	&= \e^{-2|x|}\e^{-2t} \int_0^\infty \!\!\dd{z} I_0\big(2\sqrt{2t (|x|+z)}\big)I_0\big(2\sqrt{2t z}\big)4\e^{-4z}\nonumber.  
\end{align}
Let us denote the integral on the right in the above by $\mathcal I = \mathcal I(x;t)$. As Bessel function $I_0$ increases monotonically, but slower than exponentially, $\mathcal I(x;t)$ is also increases slower than exponentially with respect to $t$ and $x$. The dominating factor of the tails is thus $\exp(-2|x|)$ which is completely static and is inherited from a single jump PDF, the rest is some time dependent factor and second order correction with respect to $x$. These are important as they contain information about the dynamics. The rest of the section is devoted to their derivation.

 The integral $\mathcal I$ can be expanded into a series. One method is using Taylor expansion of $I_0$; this leads to a formula similar to the one shown in the previous section. However, more efficient expansion can be obtained if we integrate by parts. 

It is easier to explain the general method first: under mild regularity assumptions on $f$, an integration with exponent can be expressed as the infinite order differentiation operator
\begin{align}
	&\int_0^\infty\!\!\dd{z} f(z) a\e^{-a z} = f(0)+\int_0^\infty\!\!\dd{z} \dv{}{z}f(z) \e^{-a z}\nonumber\\
	& = f(0)+ \f{1}{a}\dv{}{z}f(z)\Big|_{z=0} + \f{1}{a}\int_0^\infty\!\!\dd{z} \dv{^2}{z^2}f(z) \e^{-a z} = \ldots\nonumber\\
	& = \sum_{n=0}^\infty \f{1}{a^n}\dv{^n}{z^n}f(z)\Big|_{z=0} = \f{1}{1-\f{1}{a}\dv{}{z}} f(z) \Big|_{z=0}.
\end{align}

 In our specific case we need to take $f(z) = I_0\big(2\sqrt{2t (|x|+z)}\big)\times I_0\big(2\sqrt{2t z}\big)$ and $a= 4$. Function $I_0\big(\sqrt{z}\big)$ is absolutely monotonic, so the obtained series consists of positive terms, i.e. we divide the probability mass into Bessel-like modes.
 
It can be further simplified if one notes that the derivatives of $ I_0\big(2\sqrt{2t z}\big)$ reduce to factors depending only on $t$. We expand each derivative using the binomial formula
\begin{equation}
	\dv{^m}{z^m}\big(f(z)g(z)\big) = \sum_{k=1}^m\binom{m}{k} \dv{^k}{z^k}f(z) \dv{^{m-k}}{z^{m-k}}g(z) 
\end{equation}
 and then rearrange the terms according to the order of derivative acting on  $I_0\big(2\sqrt{2t (|x|+z)}\big)$. The coefficient before the $n$th one is
\begin{align}
	&\sum_{k=n}^\infty \binom{k}{n} \f{1}{4^{k}} \dv{^{k-n}}{z^{k-n}} I_0\big(2\sqrt{2t z}\big)  \Big|_{z=0}\!\!=\sum_{k=n}^\infty \binom{k}{n} \f{1}{4^{k}} \f{(2t)^{k-n}}{(k-n)!}\nonumber\\
	&= \f{1}{4^{n}} {}_1F_1(1+n,1;t/2) = \f{1}{4^{n}}\e^{t/2}\sum_{k=0}^n \f{(n-k+1)_k}{(k!)^2}\lt(\f{t}{2}\rt)^k,
\end{align}
where $_1F_1$ is Kummer's function and $(x)_k$ is Pochhammer symbol. Denoting the  polynomial on the right by $q_n(t)$ the integral of interest takes the form
\begin{align}
	\mathcal I &= \e^{t/2} \sum_{n=0}^\infty \f{q_n(t)}{4^{n}} \dv{^n}{x^n} I_0\big(2\sqrt{2t|x|}\big)\\
	& = \e^{t/2} \sum_{n=0}^\infty \f{q_n(t)}{4^{n}} \lt(\f{2 t}{|x|}\rt)^{n/2} I_n\big(2\sqrt{2t|x|}\big).\nonumber
\end{align}
Alternatively, instead of using $I_n$, one can also repeatedly apply equalities $\dd I_0(z)/\dd z = I_1(z)$, $\dd I_1(z)/\dd x = I_0(z)- I_1(z)/z$ and express the result as  a mixture of $I_0$ and $I_1$ multiplied by powers of $t$ and $x$. In any case, the Bessel modes decay with respect to $n$ and have prefactors of type $t^\alpha/x^\beta$, which shows that this expansion may converge slowly for large $t$ and small $x$ but will converge quickly for small $t$ and large $x$. 
 
\begin{figure}
	\includegraphics[width=\columnwidth]{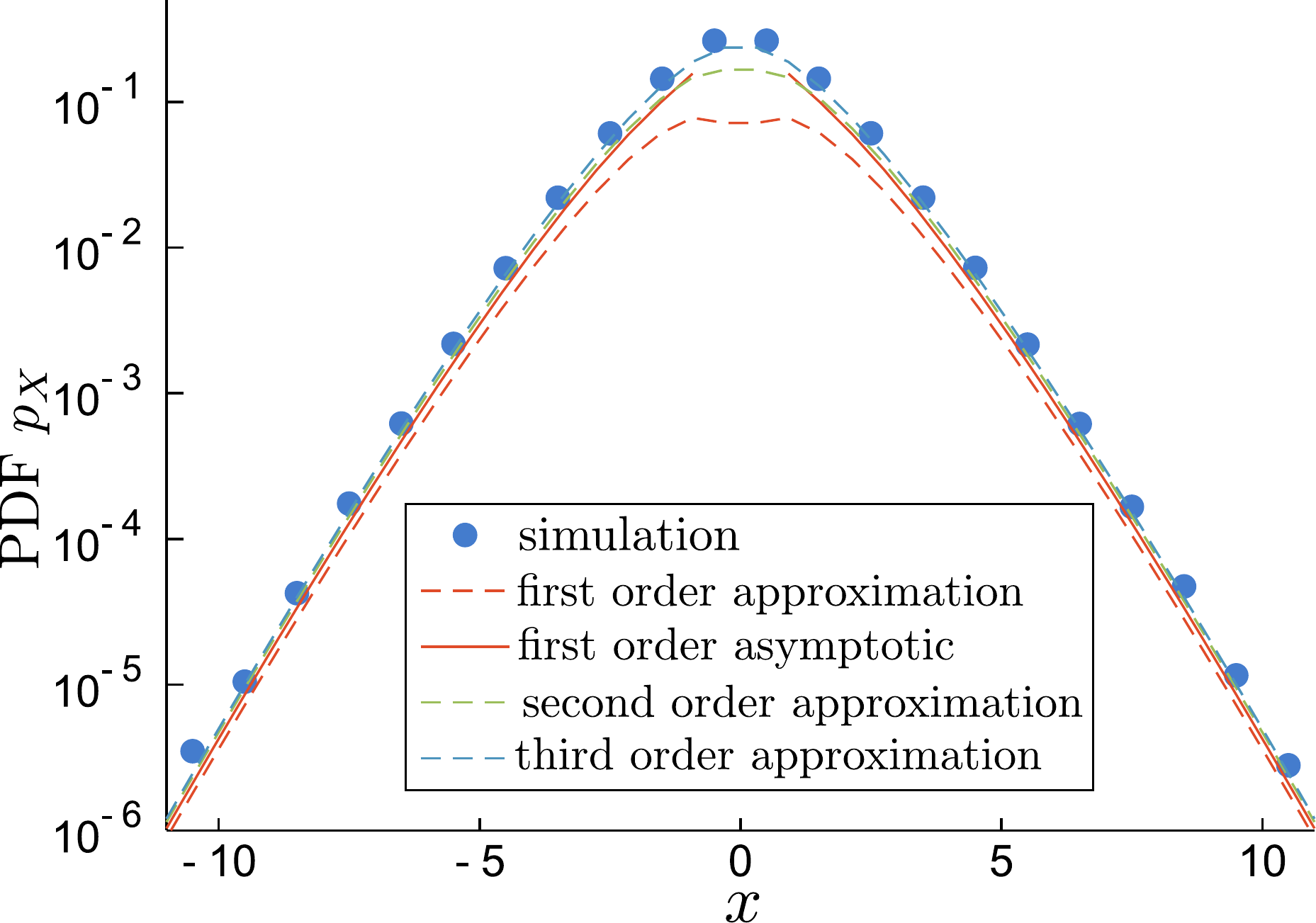}
	\caption{Comparison of the simulated compound Poisson process, expansion \reff{eq:besselAppr} and asymptotical formula \reff{eq:besselApprAsympt} for $t=2$. For smaller times $t\lesssim 1$ the fit is even better. For larger $t$, as Gaussian regime starts the exponential behaviour is pushed towards larger $|x|$.}\label{fig:besselAppr}
\end{figure}

Going back to the PDF under consideration, the expansion becomes
\begin{align}\label{eq:besselAppr}
	& p_X(x;t) =  \sum_{n=0}^\infty \f{q_n(t)}{4^{n}} \lt(\f{2 t}{|x|}\rt)^{n/2}  I_n\big(2\sqrt{2t|x|}\big) \e^{-2|x|}\e^{-3t/2}\nonumber\\
	&= I_0(2\sqrt{2t|x|})  \e^{-2|x|}\e^{-3t/2}\\
	&+ \f{1}{4}(1+t/2)\sqrt{\f{2 t}{|x|}}I_1\big(2\sqrt{2t|x|}\big)\e^{-2|x|}\e^{-3 t/2} +\ldots \nonumber
\end{align}
Using the tail asymptotic $I_n(z)\sim \e^{z}/\sqrt{2\pi z}$ we get the leading behaviour for large $|x|$
\begin{equation}\label{eq:besselApprAsympt}
	p_X(x;t)\sim \big(4\pi \sqrt{2t|x|}\big)^{-1/2}\e^{2\sqrt{2t|x|}}\e^{-2|x|}\e^{-3t/2}
\end{equation}
confirming the exponential decay of the PDF visible from the simulations, see Fig. \ref{fig:besselAppr} for a comparison.

Equation \reff{eq:besselApprAsympt} is similar to the large deviation property. We have shown that $\ln p_X(x;t) \sim -t f(|x|/t)$ with the rate function $f(y) = 3/2+2y-2\sqrt{2y}$. This works for large $|x|$ and any $t$, but for the larger $t$ the convergence becomes progressively slower; this is caused by the coefficients $q_n(t)$ increasing with $t$. Nonetheless, it is not always necessarily the best formulation of the problem. For small $t$ the rate function $f(|x|/t)$ diverges, but we can still use Eq. \reff{eq:besselAppr} because it is exact. For example, if $|x|\to \infty$ and $t|x|\to\infty$ the result is the same, but if we go with $t\to 0$ fast enough such that $t|x|\to 0$ we get $\ln p_X(x;t)\sim 2t|x|-2|x|-3t/2$.

In any case, the most important factor is the exponential decay $\exp(-2|x|)\exp(-3t/2)$ which dominates the shape of the PDF in the semi-log scale. The rest are corrections which mostly move the straight line $-2|x|-3t/2$ around the coordinate frame. This non-Gaussian Laplace behaviour is observed in the ``interim'' time range in which the particle explores the neighbourhood of several domains and the effective diffusivity has already been reduced close to its long-time limit; it is the middle part of the range shown in Figs \ref{fig:Deff} and \ref{fig:loglogMSD}.

\section{Discussion}

A robust and growing class of observations is exhibiting linear or power-law mean-square displacement joined by the pronouncedly non-Gaussian probability density, which most commonly has the exponential (Laplace) shape \cite{wangNG,fair, bhattacharya,hapca,lampo17,wang2} (see also the list of references in \cite{diffDiffChechkin}). Explaining this phenomena related to \textit{Brownian, yet non-Gaussian diffusion} is crucial for understanding transport within these media and their biochemical properties, a timely issue especially in biology and medicine. In spite of this demand, most of the studied analytical models are very case-specific \cite{kurtuldu,menzel} or do not provide description which would reflect the real structure of the considered material, the most prominent examples being the superstatistical \cite{beck2,beck3,beck4, superstatLang} and the diffusing diffusivity \cite{diffDiff, diffDiffReact, diffDiffChechkin,diffDiffSposini} approaches. The line of research presented here follows the widespread conviction that the physical origin of these observations lays in the heterogeneity of the studied media and try to emulate it by randomizing the parameters which appear in the dynamical equations of the diffusion.

In this work we solve a microscopical model of the heterogeneous medium which is compartmentalised by the thick and thin barriers that strongly impede diffusion and cause the transient-confinement of the diffusing particles. This choice of model is motivated by the ubiquity of measurements showing this so-called \textit{hop diffusion} observed in biology \cite{confDiff,krapfComp,flow, kasumi, krapfComp2,fujiwara,murase} and the \textit{cage effect} in chemistry \cite{weeksCages,chaudhuri, weeks2}. Non-Gaussian diffusion was  observed together with these phenomena in multiple works, see the review \cite{hoefling}. In many other experiments the researchers were restricting their attention to only one of these two aspects of transport, which leaves an open and intriguing possibility of this link being much more prominent.

In our model the exponential tails of the probability density essentially stem from the lack of spatial correlations in the random medium. For media with more ordered microscopic structure, in particular with the periodically placed barriers, the observed probability density is expected to be much more Gaussian \cite{diffEscape} (see \cite{tanner} for theoretical discussion). However, if the barriers are placed independently, the distances between them are exponential. Any given temporarily trapped particle then has uniform distribution over its current domain, but the whole ensemble as a mixture of those exhibits exponential tails.

The simplicity of this argument makes it quite general, but it is not enough to specify the dynamics. For this we show how the local behaviour of the particle is equivalent to a classical heat transfer problem and  then derive the escape times from the domains.

With the known domain sizes and transitions between them it becomes straightforward to interpret the ``hops'' of the \textit{hop diffusion} as the jumps of a random walk. Monte Carlo simulations then show that this approximation is indeed very close to the original diffusion and both are the cases of non-Gaussian diffusion which at long times reaches Brownian limit.

This process is Fickian in both short and long timescales, but they differ by the diffusivity, which transitions from the initial ``unhindered'' short-time $D$ to the lower $D_\text{eff}=\expval{\kappa_B}D$; Such decay of diffusivity was reported in numerous experimental works \cite{murase}. The mean Biot number $\expval{\kappa_B}$ is determined by distances between barriers and their permeability, $\expval{\kappa_B} = \kappa\expval{L}$  and our results extends the analytical proofs for the systems with periodically placed barriers \cite{tanner, powles}. This link between micro- and macroscopic properties of the system may be verified experimentally by comparing the estimated diffusivity to the measured parameters of the barriers and their distribution in space.

It also causes the mean square displacement to be a convex function and the diffusion becomes subdiffusive in the transient time range of the diffusivity decrease. This is a well known phenomenon \cite{hoefling}, but our model also suggest that this behaviour is linked to the non-Gaussian probability density with exponential tails which may be observed in the same time regime. Subdiffusive non-Gaussian observations were reported in \cite{lampo17,weeksCages,chaudhuri,weeks2}. The derived leading behaviour of the probability density \reff{eq:besselApprAsympt}, formulas for kurtosis \reff{eq:4momAppr} and logarithm of the characteristic function \reff{eq:lcfAppr} may be of use in analysing this type of experimental data.

The model which we consider is one-dimensional so one must be careful with applying it to systems with higher dimensionality. Nevertheless, it provides strong insights for the crucial aspects of the dynamics applicable to the less idealised cases: the random walk behaviour should be observed for more general systems divided into the localised, strongly confining domains. What changes are the domain sizes and escape times. But, during our analytical calculations of the non-Gaussianity measures and the probability density we introduced few subsequent simplifications to the model, removing the dependence of the environment, correlation and even approximating transition times. Yet, all the obtained random walks were remarkably similar, showing a surprising level of universality for our results.

It should be stressed that by this argument we actually establish not one, but three connections, each remarkable on its own: between systems with barriers and random walks, between random walks and \textit{Brownian, yet non Gaussian diffusion}, and finally, between non Gaussian diffusion and systems with barriers. This approach opens new research possibilities in each of these areas, but also offers immediate scientific benefits: we manage to express the probability density of the displacements, non-Gaussianity measures and the effective diffusion coefficients using three parameters which describe the medium at microscopic level: in-domain diffusion coefficient, the average distance between barriers and their permeability. These relations form a valuable link between the molecular structure of the systems in question and the measurements of the non-Gaussian diffusion.

\newpage

\section*{Appendix: Notation}
\begin{itemize}[leftmargin=*]
	\item i.i.d. independent and identically distributed
	\item $\deq$ equal in distribution, has a distribution
	\item $\mathcal Unif(a,b)$ uniform distribution, $p_{\mathcal Unif(a,b)}(x) = \boldsymbol 1_{(a,b)}(x)/(b-a)$
	\item $\mathcal G am(\alpha,\lambda)$ gamma distribution, $p_{\mathcal G am(\alpha,\lambda)}(x) = \f{\lambda^\alpha}{\Gamma(\alpha)} x^{\alpha-1}\e^{-\lambda x}$; it has mean $\alpha/\lambda$
	\item $\mathcal Exp(\lambda)$ exponential distribution, $\mathcal Exp(\lambda) = \mathcal G(1,\lambda)$; it has mean $1/\lambda$ thus $\lambda$ is decay rate.
	\item $\mathcal Lap(\mu,s)$ Laplace distribution, $p_{\mathcal Lap(\mu,s)}(x) = \f{1}{2s}\e^{-|x-\mu|/s} $; it has variance $2s^2$
	\item $I_n(x),K_n(x)$ modified Bessel functions of the second kind
\end{itemize}

\begin{acknowledgements}
	This work was supported by the Pazy foundation grant 61139927.
\end{acknowledgements}

\section*{References}
\bibliography{barriersBib}

\begin{thebibliography}{82}%
\makeatletter
\providecommand \@ifxundefined [1]{%
 \@ifx{#1\undefined}
}%
\providecommand \@ifnum [1]{%
 \ifnum #1\expandafter \@firstoftwo
 \else \expandafter \@secondoftwo
 \fi
}%
\providecommand \@ifx [1]{%
 \ifx #1\expandafter \@firstoftwo
 \else \expandafter \@secondoftwo
 \fi
}%
\providecommand \natexlab [1]{#1}%
\providecommand \enquote  [1]{``#1''}%
\providecommand \bibnamefont  [1]{#1}%
\providecommand \bibfnamefont [1]{#1}%
\providecommand \citenamefont [1]{#1}%
\providecommand \href@noop [0]{\@secondoftwo}%
\providecommand \href [0]{\begingroup \@sanitize@url \@href}%
\providecommand \@href[1]{\@@startlink{#1}\@@href}%
\providecommand \@@href[1]{\endgroup#1\@@endlink}%
\providecommand \@sanitize@url [0]{\catcode `\\12\catcode `\$12\catcode
  `\&12\catcode `\#12\catcode `\^12\catcode `\_12\catcode `\%12\relax}%
\providecommand \@@startlink[1]{}%
\providecommand \@@endlink[0]{}%
\providecommand \url  [0]{\begingroup\@sanitize@url \@url }%
\providecommand \@url [1]{\endgroup\@href {#1}{\urlprefix }}%
\providecommand \urlprefix  [0]{URL }%
\providecommand \Eprint [0]{\href }%
\providecommand \doibase [0]{https://doi.org/}%
\providecommand \selectlanguage [0]{\@gobble}%
\providecommand \bibinfo  [0]{\@secondoftwo}%
\providecommand \bibfield  [0]{\@secondoftwo}%
\providecommand \translation [1]{[#1]}%
\providecommand \BibitemOpen [0]{}%
\providecommand \bibitemStop [0]{}%
\providecommand \bibitemNoStop [0]{.\EOS\space}%
\providecommand \EOS [0]{\spacefactor3000\relax}%
\providecommand \BibitemShut  [1]{\csname bibitem#1\endcsname}%
\let\auto@bib@innerbib\@empty
\bibitem [{\citenamefont {Feynman}(1964)}]{feynmanBM}%
  \BibitemOpen
  \bibfield  {author} {\bibinfo {author} {\bibfnamefont {R.~P.}\ \bibnamefont
  {Feynman}},\ }\bibfield  {title} {\bibinfo {title} {The {Brownian}
  movement},\ }in\ \href@noop {} {\emph {\bibinfo {booktitle} {Feynman Lectures
  of Physics}}}\ (\bibinfo  {publisher} {Addison-Wesley},\ \bibinfo {address}
  {Boston},\ \bibinfo {year} {1964})\ Chap.~\bibinfo {chapter} {6}, pp.\
  \bibinfo {pages} {41--45}\BibitemShut {NoStop}%
\bibitem [{\citenamefont {Humphries}\ \emph {et~al.}(2010)\citenamefont
  {Humphries}, \citenamefont {Queiroz}, \citenamefont {Dyer}, \citenamefont
  {Pade}, \citenamefont {Musyl}, \citenamefont {Schaefer}, \citenamefont
  {Fuller}, \citenamefont {Brunnschweiler}, \citenamefont {Doyle},
  \citenamefont {Houghton}, \citenamefont {Hays}, \citenamefont {Jones},
  \citenamefont {Noble}, \citenamefont {Wearmouth}, \citenamefont {Southall},\
  and\ \citenamefont {Sims}}]{searchBM}%
  \BibitemOpen
  \bibfield  {author} {\bibinfo {author} {\bibfnamefont {N.~E.}\ \bibnamefont
  {Humphries}}, \bibinfo {author} {\bibfnamefont {N.}~\bibnamefont {Queiroz}},
  \bibinfo {author} {\bibfnamefont {J.~R.~M.}\ \bibnamefont {Dyer}}, \bibinfo
  {author} {\bibfnamefont {N.~G.}\ \bibnamefont {Pade}}, \bibinfo {author}
  {\bibfnamefont {M.~K.}\ \bibnamefont {Musyl}}, \bibinfo {author}
  {\bibfnamefont {K.~M.}\ \bibnamefont {Schaefer}}, \bibinfo {author}
  {\bibfnamefont {D.~W.}\ \bibnamefont {Fuller}}, \bibinfo {author}
  {\bibfnamefont {J.~M.}\ \bibnamefont {Brunnschweiler}}, \bibinfo {author}
  {\bibfnamefont {T.~K.}\ \bibnamefont {Doyle}}, \bibinfo {author}
  {\bibfnamefont {J.~D.~R.}\ \bibnamefont {Houghton}}, \bibinfo {author}
  {\bibfnamefont {G.~C.}\ \bibnamefont {Hays}}, \bibinfo {author}
  {\bibfnamefont {C.~S.}\ \bibnamefont {Jones}}, \bibinfo {author}
  {\bibfnamefont {L.~R.}\ \bibnamefont {Noble}}, \bibinfo {author}
  {\bibfnamefont {V.~J.}\ \bibnamefont {Wearmouth}}, \bibinfo {author}
  {\bibfnamefont {E.~J.}\ \bibnamefont {Southall}},\ and\ \bibinfo {author}
  {\bibfnamefont {D.~W.}\ \bibnamefont {Sims}},\ }\bibfield  {title} {\bibinfo
  {title} {Environmental context explains {L{\'e}vy} and {Brownian} movement
  patterns of marine predators},\ }\href {https://doi.org/10.1038/nature09116}
  {\bibfield  {journal} {\bibinfo  {journal} {Nature}\ }\textbf {\bibinfo
  {volume} {465}},\ \bibinfo {pages} {1066} (\bibinfo {year}
  {2010})}\BibitemShut {NoStop}%
\bibitem [{\citenamefont {Merton}(1971)}]{merton}%
  \BibitemOpen
  \bibfield  {author} {\bibinfo {author} {\bibfnamefont {R.~C.}\ \bibnamefont
  {Merton}},\ }\bibfield  {title} {\bibinfo {title} {Optimum consumption and
  portfolio rules in a continuous-time model},\ }\href
  {https://doi.org/10.1016/0022-0531(71)90038-X} {\bibfield  {journal}
  {\bibinfo  {journal} {J. Econ. Theory}\ }\textbf {\bibinfo {volume} {3}},\
  \bibinfo {pages} {373 } (\bibinfo {year} {1971})}\BibitemShut {NoStop}%
\bibitem [{\citenamefont {Einstein}(1905)}]{einstein}%
  \BibitemOpen
  \bibfield  {author} {\bibinfo {author} {\bibfnamefont {A.}~\bibnamefont
  {Einstein}},\ }\bibfield  {title} {\bibinfo {title} {{{\"U}ber die von der
  molekularkinetischen Theorie der W{\"a}rme geforderte Bewegung von in
  ruhenden Fl{\"u}ssigkeiten suspendierten Teilchen}},\ }\href
  {https://doi.org/10.1002/andp.19053220806} {\bibfield  {journal} {\bibinfo
  {journal} {Ann. der Physik}\ }\textbf {\bibinfo {volume} {322}},\ \bibinfo
  {pages} {549} (\bibinfo {year} {1905})}\BibitemShut {NoStop}%
\bibitem [{\citenamefont {Billingsley}(1999)}]{billingsley}%
  \BibitemOpen
  \bibfield  {author} {\bibinfo {author} {\bibfnamefont {P.}~\bibnamefont
  {Billingsley}},\ }\href@noop {} {\emph {\bibinfo {title} {Convergence of
  Probability Measures}}},\ \bibinfo {edition} {2nd}\ ed.\ (\bibinfo
  {publisher} {Wiley},\ \bibinfo {year} {1999})\BibitemShut {NoStop}%
\bibitem [{\citenamefont {Shlesinger}\ \emph {et~al.}(1999)\citenamefont
  {Shlesinger}, \citenamefont {Klafter},\ and\ \citenamefont
  {Zumofen}}]{shlesinger}%
  \BibitemOpen
  \bibfield  {author} {\bibinfo {author} {\bibfnamefont {M.~F.}\ \bibnamefont
  {Shlesinger}}, \bibinfo {author} {\bibfnamefont {J.}~\bibnamefont
  {Klafter}},\ and\ \bibinfo {author} {\bibfnamefont {G.}~\bibnamefont
  {Zumofen}},\ }\bibfield  {title} {\bibinfo {title} {Above, below and beyond
  {Brownian} motion},\ }\href {https://doi.org/10.1119/1.19112} {\bibfield
  {journal} {\bibinfo  {journal} {Am. J. Phys.}\ }\textbf {\bibinfo {volume}
  {67}},\ \bibinfo {pages} {1253} (\bibinfo {year} {1999})}\BibitemShut
  {NoStop}%
\bibitem [{\citenamefont {Metzler}\ and\ \citenamefont
  {Klafter}(2000)}]{guide}%
  \BibitemOpen
  \bibfield  {author} {\bibinfo {author} {\bibfnamefont {R.}~\bibnamefont
  {Metzler}}\ and\ \bibinfo {author} {\bibfnamefont {J.}~\bibnamefont
  {Klafter}},\ }\bibfield  {title} {\bibinfo {title} {The random walk's guide
  to anomalous diffusion: a fractional dynamics approach},\ }\href
  {https://doi.org/10.1016/S0370-1573(00)00070-3} {\bibfield  {journal}
  {\bibinfo  {journal} {Phys. Rep.}\ }\textbf {\bibinfo {volume} {339}},\
  \bibinfo {pages} {1 } (\bibinfo {year} {2000})}\BibitemShut {NoStop}%
\bibitem [{\citenamefont {Coffey}\ \emph {et~al.}(1996)\citenamefont {Coffey},
  \citenamefont {Kalmykov},\ and\ \citenamefont {Waldron}}]{coffey}%
  \BibitemOpen
  \bibfield  {author} {\bibinfo {author} {\bibfnamefont {W.~T.}\ \bibnamefont
  {Coffey}}, \bibinfo {author} {\bibfnamefont {Y.~P.}\ \bibnamefont
  {Kalmykov}},\ and\ \bibinfo {author} {\bibfnamefont {J.~T.}\ \bibnamefont
  {Waldron}},\ }\href {https://doi.org/10.1142/2256} {\emph {\bibinfo {title}
  {The Langevin Equation}}}\ (\bibinfo  {publisher} {Word Scientific},\
  \bibinfo {year} {1996})\BibitemShut {NoStop}%
\bibitem [{\citenamefont {Wang}\ \emph {et~al.}(2012)\citenamefont {Wang},
  \citenamefont {Kuo}, \citenamefont {Bae},\ and\ \citenamefont
  {Granick}}]{wangNG}%
  \BibitemOpen
  \bibfield  {author} {\bibinfo {author} {\bibfnamefont {B.}~\bibnamefont
  {Wang}}, \bibinfo {author} {\bibfnamefont {J.}~\bibnamefont {Kuo}}, \bibinfo
  {author} {\bibfnamefont {S.~C.}\ \bibnamefont {Bae}},\ and\ \bibinfo {author}
  {\bibfnamefont {S.}~\bibnamefont {Granick}},\ }\bibfield  {title} {\bibinfo
  {title} {When {Brownian} diffusion is not {Gaussian}},\ }\href
  {https://doi.org/10.1038/nmat3308} {\bibfield  {journal} {\bibinfo  {journal}
  {Nat. Mat.}\ }\textbf {\bibinfo {volume} {11}},\ \bibinfo {pages} {481}
  (\bibinfo {year} {2012})}\BibitemShut {NoStop}%
\bibitem [{\citenamefont {Metzler}(2017)}]{fair}%
  \BibitemOpen
  \bibfield  {author} {\bibinfo {author} {\bibfnamefont {R.}~\bibnamefont
  {Metzler}},\ }\bibfield  {title} {\bibinfo {title} {Gaussianity fair: The
  riddle of anomalous yet non-{Gaussian} diffusion},\ }\href
  {https://doi.org/10.1016/j.bpj.2016.12.019} {\bibfield  {journal} {\bibinfo
  {journal} {Biophys. J.}\ }\textbf {\bibinfo {volume} {112}},\ \bibinfo
  {pages} {413} (\bibinfo {year} {2017})}\BibitemShut {NoStop}%
\bibitem [{\citenamefont {Bhattacharya}\ \emph {et~al.}(2013)\citenamefont
  {Bhattacharya}, \citenamefont {Sharma}, \citenamefont {Saurabh},
  \citenamefont {De}, \citenamefont {Sain}, \citenamefont {Nandi},\ and\
  \citenamefont {Chowdhury}}]{bhattacharya}%
  \BibitemOpen
  \bibfield  {author} {\bibinfo {author} {\bibfnamefont {S.}~\bibnamefont
  {Bhattacharya}}, \bibinfo {author} {\bibfnamefont {D.}~\bibnamefont
  {Sharma}}, \bibinfo {author} {\bibfnamefont {S.}~\bibnamefont {Saurabh}},
  \bibinfo {author} {\bibfnamefont {S.}~\bibnamefont {De}}, \bibinfo {author}
  {\bibfnamefont {A.}~\bibnamefont {Sain}}, \bibinfo {author} {\bibfnamefont
  {A.}~\bibnamefont {Nandi}},\ and\ \bibinfo {author} {\bibfnamefont
  {A.}~\bibnamefont {Chowdhury}},\ }\bibfield  {title} {\bibinfo {title}
  {Plasticization of poly(vinylpyrrolidone) thin films under ambient humidity:
  Insight from single-molecule tracer diffusion dynamics},\ }\href
  {https://doi.org/10.1021/jp401704e} {\bibfield  {journal} {\bibinfo
  {journal} {J. Phys. Chem. B}\ }\textbf {\bibinfo {volume} {117}},\ \bibinfo
  {pages} {7771} (\bibinfo {year} {2013})}\BibitemShut {NoStop}%
\bibitem [{\citenamefont {Hapca}\ \emph {et~al.}(2009)\citenamefont {Hapca},
  \citenamefont {Crawford},\ and\ \citenamefont {Young}}]{hapca}%
  \BibitemOpen
  \bibfield  {author} {\bibinfo {author} {\bibfnamefont {S.}~\bibnamefont
  {Hapca}}, \bibinfo {author} {\bibfnamefont {J.~W.}\ \bibnamefont
  {Crawford}},\ and\ \bibinfo {author} {\bibfnamefont {I.~M.}\ \bibnamefont
  {Young}},\ }\bibfield  {title} {\bibinfo {title} {Anomalous diffusion of
  heterogeneous populations characterized by normal diffusion at the individual
  level},\ }\href {https://doi.org/10.1098/rsif.2008.0261} {\bibfield
  {journal} {\bibinfo  {journal} {J. Roy. Soc. Interface}\ }\textbf {\bibinfo
  {volume} {6}},\ \bibinfo {pages} {111} (\bibinfo {year} {2009})}\BibitemShut
  {NoStop}%
\bibitem [{\citenamefont {Lampo}\ \emph {et~al.}(2017)\citenamefont {Lampo},
  \citenamefont {Stylianidou}, \citenamefont {Backlund}, \citenamefont
  {Wiggins},\ and\ \citenamefont {Spakowitz}}]{lampo17}%
  \BibitemOpen
  \bibfield  {author} {\bibinfo {author} {\bibfnamefont {T.~J.}\ \bibnamefont
  {Lampo}}, \bibinfo {author} {\bibfnamefont {S.}~\bibnamefont {Stylianidou}},
  \bibinfo {author} {\bibfnamefont {M.~P.}\ \bibnamefont {Backlund}}, \bibinfo
  {author} {\bibfnamefont {P.~A.}\ \bibnamefont {Wiggins}},\ and\ \bibinfo
  {author} {\bibfnamefont {A.~J.}\ \bibnamefont {Spakowitz}},\ }\bibfield
  {title} {\bibinfo {title} {Cytoplasmic {RNA}-protein particles exhibit
  non-{Gaussian} subdiffusive behavior},\ }\href
  {https://doi.org/10.1016/j.bpj.2016.11.3208} {\bibfield  {journal} {\bibinfo
  {journal} {Biophys. J.}\ }\textbf {\bibinfo {volume} {112}},\ \bibinfo
  {pages} {532 } (\bibinfo {year} {2017})}\BibitemShut {NoStop}%
\bibitem [{\citenamefont {Wang}\ \emph {et~al.}(2009)\citenamefont {Wang},
  \citenamefont {Anthony}, \citenamefont {Bae},\ and\ \citenamefont
  {Granick}}]{wang2}%
  \BibitemOpen
  \bibfield  {author} {\bibinfo {author} {\bibfnamefont {B.}~\bibnamefont
  {Wang}}, \bibinfo {author} {\bibfnamefont {S.~M.}\ \bibnamefont {Anthony}},
  \bibinfo {author} {\bibfnamefont {S.~C.}\ \bibnamefont {Bae}},\ and\ \bibinfo
  {author} {\bibfnamefont {S.}~\bibnamefont {Granick}},\ }\bibfield  {title}
  {\bibinfo {title} {Anomalous yet {Brownian}},\ }\href
  {https://doi.org/10.1073/pnas.0903554106} {\bibfield  {journal} {\bibinfo
  {journal} {Proc. Natl. Acad. Sci. U.S.A.}\ }\textbf {\bibinfo {volume}
  {106}},\ \bibinfo {pages} {15160} (\bibinfo {year} {2009})}\BibitemShut
  {NoStop}%
\bibitem [{\citenamefont {Beck}\ and\ \citenamefont {Cohen}(2003)}]{beck2}%
  \BibitemOpen
  \bibfield  {author} {\bibinfo {author} {\bibfnamefont {C.}~\bibnamefont
  {Beck}}\ and\ \bibinfo {author} {\bibfnamefont {E.~G.}\ \bibnamefont
  {Cohen}},\ }\bibfield  {title} {\bibinfo {title} {Superstatistics},\ }\href
  {https://doi.org/10.1016/S0378-4371(03)00019-0} {\bibfield  {journal}
  {\bibinfo  {journal} {Physica A}\ }\textbf {\bibinfo {volume} {322}},\
  \bibinfo {pages} {267 } (\bibinfo {year} {2003})}\BibitemShut {NoStop}%
\bibitem [{\citenamefont {Beck}(2006)}]{beck3}%
  \BibitemOpen
  \bibfield  {author} {\bibinfo {author} {\bibfnamefont {C.}~\bibnamefont
  {Beck}},\ }\bibfield  {title} {\bibinfo {title} {Superstatistical {Brownian}
  motion},\ }\href {https://doi.org/10.1143/PTPS.162.29} {\bibfield  {journal}
  {\bibinfo  {journal} {Prog. Theor. Phys. Suppl.}\ }\textbf {\bibinfo {volume}
  {162}},\ \bibinfo {pages} {29} (\bibinfo {year} {2006})}\BibitemShut
  {NoStop}%
\bibitem [{\citenamefont {Beck}(2011)}]{beck4}%
  \BibitemOpen
  \bibfield  {author} {\bibinfo {author} {\bibfnamefont {C.}~\bibnamefont
  {Beck}},\ }\bibfield  {title} {\bibinfo {title} {Generalized statistical
  mechanics for superstatistical systems},\ }\href
  {https://doi.org/10.1098/rsta.2010.0280} {\bibfield  {journal} {\bibinfo
  {journal} {Philos. T. Roy. Soc. A}\ }\textbf {\bibinfo {volume} {369}},\
  \bibinfo {pages} {453} (\bibinfo {year} {2011})}\BibitemShut {NoStop}%
\bibitem [{\citenamefont {{\'S}l\k{e}zak}\ \emph {et~al.}(2018)\citenamefont
  {{\'S}l\k{e}zak}, \citenamefont {Metzler},\ and\ \citenamefont
  {Magdziarz}}]{superstatLang}%
  \BibitemOpen
  \bibfield  {author} {\bibinfo {author} {\bibfnamefont {J.}~\bibnamefont
  {{\'S}l\k{e}zak}}, \bibinfo {author} {\bibfnamefont {R.}~\bibnamefont
  {Metzler}},\ and\ \bibinfo {author} {\bibfnamefont {M.}~\bibnamefont
  {Magdziarz}},\ }\bibfield  {title} {\bibinfo {title} {Superstatistical
  generalised {Langevin} equation: non-{Gaussian} viscoelastic anomalous
  diffusion},\ }\href {https://doi.org/10.1088/1367-2630/aaa3d4} {\bibfield
  {journal} {\bibinfo  {journal} {New J. Phys.}\ }\textbf {\bibinfo {volume}
  {20}},\ \bibinfo {pages} {023026} (\bibinfo {year} {2018})}\BibitemShut
  {NoStop}%
\bibitem [{\citenamefont {Chubynsky}\ and\ \citenamefont
  {Slater}(2014)}]{diffDiff}%
  \BibitemOpen
  \bibfield  {author} {\bibinfo {author} {\bibfnamefont {M.~V.}\ \bibnamefont
  {Chubynsky}}\ and\ \bibinfo {author} {\bibfnamefont {G.~W.}\ \bibnamefont
  {Slater}},\ }\bibfield  {title} {\bibinfo {title} {Diffusing diffusivity: A
  model for anomalous, yet {Brownian}, diffusion},\ }\href
  {https://doi.org/10.1103/PhysRevLett.113.098302} {\bibfield  {journal}
  {\bibinfo  {journal} {Phys. Rev. Lett.}\ }\textbf {\bibinfo {volume} {113}},\
  \bibinfo {pages} {098302} (\bibinfo {year} {2014})}\BibitemShut {NoStop}%
\bibitem [{\citenamefont {Lanoisel{\'e}e}\ \emph {et~al.}(2018)\citenamefont
  {Lanoisel{\'e}e}, \citenamefont {Moutal},\ and\ \citenamefont
  {Grebenkov}}]{diffDiffReact}%
  \BibitemOpen
  \bibfield  {author} {\bibinfo {author} {\bibfnamefont {Y.}~\bibnamefont
  {Lanoisel{\'e}e}}, \bibinfo {author} {\bibfnamefont {N.}~\bibnamefont
  {Moutal}},\ and\ \bibinfo {author} {\bibfnamefont {D.~S.}\ \bibnamefont
  {Grebenkov}},\ }\bibfield  {title} {\bibinfo {title} {Diffusion-limited
  reactions in dynamic heterogeneous media},\ }\href
  {https://doi.org/10.1038/s41467-018-06610-6} {\bibfield  {journal} {\bibinfo
  {journal} {Nat. Commun.}\ }\textbf {\bibinfo {volume} {9}},\ \bibinfo {pages}
  {4398} (\bibinfo {year} {2018})}\BibitemShut {NoStop}%
\bibitem [{\citenamefont {Chechkin}\ \emph {et~al.}(2017)\citenamefont
  {Chechkin}, \citenamefont {Seno}, \citenamefont {Metzler},\ and\
  \citenamefont {Sokolov}}]{diffDiffChechkin}%
  \BibitemOpen
  \bibfield  {author} {\bibinfo {author} {\bibfnamefont {A.~V.}\ \bibnamefont
  {Chechkin}}, \bibinfo {author} {\bibfnamefont {F.}~\bibnamefont {Seno}},
  \bibinfo {author} {\bibfnamefont {R.}~\bibnamefont {Metzler}},\ and\ \bibinfo
  {author} {\bibfnamefont {I.~M.}\ \bibnamefont {Sokolov}},\ }\bibfield
  {title} {\bibinfo {title} {Brownian yet non-{Gaussian} diffusion: From
  superstatistics to subordination of diffusing diffusivities},\ }\href
  {https://doi.org/10.1103/PhysRevX.7.021002} {\bibfield  {journal} {\bibinfo
  {journal} {Phys. Rev. X}\ }\textbf {\bibinfo {volume} {7}},\ \bibinfo {pages}
  {021002} (\bibinfo {year} {2017})}\BibitemShut {NoStop}%
\bibitem [{\citenamefont {Ediger}(2000)}]{hetero}%
  \BibitemOpen
  \bibfield  {author} {\bibinfo {author} {\bibfnamefont {M.~D.}\ \bibnamefont
  {Ediger}},\ }\bibfield  {title} {\bibinfo {title} {Spatially heterogeneous
  dynamics in supercooled liquids},\ }\href
  {https://doi.org/10.1146/annurev.physchem.51.1.99} {\bibfield  {journal}
  {\bibinfo  {journal} {Ann. Rev. Phys. Chem.}\ }\textbf {\bibinfo {volume}
  {51}},\ \bibinfo {pages} {99} (\bibinfo {year} {2000})}\BibitemShut {NoStop}%
\bibitem [{\citenamefont {He}\ \emph {et~al.}(2016)\citenamefont {He},
  \citenamefont {Song}, \citenamefont {Su}, \citenamefont {Geng}, \citenamefont
  {Ackerson}, \citenamefont {Peng},\ and\ \citenamefont {Tong}}]{he16}%
  \BibitemOpen
  \bibfield  {author} {\bibinfo {author} {\bibfnamefont {W.}~\bibnamefont
  {He}}, \bibinfo {author} {\bibfnamefont {H.}~\bibnamefont {Song}}, \bibinfo
  {author} {\bibfnamefont {Y.}~\bibnamefont {Su}}, \bibinfo {author}
  {\bibfnamefont {L.}~\bibnamefont {Geng}}, \bibinfo {author} {\bibfnamefont
  {B.~J.}\ \bibnamefont {Ackerson}}, \bibinfo {author} {\bibfnamefont {H.~B.}\
  \bibnamefont {Peng}},\ and\ \bibinfo {author} {\bibfnamefont
  {P.}~\bibnamefont {Tong}},\ }\bibfield  {title} {\bibinfo {title} {Dynamic
  heterogeneity and non-{Gaussian} statistics for acetylcholine receptors on
  live cell membrane},\ }\href {https://doi.org/10.1038/ncomms11701} {\bibfield
   {journal} {\bibinfo  {journal} {Nat Commun.}\ }\textbf {\bibinfo {volume}
  {7}},\ \bibinfo {pages} {178} (\bibinfo {year} {2016})}\BibitemShut {NoStop}%
\bibitem [{\citenamefont {Sokolov}(2010{\natexlab{a}})}]{itoInterp}%
  \BibitemOpen
  \bibfield  {author} {\bibinfo {author} {\bibfnamefont {I.~M.}\ \bibnamefont
  {Sokolov}},\ }\bibfield  {title} {\bibinfo {title} {{Ito, Stratonovich,
  H{\"a}nggi} and all the rest: the thermodynamics of interpretation .},\
  }\href@noop {} {\bibfield  {journal} {\bibinfo  {journal} {Chem. Phys.}\
  }\textbf {\bibinfo {volume} {375}} (\bibinfo {year}
  {2010}{\natexlab{a}})}\BibitemShut {NoStop}%
\bibitem [{\citenamefont {Postnikov}\ \emph {et~al.}()\citenamefont
  {Postnikov}, \citenamefont {Chechkin},\ and\ \citenamefont
  {Sokolov}}]{BnGPostnikov}%
  \BibitemOpen
  \bibfield  {author} {\bibinfo {author} {\bibfnamefont {E.~B.}\ \bibnamefont
  {Postnikov}}, \bibinfo {author} {\bibfnamefont {A.}~\bibnamefont
  {Chechkin}},\ and\ \bibinfo {author} {\bibfnamefont {I.~M.}\ \bibnamefont
  {Sokolov}},\ }\bibfield  {title} {\bibinfo {title} {Brownian yet
  non-{Gaussian} diffusion in heterogeneous media: from superstatistics to
  homogenization},\ }\href@noop {} {\bibfield  {journal} {\bibinfo  {journal}
  {New J. Phys.}\ }\textbf {\bibinfo {volume} {22}},\ \bibinfo {pages}
  {063046}}\BibitemShut {NoStop}%
\bibitem [{\citenamefont {Novikov}\ \emph {et~al.}(2011)\citenamefont
  {Novikov}, \citenamefont {Fieremans}, \citenamefont {Jensen},\ and\
  \citenamefont {Helpern}}]{novikov}%
  \BibitemOpen
  \bibfield  {author} {\bibinfo {author} {\bibfnamefont {D.~S.}\ \bibnamefont
  {Novikov}}, \bibinfo {author} {\bibfnamefont {E.}~\bibnamefont {Fieremans}},
  \bibinfo {author} {\bibfnamefont {J.~H.}\ \bibnamefont {Jensen}},\ and\
  \bibinfo {author} {\bibfnamefont {J.~A.}\ \bibnamefont {Helpern}},\
  }\bibfield  {title} {\bibinfo {title} {Random walks with barriers},\ }\href
  {https://doi.org/10.1038/nphys1936} {\bibfield  {journal} {\bibinfo
  {journal} {Nat. Phys.}\ }\textbf {\bibinfo {volume} {7}},\ \bibinfo {pages}
  {508} (\bibinfo {year} {2011})}\BibitemShut {NoStop}%
\bibitem [{\citenamefont {Song}\ \emph {et~al.}(2000)\citenamefont {Song},
  \citenamefont {Ryu},\ and\ \citenamefont {Sen}}]{song}%
  \BibitemOpen
  \bibfield  {author} {\bibinfo {author} {\bibfnamefont {Y.-Q.}\ \bibnamefont
  {Song}}, \bibinfo {author} {\bibfnamefont {S.}~\bibnamefont {Ryu}},\ and\
  \bibinfo {author} {\bibfnamefont {P.~N.}\ \bibnamefont {Sen}},\ }\bibfield
  {title} {\bibinfo {title} {Determining multiple length scales in rocks},\
  }\href {https://doi.org/10.1038/35018057d} {\bibfield  {journal} {\bibinfo
  {journal} {Nature}\ }\textbf {\bibinfo {volume} {406}},\ \bibinfo {pages}
  {11701} (\bibinfo {year} {2000})}\BibitemShut {NoStop}%
\bibitem [{\citenamefont {Mair}\ \emph {et~al.}(1999)\citenamefont {Mair},
  \citenamefont {Wong}, \citenamefont {Hoffmann}, \citenamefont {H\"urlimann},
  \citenamefont {Patz}, \citenamefont {Schwartz},\ and\ \citenamefont
  {Walsworth}}]{mair}%
  \BibitemOpen
  \bibfield  {author} {\bibinfo {author} {\bibfnamefont {R.~W.}\ \bibnamefont
  {Mair}}, \bibinfo {author} {\bibfnamefont {G.~P.}\ \bibnamefont {Wong}},
  \bibinfo {author} {\bibfnamefont {D.}~\bibnamefont {Hoffmann}}, \bibinfo
  {author} {\bibfnamefont {M.~D.}\ \bibnamefont {H\"urlimann}}, \bibinfo
  {author} {\bibfnamefont {S.}~\bibnamefont {Patz}}, \bibinfo {author}
  {\bibfnamefont {L.~M.}\ \bibnamefont {Schwartz}},\ and\ \bibinfo {author}
  {\bibfnamefont {R.~L.}\ \bibnamefont {Walsworth}},\ }\bibfield  {title}
  {\bibinfo {title} {Probing porous media with gas diffusion {NMR}},\ }\href
  {https://doi.org/10.1103/PhysRevLett.83.3324} {\bibfield  {journal} {\bibinfo
   {journal} {Phys. Rev. Lett.}\ }\textbf {\bibinfo {volume} {83}},\ \bibinfo
  {pages} {3324} (\bibinfo {year} {1999})}\BibitemShut {NoStop}%
\bibitem [{\citenamefont {Fujiwara}\ \emph {et~al.}(2016)\citenamefont
  {Fujiwara}, \citenamefont {Iwasawa}, \citenamefont {Kalay}, \citenamefont
  {Tsunoyama}, \citenamefont {Watanabe}, \citenamefont {Umemura}, \citenamefont
  {Murakoshi}, \citenamefont {Suzuki}, \citenamefont {Nemoto}, \citenamefont
  {Morone},\ and\ \citenamefont {Kusumi}}]{confDiff}%
  \BibitemOpen
  \bibfield  {author} {\bibinfo {author} {\bibfnamefont {T.~K.}\ \bibnamefont
  {Fujiwara}}, \bibinfo {author} {\bibfnamefont {K.}~\bibnamefont {Iwasawa}},
  \bibinfo {author} {\bibfnamefont {Z.}~\bibnamefont {Kalay}}, \bibinfo
  {author} {\bibfnamefont {T.~A.}\ \bibnamefont {Tsunoyama}}, \bibinfo {author}
  {\bibfnamefont {Y.}~\bibnamefont {Watanabe}}, \bibinfo {author}
  {\bibfnamefont {Y.~M.}\ \bibnamefont {Umemura}}, \bibinfo {author}
  {\bibfnamefont {H.}~\bibnamefont {Murakoshi}}, \bibinfo {author}
  {\bibfnamefont {K.~G.~N.}\ \bibnamefont {Suzuki}}, \bibinfo {author}
  {\bibfnamefont {Y.~L.}\ \bibnamefont {Nemoto}}, \bibinfo {author}
  {\bibfnamefont {N.}~\bibnamefont {Morone}},\ and\ \bibinfo {author}
  {\bibfnamefont {A.}~\bibnamefont {Kusumi}},\ }\bibfield  {title} {\bibinfo
  {title} {Confined diffusion of transmembrane proteins and lipids induced by
  the same actin meshwork lining the plasma membrane},\ }\href
  {https://doi.org/10.1091/mbc.E15-04-0186} {\bibfield  {journal} {\bibinfo
  {journal} {Mol. Biol. Cell}\ }\textbf {\bibinfo {volume} {27}},\ \bibinfo
  {pages} {1101} (\bibinfo {year} {2016})}\BibitemShut {NoStop}%
\bibitem [{\citenamefont {Sadegh}\ \emph {et~al.}(2017)\citenamefont {Sadegh},
  \citenamefont {Higgins}, \citenamefont {Mannion}, \citenamefont {Tamkun},\
  and\ \citenamefont {Krapf}}]{krapfComp}%
  \BibitemOpen
  \bibfield  {author} {\bibinfo {author} {\bibfnamefont {S.}~\bibnamefont
  {Sadegh}}, \bibinfo {author} {\bibfnamefont {J.~L.}\ \bibnamefont {Higgins}},
  \bibinfo {author} {\bibfnamefont {P.~C.}\ \bibnamefont {Mannion}}, \bibinfo
  {author} {\bibfnamefont {M.~M.}\ \bibnamefont {Tamkun}},\ and\ \bibinfo
  {author} {\bibfnamefont {D.}~\bibnamefont {Krapf}},\ }\bibfield  {title}
  {\bibinfo {title} {Plasma membrane is compartmentalized by a self-similar
  cortical actin meshwork},\ }\href {https://doi.org/10.1103/PhysRevX.7.011031}
  {\bibfield  {journal} {\bibinfo  {journal} {Phys. Rev. X}\ }\textbf {\bibinfo
  {volume} {7}},\ \bibinfo {pages} {011031} (\bibinfo {year}
  {2017})}\BibitemShut {NoStop}%
\bibitem [{\citenamefont {Chein}\ \emph {et~al.}(2019)\citenamefont {Chein},
  \citenamefont {Perlson},\ and\ \citenamefont {Roichman}}]{flow}%
  \BibitemOpen
  \bibfield  {author} {\bibinfo {author} {\bibfnamefont {M.}~\bibnamefont
  {Chein}}, \bibinfo {author} {\bibfnamefont {E.}~\bibnamefont {Perlson}},\
  and\ \bibinfo {author} {\bibfnamefont {Y.}~\bibnamefont {Roichman}},\
  }\bibfield  {title} {\bibinfo {title} {Flow arrest in the plasma membrane},\
  }\href {https://doi.org/10.1016/j.bpj.2019.07.001} {\bibfield  {journal}
  {\bibinfo  {journal} {Biophys. J.}\ }\textbf {\bibinfo {volume} {117}},\
  \bibinfo {pages} {1 } (\bibinfo {year} {2019})}\BibitemShut {NoStop}%
\bibitem [{\citenamefont {Kusumi}\ \emph {et~al.}(2005)\citenamefont {Kusumi},
  \citenamefont {Nakada}, \citenamefont {Ritchie}, \citenamefont {Murase},
  \citenamefont {Suzuki}, \citenamefont {Murakoshi}, \citenamefont {Kasai},
  \citenamefont {Kondo},\ and\ \citenamefont {Fujiwara}}]{kasumi}%
  \BibitemOpen
  \bibfield  {author} {\bibinfo {author} {\bibfnamefont {A.}~\bibnamefont
  {Kusumi}}, \bibinfo {author} {\bibfnamefont {C.}~\bibnamefont {Nakada}},
  \bibinfo {author} {\bibfnamefont {K.}~\bibnamefont {Ritchie}}, \bibinfo
  {author} {\bibfnamefont {K.}~\bibnamefont {Murase}}, \bibinfo {author}
  {\bibfnamefont {K.}~\bibnamefont {Suzuki}}, \bibinfo {author} {\bibfnamefont
  {H.}~\bibnamefont {Murakoshi}}, \bibinfo {author} {\bibfnamefont {R.~S.}\
  \bibnamefont {Kasai}}, \bibinfo {author} {\bibfnamefont {J.}~\bibnamefont
  {Kondo}},\ and\ \bibinfo {author} {\bibfnamefont {T.}~\bibnamefont
  {Fujiwara}},\ }\bibfield  {title} {\bibinfo {title} {Paradigm shift of the
  plasma membrane concept from the two-dimensional continuum fluid to the
  partitioned fluid: High-speed single-molecule tracking of membrane
  molecules},\ }\href
  {https://doi.org/10.1146/annurev.biophys.34.040204.144637} {\bibfield
  {journal} {\bibinfo  {journal} {Annu. Rev. Biophys. Biomol. Struct.}\
  }\textbf {\bibinfo {volume} {34}},\ \bibinfo {pages} {351} (\bibinfo {year}
  {2005})}\BibitemShut {NoStop}%
\bibitem [{\citenamefont {Sykov{\'a}}\ and\ \citenamefont
  {Nicholson}(2008)}]{sykova}%
  \BibitemOpen
  \bibfield  {author} {\bibinfo {author} {\bibfnamefont {E.}~\bibnamefont
  {Sykov{\'a}}}\ and\ \bibinfo {author} {\bibfnamefont {C.}~\bibnamefont
  {Nicholson}},\ }\bibfield  {title} {\bibinfo {title} {Diffusion in brain
  extracellular space},\ }\href {https://doi.org/10.1152/physrev.00027.2007}
  {\bibfield  {journal} {\bibinfo  {journal} {Physiol. Rev.}\ }\textbf
  {\bibinfo {volume} {88}},\ \bibinfo {pages} {1277} (\bibinfo {year}
  {2008})}\BibitemShut {NoStop}%
\bibitem [{\citenamefont {Cory}\ and\ \citenamefont {Garroway}(1990)}]{cory}%
  \BibitemOpen
  \bibfield  {author} {\bibinfo {author} {\bibfnamefont {D.~G.}\ \bibnamefont
  {Cory}}\ and\ \bibinfo {author} {\bibfnamefont {A.~N.}\ \bibnamefont
  {Garroway}},\ }\bibfield  {title} {\bibinfo {title} {Measurement of
  translational displacement probabilities by {NMR}: an indicator of
  compartmentation.},\ }\href@noop {} {\bibfield  {journal} {\bibinfo
  {journal} {Magn. Reson. Med.}\ }\textbf {\bibinfo {volume} {14}},\ \bibinfo
  {pages} {435} (\bibinfo {year} {1990})}\BibitemShut {NoStop}%
\bibitem [{\citenamefont {Yablonskiy}\ \emph {et~al.}(2002)\citenamefont
  {Yablonskiy}, \citenamefont {Sukstanskii}, \citenamefont {Leawoods},
  \citenamefont {Gierada}, \citenamefont {Bretthorst}, \citenamefont {Lefrak},
  \citenamefont {Cooper},\ and\ \citenamefont {Conradi}}]{yablonskiy}%
  \BibitemOpen
  \bibfield  {author} {\bibinfo {author} {\bibfnamefont {D.~A.}\ \bibnamefont
  {Yablonskiy}}, \bibinfo {author} {\bibfnamefont {A.~L.}\ \bibnamefont
  {Sukstanskii}}, \bibinfo {author} {\bibfnamefont {J.~C.}\ \bibnamefont
  {Leawoods}}, \bibinfo {author} {\bibfnamefont {D.~S.}\ \bibnamefont
  {Gierada}}, \bibinfo {author} {\bibfnamefont {G.~L.}\ \bibnamefont
  {Bretthorst}}, \bibinfo {author} {\bibfnamefont {S.~S.}\ \bibnamefont
  {Lefrak}}, \bibinfo {author} {\bibfnamefont {J.~D.}\ \bibnamefont {Cooper}},\
  and\ \bibinfo {author} {\bibfnamefont {M.~S.}\ \bibnamefont {Conradi}},\
  }\bibfield  {title} {\bibinfo {title} {Quantitative in vivo assessment of
  lung microstructure at the alveolar level with hyperpolarized $^3${He}
  diffusion {MRI}},\ }\href {https://doi.org/10.1073/pnas.052594699} {\bibfield
   {journal} {\bibinfo  {journal} {Proc. Natl. Acad. Sci. U.S.A.}\ }\textbf
  {\bibinfo {volume} {99}},\ \bibinfo {pages} {3111} (\bibinfo {year}
  {2002})}\BibitemShut {NoStop}%
\bibitem [{\citenamefont {Krapf}(2018)}]{krapfComp2}%
  \BibitemOpen
  \bibfield  {author} {\bibinfo {author} {\bibfnamefont {D.}~\bibnamefont
  {Krapf}},\ }\bibfield  {title} {\bibinfo {title} {Compartmentalization of the
  plasma membrane},\ }\href {https://doi.org/10.1016/j.ceb.2018.04.002}
  {\bibfield  {journal} {\bibinfo  {journal} {Curr. Opin. Cell Biol.}\ }\textbf
  {\bibinfo {volume} {53}},\ \bibinfo {pages} {15 } (\bibinfo {year}
  {2018})}\BibitemShut {NoStop}%
\bibitem [{\citenamefont {Fujiwara}\ \emph {et~al.}(2002)\citenamefont
  {Fujiwara}, \citenamefont {Ritchie}, \citenamefont {Murakoshi}, \citenamefont
  {Jacobson},\ and\ \citenamefont {Kusumi}}]{fujiwara}%
  \BibitemOpen
  \bibfield  {author} {\bibinfo {author} {\bibfnamefont {T.}~\bibnamefont
  {Fujiwara}}, \bibinfo {author} {\bibfnamefont {K.}~\bibnamefont {Ritchie}},
  \bibinfo {author} {\bibfnamefont {H.}~\bibnamefont {Murakoshi}}, \bibinfo
  {author} {\bibfnamefont {K.}~\bibnamefont {Jacobson}},\ and\ \bibinfo
  {author} {\bibfnamefont {A.}~\bibnamefont {Kusumi}},\ }\bibfield  {title}
  {\bibinfo {title} {Phospholipids undergo hop diffusion in compartmentalized
  cell membrane},\ }\href {https://doi.org/10.1083/jcb.200202050} {\bibfield
  {journal} {\bibinfo  {journal} {J. Cell Biol.}\ }\textbf {\bibinfo {volume}
  {157}},\ \bibinfo {pages} {1071} (\bibinfo {year} {2002})}\BibitemShut
  {NoStop}%
\bibitem [{\citenamefont {Murase}\ \emph {et~al.}(2004)\citenamefont {Murase},
  \citenamefont {Fujiwara}, \citenamefont {Yasuhiro}, \citenamefont {Suzuki},
  \citenamefont {Iino}, \citenamefont {Yamashita}, \citenamefont {Saito},
  \citenamefont {Murakoshi}, \citenamefont {Ritchie},\ and\ \citenamefont
  {Kusumi}}]{murase}%
  \BibitemOpen
  \bibfield  {author} {\bibinfo {author} {\bibfnamefont {K.}~\bibnamefont
  {Murase}}, \bibinfo {author} {\bibfnamefont {T.}~\bibnamefont {Fujiwara}},
  \bibinfo {author} {\bibfnamefont {U.}~\bibnamefont {Yasuhiro}}, \bibinfo
  {author} {\bibfnamefont {K.}~\bibnamefont {Suzuki}}, \bibinfo {author}
  {\bibfnamefont {R.}~\bibnamefont {Iino}}, \bibinfo {author} {\bibfnamefont
  {H.}~\bibnamefont {Yamashita}}, \bibinfo {author} {\bibfnamefont
  {M.}~\bibnamefont {Saito}}, \bibinfo {author} {\bibfnamefont
  {H.}~\bibnamefont {Murakoshi}}, \bibinfo {author} {\bibfnamefont
  {K.}~\bibnamefont {Ritchie}},\ and\ \bibinfo {author} {\bibfnamefont
  {A.}~\bibnamefont {Kusumi}},\ }\bibfield  {title} {\bibinfo {title}
  {Ultrafine membrane compartments for molecular diffusion as revealed by
  single molecule techniques},\ }\href
  {https://doi.org/10.1529/biophysj.103.035717} {\bibfield  {journal} {\bibinfo
   {journal} {Biophys. J.}\ }\textbf {\bibinfo {volume} {86}},\ \bibinfo
  {pages} {4075 } (\bibinfo {year} {2004})}\BibitemShut {NoStop}%
\bibitem [{\citenamefont {Latour}\ \emph {et~al.}(1994)\citenamefont {Latour},
  \citenamefont {Svoboda}, \citenamefont {Mitra},\ and\ \citenamefont
  {H.}}]{latour}%
  \BibitemOpen
  \bibfield  {author} {\bibinfo {author} {\bibfnamefont {L.~L.}\ \bibnamefont
  {Latour}}, \bibinfo {author} {\bibfnamefont {K.}~\bibnamefont {Svoboda}},
  \bibinfo {author} {\bibfnamefont {P.~P.}\ \bibnamefont {Mitra}},\ and\
  \bibinfo {author} {\bibfnamefont {S.~C.}\ \bibnamefont {H.}},\ }\bibfield
  {title} {\bibinfo {title} {Time-dependent diffusion of water in a biological
  model system},\ }\href {https://doi.org/10.1073/pnas.91.4.1229} {\bibfield
  {journal} {\bibinfo  {journal} {Proc. Natl. Acad. Sci. U.S.A.}\ }\textbf
  {\bibinfo {volume} {91}},\ \bibinfo {pages} {1229} (\bibinfo {year}
  {1994})}\BibitemShut {NoStop}%
\bibitem [{\citenamefont {Sen}(2003)}]{sen}%
  \BibitemOpen
  \bibfield  {author} {\bibinfo {author} {\bibfnamefont {P.~N.}\ \bibnamefont
  {Sen}},\ }\bibfield  {title} {\bibinfo {title} {Time-dependent diffusion
  coefficient as a probe of the permeability of the pore wall},\ }\href
  {https://doi.org/10.1063/1.1611477} {\bibfield  {journal} {\bibinfo
  {journal} {J. Chem. Phys.}\ }\textbf {\bibinfo {volume} {119}},\ \bibinfo
  {pages} {9871} (\bibinfo {year} {2003})}\BibitemShut {NoStop}%
\bibitem [{\citenamefont {Wang}\ \emph {et~al.}(2019)\citenamefont {Wang},
  \citenamefont {Wu}, \citenamefont {Liu}, \citenamefont {Chen},\ and\
  \citenamefont {Schwartz}}]{diffEscape}%
  \BibitemOpen
  \bibfield  {author} {\bibinfo {author} {\bibfnamefont {D.}~\bibnamefont
  {Wang}}, \bibinfo {author} {\bibfnamefont {H.}~\bibnamefont {Wu}}, \bibinfo
  {author} {\bibfnamefont {L.}~\bibnamefont {Liu}}, \bibinfo {author}
  {\bibfnamefont {J.}~\bibnamefont {Chen}},\ and\ \bibinfo {author}
  {\bibfnamefont {D.~K.}\ \bibnamefont {Schwartz}},\ }\bibfield  {title}
  {\bibinfo {title} {Diffusive escape of a nanoparticle from a porous cavity},\
  }\href {https://doi.org/10.1103/PhysRevLett.123.118002} {\bibfield  {journal}
  {\bibinfo  {journal} {Phys. Rev. Lett.}\ }\textbf {\bibinfo {volume} {123}},\
  \bibinfo {pages} {118002} (\bibinfo {year} {2019})}\BibitemShut {NoStop}%
\bibitem [{\citenamefont {Weeks}\ and\ \citenamefont
  {Weitz}(2002)}]{weeksCages}%
  \BibitemOpen
  \bibfield  {author} {\bibinfo {author} {\bibfnamefont {E.~R.}\ \bibnamefont
  {Weeks}}\ and\ \bibinfo {author} {\bibfnamefont {D.}~\bibnamefont {Weitz}},\
  }\bibfield  {title} {\bibinfo {title} {Subdiffusion and the cage effect
  studied near the colloidal glass transition},\ }\href
  {https://doi.org/10.1016/S0301-0104(02)00667-5} {\bibfield  {journal}
  {\bibinfo  {journal} {Chem. Phys}\ }\textbf {\bibinfo {volume} {284}},\
  \bibinfo {pages} {361 } (\bibinfo {year} {2002})},\ \bibinfo {note} {{Strange
  Kinetics}}\BibitemShut {NoStop}%
\bibitem [{\citenamefont {Chaudhuri}\ \emph {et~al.}(2007)\citenamefont
  {Chaudhuri}, \citenamefont {Berthier},\ and\ \citenamefont
  {Kob}}]{chaudhuri}%
  \BibitemOpen
  \bibfield  {author} {\bibinfo {author} {\bibfnamefont {P.}~\bibnamefont
  {Chaudhuri}}, \bibinfo {author} {\bibfnamefont {L.}~\bibnamefont
  {Berthier}},\ and\ \bibinfo {author} {\bibfnamefont {W.}~\bibnamefont
  {Kob}},\ }\bibfield  {title} {\bibinfo {title} {Universal nature of particle
  displacements close to glass and jamming transitions},\ }\href
  {https://doi.org/10.1103/PhysRevLett.99.060604} {\bibfield  {journal}
  {\bibinfo  {journal} {Phys. Rev. Lett.}\ }\textbf {\bibinfo {volume} {99}},\
  \bibinfo {pages} {060604} (\bibinfo {year} {2007})}\BibitemShut {NoStop}%
\bibitem [{\citenamefont {Weeks}\ \emph {et~al.}(2000)\citenamefont {Weeks},
  \citenamefont {Crocker}, \citenamefont {Levitt}, \citenamefont {Schofield},\
  and\ \citenamefont {Weitz}}]{weeks2}%
  \BibitemOpen
  \bibfield  {author} {\bibinfo {author} {\bibfnamefont {E.~R.}\ \bibnamefont
  {Weeks}}, \bibinfo {author} {\bibfnamefont {J.~C.}\ \bibnamefont {Crocker}},
  \bibinfo {author} {\bibfnamefont {A.~C.}\ \bibnamefont {Levitt}}, \bibinfo
  {author} {\bibfnamefont {A.}~\bibnamefont {Schofield}},\ and\ \bibinfo
  {author} {\bibfnamefont {D.~A.}\ \bibnamefont {Weitz}},\ }\bibfield  {title}
  {\bibinfo {title} {Three-dimensional direct imaging of structural relaxation
  near the colloidal glass transition},\ }\href
  {https://doi.org/10.1126/science.287.5453.627} {\bibfield  {journal}
  {\bibinfo  {journal} {Science}\ }\textbf {\bibinfo {volume} {287}},\ \bibinfo
  {pages} {627} (\bibinfo {year} {2000})}\BibitemShut {NoStop}%
\bibitem [{\citenamefont {Ebeling}\ and\ \citenamefont
  {Sokolov}(2005)}]{statTher}%
  \BibitemOpen
  \bibfield  {author} {\bibinfo {author} {\bibfnamefont {W.}~\bibnamefont
  {Ebeling}}\ and\ \bibinfo {author} {\bibfnamefont {I.~M.}\ \bibnamefont
  {Sokolov}},\ }\href {https://doi.org/10.1142/2012} {\emph {\bibinfo {title}
  {Statistical Thermodynamics and Stochastic Theory of Nonequilibrium
  Systems}}}\ (\bibinfo  {publisher} {World Scientific},\ \bibinfo {year}
  {2005})\BibitemShut {NoStop}%
\bibitem [{\citenamefont {Kallenberg}(2017)}]{kallenberg}%
  \BibitemOpen
  \bibfield  {author} {\bibinfo {author} {\bibfnamefont {O.}~\bibnamefont
  {Kallenberg}},\ }\bibfield  {title} {\bibinfo {title} {Poisson and related
  processes},\ }in\ \href {https://doi.org/10.1007/978-3-319-41598-7_3} {\emph
  {\bibinfo {booktitle} {Random Measures, Theory and Applications.}}}\
  (\bibinfo  {publisher} {Springer},\ \bibinfo {address} {Cham},\ \bibinfo
  {year} {2017})\ Chap.~\bibinfo {chapter} {3}, pp.\ \bibinfo {pages}
  {70--108}\BibitemShut {NoStop}%
\bibitem [{\citenamefont {Cox}(1962)}]{cox}%
  \BibitemOpen
  \bibfield  {author} {\bibinfo {author} {\bibfnamefont {D.~R.}\ \bibnamefont
  {Cox}},\ }\href@noop {} {\emph {\bibinfo {title} {Renewal Theory}}}\
  (\bibinfo  {publisher} {Methuen \& Co.},\ \bibinfo {year} {1962})\BibitemShut
  {NoStop}%
\bibitem [{{\relax DLMF}()}]{DLMF}%
  \BibitemOpen
  {\relax DLMF},\ \href {http://dlmf.nist.gov/} {\bibinfo {title} {{\it NIST
  Digital Library of Mathematical Functions}}},\ \bibinfo {howpublished}
  {http://dlmf.nist.gov/, Release 1.0.21 of 2018-12-15},\ \bibinfo {note}
  {f.~W.~J. Olver, A.~B. {Olde Daalhuis}, D.~W. Lozier, B.~I. Schneider, R.~F.
  Boisvert, C.~W. Clark, B.~R. Miller and B.~V. Saunders, eds.}\BibitemShut
  {Stop}%
\bibitem [{Note1()}]{Note1}%
  \BibitemOpen
  \bibinfo {note} {In this more realistic scenario the left end of the initial
  domain lays at $\Theta L$ and the right one at $(\Theta -1) L$; variable
  $\Theta $ has the uniform distribution $\protect \mathcal Unif(0,1)$.
  Calculation analogical to (\ref {eq:pdfElementary}) yields incomplete gamma
  PDF $p_X(x) = \Gamma (0,|x|)/(2|x|)$ which has tails $\sim \exp
  (-|x|)/(2|x|)$, they are even thicker than $\exp (-2|x|)$. On the other hand
  this distribution has logarithmic singularity at $x=0$ which make the motion
  more constrained at short distances.}\BibitemShut {Stop}%
\bibitem [{\citenamefont {Sokolov}(2010{\natexlab{b}})}]{sokolovInterp}%
  \BibitemOpen
  \bibfield  {author} {\bibinfo {author} {\bibfnamefont {I.~M.}\ \bibnamefont
  {Sokolov}},\ }\bibfield  {title} {\bibinfo {title} {{Ito, Stratonovich,
  H{\"a}nggi} and all the rest: The thermodynamics of interpretation},\ }\href
  {https://doi.org/10.1016/j.chemphys.2010.07.024} {\bibfield  {journal}
  {\bibinfo  {journal} {Chem. Phys.}\ }\textbf {\bibinfo {volume} {375}},\
  \bibinfo {pages} {359 } (\bibinfo {year} {2010}{\natexlab{b}})},\ \bibinfo
  {note} {{Stochastic processes in Physics and Chemistry (in honor of Peter
  H{\"a}nggi)}}\BibitemShut {NoStop}%
\bibitem [{\citenamefont {S{\'a}nchez-Palencia}(1980)}]{vibrTher}%
  \BibitemOpen
  \bibfield  {author} {\bibinfo {author} {\bibfnamefont {E.}~\bibnamefont
  {S{\'a}nchez-Palencia}},\ }\href@noop {} {\emph {\bibinfo {title}
  {Non-Homogeneous Media and Vibration Theory}}},\ \bibinfo {series} {Lecture
  Notes in Physics}, Vol.\ \bibinfo {volume} {127}\ (\bibinfo  {publisher}
  {Springer},\ \bibinfo {year} {1980})\BibitemShut {NoStop}%
\bibitem [{\citenamefont {Mandrekar}\ and\ \citenamefont
  {Pilipenko}(2016)}]{hardMem}%
  \BibitemOpen
  \bibfield  {author} {\bibinfo {author} {\bibfnamefont {V.}~\bibnamefont
  {Mandrekar}}\ and\ \bibinfo {author} {\bibfnamefont {A.}~\bibnamefont
  {Pilipenko}},\ }\bibfield  {title} {\bibinfo {title} {On a {Brownian motion}
  with a hard membrane},\ }\href {https://doi.org/10.1016/j.spl.2016.02.005}
  {\bibfield  {journal} {\bibinfo  {journal} {Stat. Probil. Lett.}\ }\textbf
  {\bibinfo {volume} {113}},\ \bibinfo {pages} {62 } (\bibinfo {year}
  {2016})}\BibitemShut {NoStop}%
\bibitem [{\citenamefont {Erban}\ and\ \citenamefont {Chapman}(2007)}]{erban}%
  \BibitemOpen
  \bibfield  {author} {\bibinfo {author} {\bibfnamefont {R.}~\bibnamefont
  {Erban}}\ and\ \bibinfo {author} {\bibfnamefont {S.~J.}\ \bibnamefont
  {Chapman}},\ }\bibfield  {title} {\bibinfo {title} {Reactive boundary
  conditions for stochastic simulations of reaction{\textendash}diffusion
  processes},\ }\href {https://doi.org/10.1088/1478-3975/4/1/003} {\bibfield
  {journal} {\bibinfo  {journal} {Phys. Biol.}\ }\textbf {\bibinfo {volume}
  {4}},\ \bibinfo {pages} {16} (\bibinfo {year} {2007})}\BibitemShut {NoStop}%
\bibitem [{\citenamefont {Andrews}(2009)}]{andrews}%
  \BibitemOpen
  \bibfield  {author} {\bibinfo {author} {\bibfnamefont {S.~S.}\ \bibnamefont
  {Andrews}},\ }\bibfield  {title} {\bibinfo {title} {Accurate particle-based
  simulation of adsorption, desorption and partial transmission},\ }\href
  {https://doi.org/10.1088/1478-3975/6/4/046015} {\bibfield  {journal}
  {\bibinfo  {journal} {Phys. Biol.}\ }\textbf {\bibinfo {volume} {6}},\
  \bibinfo {pages} {046015} (\bibinfo {year} {2009})}\BibitemShut {NoStop}%
\bibitem [{Note2()}]{Note2}%
  \BibitemOpen
  \bibinfo {note} {In the overall valuable work of Powles \protect \emph {et
  al.} \cite {powles} a system with the regularly placed barriers at $x_k \in
  \{ \protect \ldots , -1,0,1,2,\protect \ldots \}$ is considered. They claim
  they obtained the exact PDF for this particular case but no proof is provided
  and the result may be doubted.}\BibitemShut {Stop}%
\bibitem [{\citenamefont {Lejay}(2006)}]{lejaySkew}%
  \BibitemOpen
  \bibfield  {author} {\bibinfo {author} {\bibfnamefont {A.}~\bibnamefont
  {Lejay}},\ }\bibfield  {title} {\bibinfo {title} {On the constructions of the
  skew {Brownian} motion},\ }\href {https://doi.org/10.1214/154957807000000013}
  {\bibfield  {journal} {\bibinfo  {journal} {Probab. Surv.}\ }\textbf
  {\bibinfo {volume} {3}},\ \bibinfo {pages} {413} (\bibinfo {year}
  {2006})}\BibitemShut {NoStop}%
\bibitem [{\citenamefont {Lejay}(2016)}]{lejay}%
  \BibitemOpen
  \bibfield  {author} {\bibinfo {author} {\bibfnamefont {A.}~\bibnamefont
  {Lejay}},\ }\bibfield  {title} {\bibinfo {title} {The snapping out {Brownian}
  motion},\ }\href {https://doi.org/10.1214/15-AAP1131} {\bibfield  {journal}
  {\bibinfo  {journal} {Ann. Appl. Probab.}\ }\textbf {\bibinfo {volume}
  {26}},\ \bibinfo {pages} {1727} (\bibinfo {year} {2016})}\BibitemShut
  {NoStop}%
\bibitem [{\citenamefont {Lejay}(2018)}]{lejayMC}%
  \BibitemOpen
  \bibfield  {author} {\bibinfo {author} {\bibfnamefont {A.}~\bibnamefont
  {Lejay}},\ }\bibfield  {title} {\bibinfo {title} {A {Monte Carlo} estimation
  of the mean residence time in cells surrounded by thin layers},\ }\href
  {https://doi.org/10.1016/j.matcom.2017.05.008} {\bibfield  {journal}
  {\bibinfo  {journal} {Math. Comput. Simulat.}\ }\textbf {\bibinfo {volume}
  {143}},\ \bibinfo {pages} {65} (\bibinfo {year} {2018})},\ \bibinfo {note}
  {10th IMACS Seminar on Monte Carlo Methods}\BibitemShut {NoStop}%
\bibitem [{\citenamefont {Nagylaki}(1976)}]{nag}%
  \BibitemOpen
  \bibfield  {author} {\bibinfo {author} {\bibfnamefont {T.}~\bibnamefont
  {Nagylaki}},\ }\bibfield  {title} {\bibinfo {title} {Clines with variable
  migration},\ }\href@noop {} {\bibfield  {journal} {\bibinfo  {journal}
  {Genetics}\ }\textbf {\bibinfo {volume} {83}},\ \bibinfo {pages} {867}
  (\bibinfo {year} {1976})}\BibitemShut {NoStop}%
\bibitem [{\citenamefont {Barton}(2008)}]{barton}%
  \BibitemOpen
  \bibfield  {author} {\bibinfo {author} {\bibfnamefont {N.~H.}\ \bibnamefont
  {Barton}},\ }\bibfield  {title} {\bibinfo {title} {The effect of a barrier to
  gene flow on patterns of geographic variation},\ }\href
  {https://doi.org/10.1017/S0016672307009081} {\bibfield  {journal} {\bibinfo
  {journal} {Genet. Res.}\ }\textbf {\bibinfo {volume} {90}},\ \bibinfo {pages}
  {139–149} (\bibinfo {year} {2008})}\BibitemShut {NoStop}%
\bibitem [{\citenamefont {Borodin}(2017)}]{localTime}%
  \BibitemOpen
  \bibfield  {author} {\bibinfo {author} {\bibfnamefont {A.~N.}\ \bibnamefont
  {Borodin}},\ }\bibfield  {title} {\bibinfo {title} {Brownian local time},\
  }in\ \href {https://doi.org/10.1007/978-3-319-62310-8_5} {\emph {\bibinfo
  {booktitle} {Stochastic Processes.}}}\ (\bibinfo  {publisher}
  {Birkh\"auser},\ \bibinfo {year} {2017})\ pp.\ \bibinfo {pages}
  {359--438}\BibitemShut {NoStop}%
\bibitem [{\citenamefont {Barthelemy}\ \emph {et~al.}(2008)\citenamefont
  {Barthelemy}, \citenamefont {Bertolotti},\ and\ \citenamefont
  {Wiersma}}]{levyFlight}%
  \BibitemOpen
  \bibfield  {author} {\bibinfo {author} {\bibfnamefont {P.}~\bibnamefont
  {Barthelemy}}, \bibinfo {author} {\bibfnamefont {J.}~\bibnamefont
  {Bertolotti}},\ and\ \bibinfo {author} {\bibfnamefont {D.~A.}\ \bibnamefont
  {Wiersma}},\ }\bibfield  {title} {\bibinfo {title} {A l{\'e}vy flight for
  light},\ }\href {https://doi.org/10.1038/nature06948} {\bibfield  {journal}
  {\bibinfo  {journal} {Nature}\ }\textbf {\bibinfo {volume} {453}},\ \bibinfo
  {pages} {495} (\bibinfo {year} {2008})}\BibitemShut {NoStop}%
\bibitem [{\citenamefont {Burioni}\ \emph {et~al.}(2010)\citenamefont
  {Burioni}, \citenamefont {Caniparoli},\ and\ \citenamefont
  {Vezzani}}]{LWquenched}%
  \BibitemOpen
  \bibfield  {author} {\bibinfo {author} {\bibfnamefont {R.}~\bibnamefont
  {Burioni}}, \bibinfo {author} {\bibfnamefont {L.}~\bibnamefont
  {Caniparoli}},\ and\ \bibinfo {author} {\bibfnamefont {A.}~\bibnamefont
  {Vezzani}},\ }\bibfield  {title} {\bibinfo {title} {L\'evy walks and scaling
  in quenched disordered media},\ }\href
  {https://doi.org/10.1103/PhysRevE.81.060101} {\bibfield  {journal} {\bibinfo
  {journal} {Phys. Rev. E}\ }\textbf {\bibinfo {volume} {81}},\ \bibinfo
  {pages} {060101} (\bibinfo {year} {2010})}\BibitemShut {NoStop}%
\bibitem [{\citenamefont {Burioni}\ \emph {et~al.}(2014)\citenamefont
  {Burioni}, \citenamefont {Ubaldi},\ and\ \citenamefont
  {Vezzani}}]{superfiffQD}%
  \BibitemOpen
  \bibfield  {author} {\bibinfo {author} {\bibfnamefont {R.}~\bibnamefont
  {Burioni}}, \bibinfo {author} {\bibfnamefont {E.}~\bibnamefont {Ubaldi}},\
  and\ \bibinfo {author} {\bibfnamefont {A.}~\bibnamefont {Vezzani}},\
  }\bibfield  {title} {\bibinfo {title} {Superdiffusion and transport in
  two-dimensional systems with l\'evy-like quenched disorder},\ }\href
  {https://doi.org/10.1103/PhysRevE.89.022135} {\bibfield  {journal} {\bibinfo
  {journal} {Phys. Rev. E}\ }\textbf {\bibinfo {volume} {89}},\ \bibinfo
  {pages} {022135} (\bibinfo {year} {2014})}\BibitemShut {NoStop}%
\bibitem [{\citenamefont {Grebenkov}(2006)}]{grebenkovPart}%
  \BibitemOpen
  \bibfield  {author} {\bibinfo {author} {\bibfnamefont {D.~S.}\ \bibnamefont
  {Grebenkov}},\ }\bibfield  {title} {\bibinfo {title} {Partially reflected
  {Brownian} motion: A stochastic approach to transport phenomena},\ }in\
  \href@noop {} {\emph {\bibinfo {booktitle} {Focus on Probability Theory}}}\
  (\bibinfo  {publisher} {Nova Science Publishers},\ \bibinfo {year} {2006})\
  pp.\ \bibinfo {pages} {135--169}\BibitemShut {NoStop}%
\bibitem [{\citenamefont {Bouchaud}\ and\ \citenamefont
  {Georges}(1990)}]{bouchaud}%
  \BibitemOpen
  \bibfield  {author} {\bibinfo {author} {\bibfnamefont {J.-P.}\ \bibnamefont
  {Bouchaud}}\ and\ \bibinfo {author} {\bibfnamefont {A.}~\bibnamefont
  {Georges}},\ }\bibfield  {title} {\bibinfo {title} {Anomalous diffusion in
  disordered media: Statistical mechanisms, models and physical applications},\
  }\href {https://doi.org/10.1016/0370-1573(90)90099-N} {\bibfield  {journal}
  {\bibinfo  {journal} {Phys. Rep.}\ }\textbf {\bibinfo {volume} {195}},\
  \bibinfo {pages} {127 } (\bibinfo {year} {1990})}\BibitemShut {NoStop}%
\bibitem [{\citenamefont {Magdziarz}\ \emph {et~al.}(2015)\citenamefont
  {Magdziarz}, \citenamefont {Scheffler}, \citenamefont {Straka},\ and\
  \citenamefont {{\.Z}ebrowski}}]{limitTher}%
  \BibitemOpen
  \bibfield  {author} {\bibinfo {author} {\bibfnamefont {M.}~\bibnamefont
  {Magdziarz}}, \bibinfo {author} {\bibfnamefont {H.}~\bibnamefont
  {Scheffler}}, \bibinfo {author} {\bibfnamefont {P.}~\bibnamefont {Straka}},\
  and\ \bibinfo {author} {\bibfnamefont {P.}~\bibnamefont {{\.Z}ebrowski}},\
  }\bibfield  {title} {\bibinfo {title} {Limit theorems and governing equations
  for {L{\'e}vy} walks},\ }\href {https://doi.org/10.1016/j.spa.2015.05.014}
  {\bibfield  {journal} {\bibinfo  {journal} {Stoch. Proc. Appl.}\ }\textbf
  {\bibinfo {volume} {125}},\ \bibinfo {pages} {4021 } (\bibinfo {year}
  {2015})}\BibitemShut {NoStop}%
\bibitem [{\citenamefont {Tanner}(1978)}]{tanner}%
  \BibitemOpen
  \bibfield  {author} {\bibinfo {author} {\bibfnamefont {J.~E.}\ \bibnamefont
  {Tanner}},\ }\bibfield  {title} {\bibinfo {title} {Transient diffusion in a
  system partitioned by permeable barriers. {A}pplication to {NMR} measurements
  with a pulsed field gradient},\ }\href {https://doi.org/10.1063/1.436751}
  {\bibfield  {journal} {\bibinfo  {journal} {J. Chem. Phys.}\ }\textbf
  {\bibinfo {volume} {69}},\ \bibinfo {pages} {1748} (\bibinfo {year}
  {1978})}\BibitemShut {NoStop}%
\bibitem [{\citenamefont {Powles}\ \emph {et~al.}(1992)\citenamefont {Powles},
  \citenamefont {Mallett}, \citenamefont {Rickayzen},\ and\ \citenamefont
  {Evans}}]{powles}%
  \BibitemOpen
  \bibfield  {author} {\bibinfo {author} {\bibfnamefont {J.~G.}\ \bibnamefont
  {Powles}}, \bibinfo {author} {\bibfnamefont {M.~J.~D.}\ \bibnamefont
  {Mallett}}, \bibinfo {author} {\bibfnamefont {G.}~\bibnamefont {Rickayzen}},\
  and\ \bibinfo {author} {\bibfnamefont {W.~A.~B.}\ \bibnamefont {Evans}},\
  }\bibfield  {title} {\bibinfo {title} {Exact analytic solutions for diffusion
  impeded by an infinite array of partially permeable barriers},\ }\href
  {https://doi.org/10.1098/rspa.1992.0025} {\bibfield  {journal} {\bibinfo
  {journal} {Proc. Roy. Soc. A}\ }\textbf {\bibinfo {volume} {436}},\ \bibinfo
  {pages} {391} (\bibinfo {year} {1992})}\BibitemShut {NoStop}%
\bibitem [{\citenamefont {Crick}(1970)}]{crick}%
  \BibitemOpen
  \bibfield  {author} {\bibinfo {author} {\bibfnamefont {F.}~\bibnamefont
  {Crick}},\ }\bibfield  {title} {\bibinfo {title} {Diffusion in
  embryogenesis},\ }\href {https://doi.org/10.1038/225420a0} {\bibfield
  {journal} {\bibinfo  {journal} {Nature}\ }\textbf {\bibinfo {volume} {225}},\
  \bibinfo {pages} {420} (\bibinfo {year} {1970})}\BibitemShut {NoStop}%
\bibitem [{\citenamefont {Grebenkov}\ \emph {et~al.}(2014)\citenamefont
  {Grebenkov}, \citenamefont {Van~Nguyen},\ and\ \citenamefont
  {Li}}]{grebenkovBarrier}%
  \BibitemOpen
  \bibfield  {author} {\bibinfo {author} {\bibfnamefont {D.~S.}\ \bibnamefont
  {Grebenkov}}, \bibinfo {author} {\bibfnamefont {D.}~\bibnamefont
  {Van~Nguyen}},\ and\ \bibinfo {author} {\bibfnamefont {J.-R.}\ \bibnamefont
  {Li}},\ }\bibfield  {title} {\bibinfo {title} {Exploring diffusion across
  permeable barriers at high gradients. {I. Narrow} pulse approximation},\
  }\href {https://doi.org/10.1016/j.jmr.2014.07.013} {\bibfield  {journal}
  {\bibinfo  {journal} {J. Magn. Reson.}\ }\textbf {\bibinfo {volume} {248}},\
  \bibinfo {pages} {153 } (\bibinfo {year} {2014})}\BibitemShut {NoStop}%
\bibitem [{\citenamefont {H\"ofling}\ and\ \citenamefont
  {Franosch}(2013)}]{hoefling}%
  \BibitemOpen
  \bibfield  {author} {\bibinfo {author} {\bibfnamefont {F.}~\bibnamefont
  {H\"ofling}}\ and\ \bibinfo {author} {\bibfnamefont {T.}~\bibnamefont
  {Franosch}},\ }\bibfield  {title} {\bibinfo {title} {Anomalous transport in
  the crowded world of biological cells},\ }\href
  {https://doi.org/10.1088/0034-4885/76/4/046602} {\bibfield  {journal}
  {\bibinfo  {journal} {Rep. Prog. Phys.}\ }\textbf {\bibinfo {volume} {76}},\
  \bibinfo {pages} {046602} (\bibinfo {year} {2013})}\BibitemShut {NoStop}%
\bibitem [{\citenamefont {Zaburdaev}\ \emph {et~al.}(2015)\citenamefont
  {Zaburdaev}, \citenamefont {Denisov},\ and\ \citenamefont {Klafter}}]{LW}%
  \BibitemOpen
  \bibfield  {author} {\bibinfo {author} {\bibfnamefont {V.}~\bibnamefont
  {Zaburdaev}}, \bibinfo {author} {\bibfnamefont {S.}~\bibnamefont {Denisov}},\
  and\ \bibinfo {author} {\bibfnamefont {J.}~\bibnamefont {Klafter}},\
  }\bibfield  {title} {\bibinfo {title} {L\'evy walks},\ }\href
  {https://doi.org/10.1103/RevModPhys.87.483} {\bibfield  {journal} {\bibinfo
  {journal} {Rev. Mod. Phys.}\ }\textbf {\bibinfo {volume} {87}},\ \bibinfo
  {pages} {483} (\bibinfo {year} {2015})}\BibitemShut {NoStop}%
\bibitem [{\citenamefont {Jakeman}\ and\ \citenamefont {Pusey}(1978)}]{Kdist}%
  \BibitemOpen
  \bibfield  {author} {\bibinfo {author} {\bibfnamefont {E.}~\bibnamefont
  {Jakeman}}\ and\ \bibinfo {author} {\bibfnamefont {P.~N.}\ \bibnamefont
  {Pusey}},\ }\bibfield  {title} {\bibinfo {title} {Significance of {$K$}
  distributions in scattering experiments},\ }\href
  {https://doi.org/10.1103/PhysRevLett.40.546} {\bibfield  {journal} {\bibinfo
  {journal} {Phys. Rev. Lett.}\ }\textbf {\bibinfo {volume} {40}},\ \bibinfo
  {pages} {546} (\bibinfo {year} {1978})}\BibitemShut {NoStop}%
\bibitem [{\citenamefont {Kotz}\ \emph {et~al.}(2001)\citenamefont {Kotz},
  \citenamefont {Kozubowski},\ and\ \citenamefont {Podg{\'o}rski}}]{kotz}%
  \BibitemOpen
  \bibfield  {author} {\bibinfo {author} {\bibfnamefont {S.}~\bibnamefont
  {Kotz}}, \bibinfo {author} {\bibfnamefont {T.~J.}\ \bibnamefont
  {Kozubowski}},\ and\ \bibinfo {author} {\bibfnamefont {K.}~\bibnamefont
  {Podg{\'o}rski}},\ }\href@noop {} {\emph {\bibinfo {title} {The Laplace
  Distribution and Generalizations}}}\ (\bibinfo  {publisher} {Springer},\
  \bibinfo {year} {2001})\BibitemShut {NoStop}%
\bibitem [{\citenamefont {Papapantoleon}(2008)}]{levyFinance}%
  \BibitemOpen
  \bibfield  {author} {\bibinfo {author} {\bibfnamefont {A.}~\bibnamefont
  {Papapantoleon}},\ }\href@noop {} {\bibinfo {title} {An introduction to
  {L\'evy} processes with applications in finance}},\ \bibinfo {howpublished}
  {arXiv:0804.0482} (\bibinfo {year} {2008}),\ \bibinfo {note} {{Vienna
  University of Technology}, lecture notes}\BibitemShut {NoStop}%
\bibitem [{\citenamefont {{\'S}l\k{e}zak}\ \emph {et~al.}(2019)\citenamefont
  {{\'S}l\k{e}zak}, \citenamefont {Metzler},\ and\ \citenamefont
  {Magdziarz}}]{codiff}%
  \BibitemOpen
  \bibfield  {author} {\bibinfo {author} {\bibfnamefont {J.}~\bibnamefont
  {{\'S}l\k{e}zak}}, \bibinfo {author} {\bibfnamefont {R.}~\bibnamefont
  {Metzler}},\ and\ \bibinfo {author} {\bibfnamefont {M.}~\bibnamefont
  {Magdziarz}},\ }\bibfield  {title} {\bibinfo {title} {Codifference can detect
  ergodicity breaking and non-{Gaussianity}},\ }\href
  {https://doi.org/10.1088/1367-2630/ab13f3} {\bibfield  {journal} {\bibinfo
  {journal} {New J. Phys.}\ }\textbf {\bibinfo {volume} {21}},\ \bibinfo
  {pages} {053008} (\bibinfo {year} {2019})}\BibitemShut {NoStop}%
\bibitem [{Note3()}]{Note3}%
  \BibitemOpen
  \bibinfo {note} {We note that this route will bring stronger results than
  using the Fourier transform (\ref {eq:gammaPDF}) directly, as it is
  unfortunately hard to inverse analytically. One can use the method of
  steepest descent to uncover the tail behaviour of $p_X$, but the result does
  not seem to be very practical, describing only a far away range not easily
  available in experiments.}\BibitemShut {Stop}%
\bibitem [{\citenamefont {Buchak}\ and\ \citenamefont {Sakhno}(2017)}]{buchak}%
  \BibitemOpen
  \bibfield  {author} {\bibinfo {author} {\bibfnamefont {K.}~\bibnamefont
  {Buchak}}\ and\ \bibinfo {author} {\bibfnamefont {L.}~\bibnamefont
  {Sakhno}},\ }\bibfield  {title} {\bibinfo {title} {Compositions of {Poisson}
  and {Gamma} processes},\ }\href {https://doi.org/10.15559/17-VMSTA79}
  {\bibfield  {journal} {\bibinfo  {journal} {Mod. Stoch.: Theory Appl.}\
  }\textbf {\bibinfo {volume} {4}},\ \bibinfo {pages} {161} (\bibinfo {year}
  {2017})}\BibitemShut {NoStop}%
\bibitem [{\citenamefont {Kurtuldu}\ \emph {et~al.}(2011)\citenamefont
  {Kurtuldu}, \citenamefont {Guasto}, \citenamefont {Johnson},\ and\
  \citenamefont {Gollub}}]{kurtuldu}%
  \BibitemOpen
  \bibfield  {author} {\bibinfo {author} {\bibfnamefont {H.}~\bibnamefont
  {Kurtuldu}}, \bibinfo {author} {\bibfnamefont {J.~S.}\ \bibnamefont
  {Guasto}}, \bibinfo {author} {\bibfnamefont {K.~A.}\ \bibnamefont
  {Johnson}},\ and\ \bibinfo {author} {\bibfnamefont {J.~P.}\ \bibnamefont
  {Gollub}},\ }\bibfield  {title} {\bibinfo {title} {Enhancement of biomixing
  by swimming algal cells in two-dimensional films},\ }\href
  {https://doi.org/10.1073/pnas.1107046108} {\bibfield  {journal} {\bibinfo
  {journal} {Proc. Natl. Acad. Sci. U.S.A.}\ }\textbf {\bibinfo {volume}
  {108}},\ \bibinfo {pages} {10391} (\bibinfo {year} {2011})}\BibitemShut
  {NoStop}%
\bibitem [{\citenamefont {Menzel}\ and\ \citenamefont
  {Goldenfeld}(2011)}]{menzel}%
  \BibitemOpen
  \bibfield  {author} {\bibinfo {author} {\bibfnamefont {A.~M.}\ \bibnamefont
  {Menzel}}\ and\ \bibinfo {author} {\bibfnamefont {N.}~\bibnamefont
  {Goldenfeld}},\ }\bibfield  {title} {\bibinfo {title} {Effect of {Coulombic}
  friction on spatial displacement statistics},\ }\href
  {https://doi.org/10.1103/PhysRevE.84.011122} {\bibfield  {journal} {\bibinfo
  {journal} {Phys. Rev. E}\ }\textbf {\bibinfo {volume} {84}},\ \bibinfo
  {pages} {011122} (\bibinfo {year} {2011})}\BibitemShut {NoStop}%
\bibitem [{\citenamefont {Sposini}\ \emph {et~al.}(2020)\citenamefont
  {Sposini}, \citenamefont {Grebenkov}, \citenamefont {Metzler}, \citenamefont
  {Oshanin},\ and\ \citenamefont {Seno}}]{diffDiffSposini}%
  \BibitemOpen
  \bibfield  {author} {\bibinfo {author} {\bibfnamefont {V.}~\bibnamefont
  {Sposini}}, \bibinfo {author} {\bibfnamefont {D.~S.}\ \bibnamefont
  {Grebenkov}}, \bibinfo {author} {\bibfnamefont {R.}~\bibnamefont {Metzler}},
  \bibinfo {author} {\bibfnamefont {G.}~\bibnamefont {Oshanin}},\ and\ \bibinfo
  {author} {\bibfnamefont {F.}~\bibnamefont {Seno}},\ }\bibfield  {title}
  {\bibinfo {title} {Universal spectral features of different classes of
  random-diffusivity processes},\ }\href
  {https://doi.org/10.1088/1367-2630/ab9200} {\bibfield  {journal} {\bibinfo
  {journal} {New J. Phys.}\ }\textbf {\bibinfo {volume} {22}},\ \bibinfo
  {pages} {063056} (\bibinfo {year} {2020})}\BibitemShut {NoStop}%
\end{thebibliography}%

\end{document}